\documentclass{webofc}
\usepackage[varg]{txfonts}   
\begin{document}
\title{Double-diffusive processes in stellar astrophysics}
%
%

\author{\firstname{Pascale} \lastname{Garaud}\inst{1}\fnsep\thanks{\email{pgaraud@ucsc.edu}}
}

\institute{Department of Applied Mathematics, Baskin School of Engineering, University of California Santa Cruz, 1156 High St, Santa Cruz CA 95064           }

\abstract{%
The past 20 years have witnessed a renewal of interest in the subject of double-diffusive processes in astrophysics, and their impact on stellar evolution. 
This lecture aims to summarize the state of the field as of early 2019, although the reader should bear in mind that it is rapidly evolving. 
An Annual Review of Fluid Mechanics article entitled {\it Double-diffusive convection at low Prandtl number} \cite{Garaud18}  
contains a reasonably comprehensive review of the topic, up to the summer of 2017. I focus here on presenting what I hope are clear derivations of some of the 
most important results with an astrophysical audience in mind, and discuss their implications for stellar evolution, both in an observational context, and in relation to previous work on the subject.}
\maketitle

\section{Introduction}
\label{sec:intro}

\begin{figure}[h]
\centering
\includegraphics[width=\textwidth]{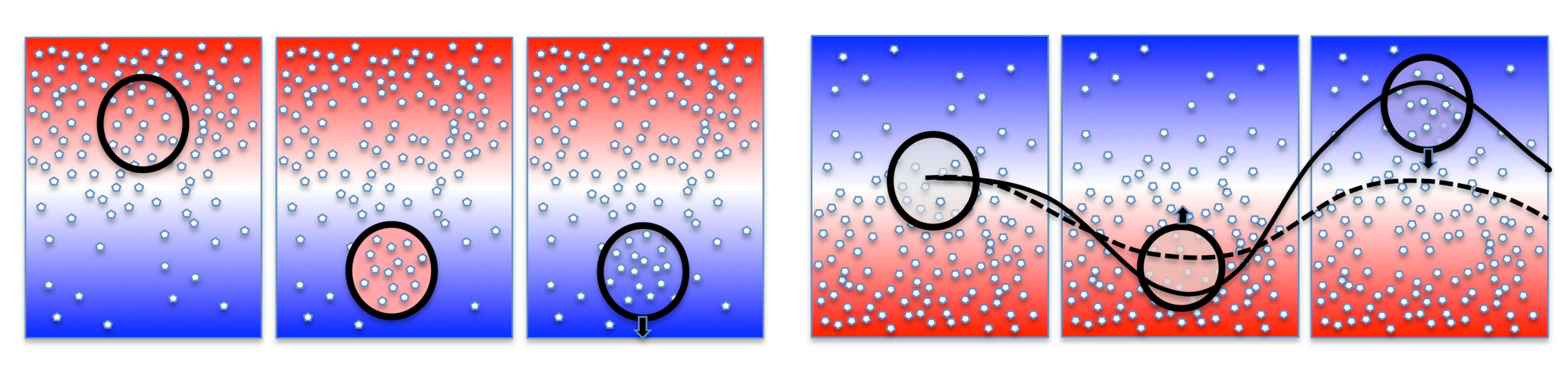}
\caption{Pictorial description of the two main double-diffusive instability mechanisms. The color represents potential temperature (or entropy) and the symbol density represents the concentration of a denser species. Left: the fingering instability. A parcel of fluid displaced downward rapidly equilibrates in temperature, but retains its high concentration. As a result it is denser than the surroundings and continues to sink. Right: the ODDC instability. In the absence of any background thermal stratification, a displaced parcel would just oscillate up and down (dashed line). Thermal diffusion at extreme phases of oscillation cools down or warms up the parcel, which amplifies the oscillation (solid line).}
\label{fig:basic}       
\end{figure}

Double-diffusive instabilities were first discovered in the context of physical oceanography by a group of scientists from the Woods Hole Oceanographic Institution \cite{Stommel1956,stern1960sfa}. 
Stern in particular realized that a region of the ocean with a stable density stratification can nevertheless undergo fluid instabilities because the water density depends on both temperature {\it and} salinity, which diffuse at different rates. To see why that may be the case, consider first a scenario in which temperature is stably stratified (with temperature increasing upward) and salinity is unstably stratified (with salinity increasing upward), as shown in Figure \ref{fig:basic}a. This could correspond for instance to the near-surface stratification of the tropical ocean. Overall, the density decreases upward because the stabilizing temperature stratification is stronger than the destabilizing salt stratification. As such, this fluid is stable to standard convection: a parcel of fluid forcibly moved downward is less dense than its surroundings, and in the absence of any diffusion, would experience a buoyancy force that pushes it back upwards. However, if the displaced parcel is small enough, its temperature rapidly adjusts to the surroundings, but its salinity does not (because salt diffuses 100 times slower than temperature). As a result, the small parcel becomes denser than its surroundings (because it has the same temperature but is saltier) and continues to sink. A similar argument can be made for small fluid parcels initially displaced upward, that continue to rise.   

This instability, called the {\it salt fingering} instability because of its tendency to produce long thin fingers of salty or fresh water, causes a net vertical transport of salt and heat downward. It is one of a few different kinds of thermohaline (or more generally thermocompositional) double-diffusive instabilities that have been discovered since the late 1950s. These include the {\it oscillatory double-diffusive instability} (ODDC), which occurs when the signs of the gradients of temperature and salinity are both reversed \cite{stern1960sfa,baines1969}, and the {\it intrusive instability}, which occurs when horizontal gradients of temperature and salinity are also present \cite{Holyer1983}.  

The idea of double-diffusive instabilities and double-diffusive mixing soon spread to the astrophysical community, thanks in part to the role of the Geophysical Fluid Dynamics summer program at the Woods Hole Oceanographic Institution, that was attended in the 1960s by astrophysicists such as Edward Spiegel, Shoji Kato, Jean-Paul Zahn, Douglas Gough, and others. In stellar interiors, the role of salt is replaced by any chemical species that has a higher mean molecular weight than its environment, as for instance He in comparison with H, or any heavier species (especially C, N, O, Si, Ni, Fe, etc. ) within a H-He mixture. In addition, whether the fluid is thermally stably stratified or not depends on the sign of the gradient of {\it potential temperature} (or equivalently, the sign of the entropy gradient) rather than on the sign of the temperature gradient alone. This accounts for adiabatic cooling or heating as the fluid parcel expands or shrinks to adapt to the local hydrostatic pressure. Fingering instabilities can take place in regions that are thermally stably stratified (i.e. radiative zones) with unstable compositional gradients. The process is usually referred to as {\it thermohaline convection} in stars. ODDC takes place in thermally unstable regions (as established by the Schwarzschild criterion) that have a sufficiently large compositional gradient to be Ledoux-stable. These regions are commonly called {\it semiconvective}, and their existence was first discussed in the late 1950s \cite{SchwarzschildHarm1958}. The relationship between semiconvection and ODDC was only later clarified by Kato \cite{Kato1966} and Spiegel \cite{Spiegel1969}.

The physics of the ODDC instability, which takes place in these semiconvective regions, are summarized in Figure \ref{fig:basic}b. If, as a thought-experiment, we ignore the unstable entropy stratification and only consider adiabatic perturbations, the fluid is stably stratified by the compositional gradient and supports internal gravity waves. As such, a displaced parcel would just bob up and down without change of amplitude owing to its own restoring buoyancy force. The temperature stratification can destabilize this oscillation, however. Thermal diffusion causes a sufficiently small parcel to adjust to the local background temperature, which then warms it up (lowering its density) when the parcel is low, and cools it down (increasing its density) when it is high. This effect enhances the buoyancy force, and causes a gradual amplification of the oscillation, thus driving the instability. 

Both fingering convection and ODDC/semiconvection were popularized in stellar astrophysics in the 1970s and 1980s. Compositional transport models were proposed for both instabilities \cite{Ulrich1972,Stevenson1977,kippenhahn80,Langer1983,Spruit1992} and many of them are still widely used in stellar evolution codes today. Until this last decade, however, none of these transport models had ever been tested, because laboratory experiments and numerical experiments at relevant parameters are prohibitively difficult to perform. As such, I would argue that any result ever obtained in the field of stellar astrophysics that strongly relies on double-diffusive mixing (including in particular fingering convection and semiconvection) should be taken with a very healthy dose of skepticism, and ought to be revisited in the light of the recent numerical and theoretical developments I will now proceed to review.

In what follows, I will first lay out for completeness and pedagogical purposes the simplest possible model setup in which to study double-diffusive instabilities, and will review its stability properties (Section \ref{sec:linear}). I will then describe in more depth recent results on mixing by the fingering instability (Section \ref{sec:fingering}) and the ODDC instability (Section \ref{sec:ODDC}), and discuss their implications. I conclude with a list of model caveats, and a summary of the more recent results obtained including the effects of shear, rotation, and magnetic fields in Section \ref{sec:ccl}. 

\section{Linear instability of double-diffusive systems}
\label{sec:linear}

\subsection{Model setup}
\label{sec:model}

In Section \ref{sec:intro}, I briefly described how the fingering and ODDC instabilities proceed, and argued that both cases involved the need for {\it small} parcels of fluids so thermal diffusion can equalize the temperature within the parcel with that of the background. The fact that double-diffusive instabilities have to be small-scale can therefore be seen as a defining property of double-diffusive convection. The lengthscale of basic double-diffusive fluid motions in both fingering and ODDC situations can usually be estimated as being of order $~10d$ \cite{stern1960sfa,schmitt1983}, where 
\begin{equation}
d =  \left( \frac{\kappa_T \nu}{|N_T^2|} \right)^{1/4} \simeq 10^3 {\rm cm} \left( \frac{\kappa_T}{ 10^7 {\rm cm}^2{\rm s}^{-1}} \right)^{1/4}  \left( \frac{\nu}{ 10 {\rm cm}^2{\rm s}^{-1}} \right)^{1/4}  \left( \frac{|N_T^2|}{10^{-4} {\rm s}^2}  \right)^{-1/4} ,
\label{eq:d}
\end{equation}
where $\kappa_T$ and $\nu$ are the thermal diffusivity and kinematic viscosity of the fluid, and $N_T^2 = - \delta g (\nabla_T - \nabla_{\rm ad}) / H_p$ is the square of the Brunt-V\"ais\"al\"a frequency associated with the entropy stratification only (i.e. ignoring compositional stratification). All the quantities used here are defined with standard notations in stellar astrophysics \cite{CoxGiuli1968,kippenhahnweigert}. Note that $N_T^2$ can be negative if the system is unstably stratified ($\nabla_T - \nabla_{\rm ad} > 0$). While  $\kappa_T$, $\nu$ and $N_T^2$ can vary a lot from star to star, and between the center and surface of a given star, the 1/4 power implies that the double-diffusive scale $10d$ itself does not vary much, taking values around 100 meters in non-degenerate regions of stars, down to about 1 centimeter in degenerate regions (assuming the fingering region extends there). We see that in all limits, $d$ is {\it much} smaller than the stellar radius.

The fact that these are small-scale diffusively driven instabilities is both a blessing and a curse from a numerical standpoint. It implies that all diffusive lengthscales must be fully resolved at all times, and that any attempt to resolve them in global hydrodynamical models of stars is futile. Double-diffusive instabilities should also {\it never} be studied with Large-Eddy Simulations or any other numerical technique involving subgrid scale parametrizations, because they rely on microscopic diffusivities to exist. On the plus side, this also implies that these instabilities are best modeled in small computational domains that can ignore complex stellar physics. In particular, the effects of curvature, compressibility and  the nonlinearity of the equation of state can almost always be neglected for such small-scale instabilities. Instead, they can adequately be studied with the Boussinesq approximation for weakly compressible gases \cite{SpiegelVeronis1960}.

\begin{figure}[h]
\centering
\sidecaption
\includegraphics[width=6cm]{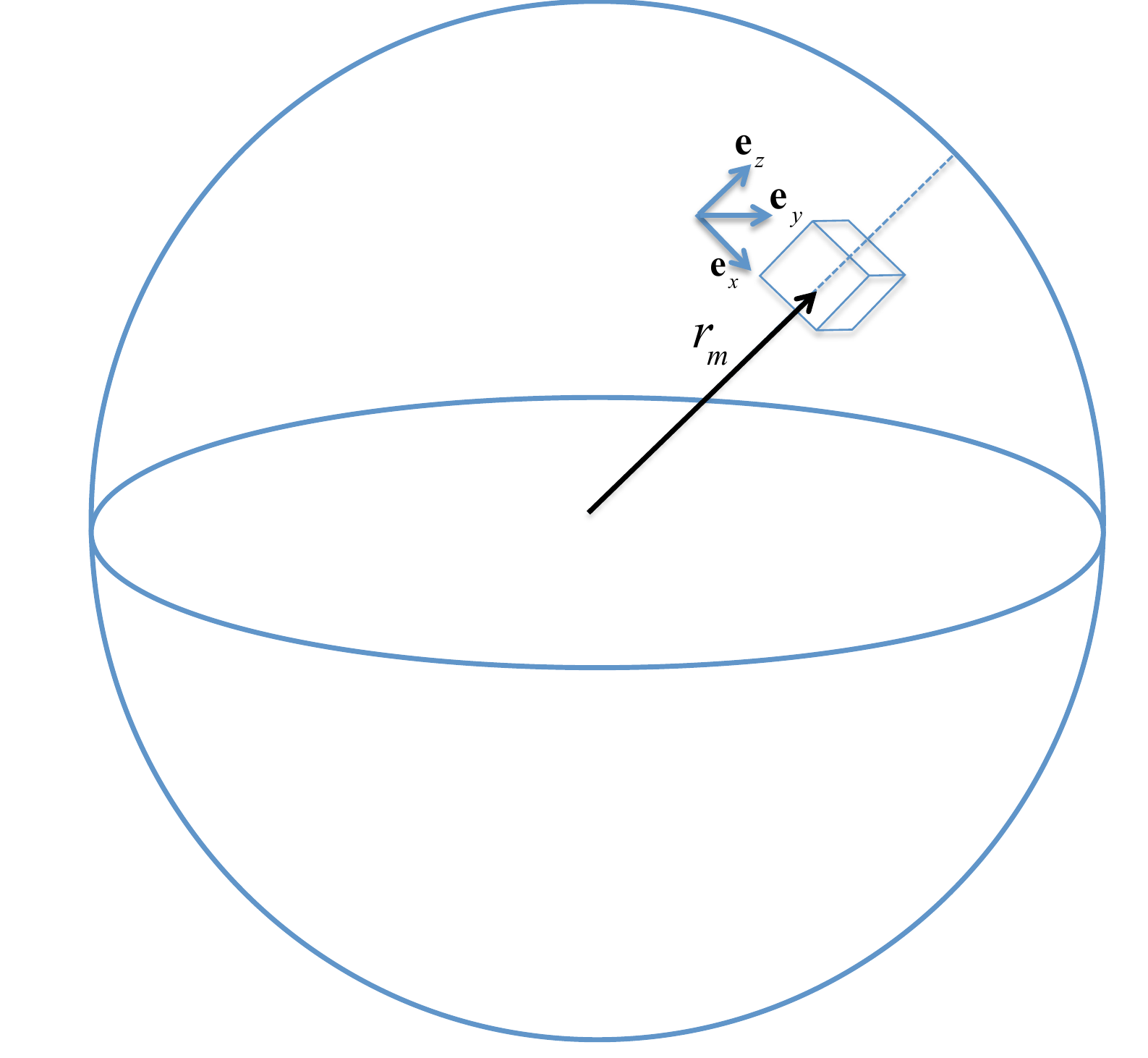}
\caption{Illustration of the model setup. A local Cartesian domain is modeled, near radius $r_m$, with the vertical defining the $z$-axis. In the absence of rotation the orientation of $x$ and $y$ are arbitrary. If the system is rotating, $x$ is aligned in the azimuthal direction.}
\label{fig:BoxStar}       
\end{figure}

To do so, let's consider from now on a small region of a star, located around a radius $r_m$, with mean temperature, density and pressure $T_m$, $\rho_m$ and $p_m$. This region is modeled in a local Cartesian reference frame with coordinates $(x,y,z)$ (see Figure \ref{fig:BoxStar}), where the local gravity defines the vertical axis: ${\bf g} = -g {\bf e}_z$. As such, we can define $z = r - r_m$, for instance. Following the Boussinesq assumptions, the computational domain size must be much smaller than any of the local pressure, density, and temperature scaleheights. This guarantees that the background temperature profile $T_0(z)$ can be linearized close to $r_m$, so that $T_0(z) = T_m + z T_{0z}$ where $T_{0z}$ is the locally constant temperature gradient\footnote{Note that all the gradients here (except for $\nabla_T$, $\nabla_{\rm ad}$ and $\nabla_{\mu}$) are taken with respect to position, not with respect to mass or pressure as it is commonly done in stellar evolution.}. Similarly, we assume that the background compositional field profile $C_0(z)$ (which could for instance taken to be the mean molecular weight, or the mass fraction of a particular chemical species), can be written as $C_0(z) = C_m + z C_{0z}$ where $C_m$ is the mean value in the domain, and $C_{0z}$ is the local gradient. Fluid motions, written as ${\bf u} = (u,v,w)$, as well as fluctuations of temperature $T$, density $\rho$ and composition $C$, evolve around that background state following the Spiegel-Veronis-Boussinesq equations \cite{SpiegelVeronis1960}, namely
\begin{eqnarray}
&& \rho_m \left( \frac{\partial {\bf u}}{\partial t} + {\bf u} \cdot \nabla {\bf u} \right) = - \nabla p - \rho g {\bf e}_z + \rho_m \nu  \nabla^2 {\bf u} ,  \label{eq:momdim} \\
&&\frac{\partial T}{\partial t} + {\bf u} \cdot \nabla T + w ( T_{0z} - T_{{\rm ad},z}) = \kappa_T \nabla^2 T , \label{eq:tempdim} \\
&& \frac{\partial C}{\partial t} + {\bf u} \cdot \nabla C + w  C_{0z}  = \kappa_C \nabla^2 C , \label{eq:compdim} \\
&& \frac{\rho}{\rho_m} = - \alpha T + \beta C \mbox{  and  } \nabla \cdot {\bf u} = 0,  \label{eq:eosdim}
\end{eqnarray}
where $\kappa_C$ is the compositional diffusivity, and the coefficients of the linearized equation of state are $\alpha = - \rho_m^{-1} (\partial \rho/\partial T)_{C_m,p_m}$ and $\beta = \rho_m^{-1} (\partial \rho / \partial C)_{T_m,p_m}$ taken at $r_m$. We see that the temperature gradient does not appear alone, but is instead combined with the adiabatic temperature gradient $T_{{\rm ad},z} = - g/c_p$ (where $c_p$ is the specific heat at constant pressure) to form the potential temperature gradient $(T_{0z} - T_{{\rm ad},z})$, which is the relevant one in stellar interiors as discussed earlier. With this model setup, we can conveniently require all perturbations ${\bf u}$, $T$, $p$ and $C$ to be triply periodic in the computational domain, which enables us to study the dynamics of double-diffusive instabilities far from any solid boundaries (that would be unphysical in a star). The set of equations (\ref{eq:momdim})--(\ref{eq:eosdim}) is commonly used to study double-diffusive convection with given background temperature and composition gradients far from boundaries \cite{baines1969,shen1995,radko2013double}.

As an illustrative side-note, these equations can very easily be used {\it as is} to look into the stability of standard (non-diffusive) multicomponent convection, and recover the well-known Ledoux criterion for instability. This is a useful warm-up exercise for the more complicated double-diffusive case. Let's first linearize the equations, substitute the equation of state into the momentum equation, and ignore all diffusive terms. We obtain (in addition to the incompressibility condition)
\begin{eqnarray}
&&\frac{\partial {\bf u}}{\partial t}  = -  \rho_m^{-1} \nabla p +  (\alpha g T - \beta g C) {\bf e}_z ,  \label{eq:momdimlin} \\
&&\frac{\partial T}{\partial t} + w ( T_{0z} - T_{{\rm ad},z}) = 0  \mbox{  and }  \frac{\partial C}{\partial t} +  w  C_{0z}  = 0.  \label{eq:compdimlin} 
\end{eqnarray}
Assuming an ansatz of the form $q(x,y,z,t) = \tilde q  \exp( i l x + i m y + i k z + \lambda t)$ for each of the variables, yields
\begin{eqnarray}
&& \lambda \tilde u  = -  i \rho_m^{-1} l \tilde p  \mbox{  ,  }  \lambda \tilde v  = -  i \rho_m^{-1} m \tilde p  \mbox{  ,  }  \lambda \tilde w  = -  i \rho_m^{-1} k \tilde p +  (\alpha g \tilde T - \beta g \tilde C)  \label{eq:momwdimlin} \\
&& \lambda \tilde T + \tilde w ( T_{0z} - T_{{\rm ad},z}) = 0  \mbox{   ,    } \lambda \tilde C+  \tilde w  C_{0z}  = 0 \mbox{   and  } l \tilde u + m \tilde v + k \tilde w = 0 .  \label{eq:divudimlin} 
\end{eqnarray}
Assuming  without loss of generality that $l \ne 0$ (otherwise switch the $x$ and $y$ variables), we can eliminate pressure from the first two equations to get $\tilde v = (m/l) \tilde {u}$, and so from incompressibility $\tilde u = -  l  k \tilde w / (l^2 + m^2  )$.  We can then eliminate pressure between the $\tilde u$ and $\tilde w$ equation, and substitute $\tilde u$, $\tilde T$ and $\tilde C$ in terms of $\tilde w$ in the result. Once completed, the process results in an equation for $\lambda$ only, because (as required from linear theory) the amplitude $\tilde w$ cancels out of the equation. We then have
\begin{equation}
\lambda^2 = - \frac{ l^2  + m^2 }{l^2 + m^2 + k^2} \left[ \alpha g (T_{0z} - T_{{\rm ad},z} ) - \beta g C_{0z} \right] =  - \frac{ l^2  + m^2 }{l^2 + m^2 + k^2} N^2 
\end{equation}
where $N^2 = N_T^2 + N_C^2$ and $N_C^2 = - \beta  g C_{0z}$ is the contribution to the Brunt-V\"ais\"al\"a frequency due to the compositional stratification. With the form of the ansatz selected above, instability occurs when solutions of this equation exist for which the real part of $\lambda$ is strictly positive. This is the case when $ \alpha g (T_{0z} - T_{{\rm ad},z} ) - \beta g C_{0z}  = N_T^2 + N_C^2 < 0$, which is equivalent to the Ledoux criterion as required. 

\subsection{Linear stability properties of homogeneous double-diffusive systems} 
\label{sec:lineargeneral}

When all the diffusion terms are taken into account to model double-diffusive instabilities, it is significantly more convenient to non-dimensionalize the equations first. Following standard practices in the field (see \cite{radko2013double} for instance), the unit lengthscale from here on is $d$, the unit timescale is the thermal diffusion timescale across $d$, namely $d^2/\kappa_T$, the unit velocity is $\kappa_T / d$, the unit temperature is $d |T_{0z} - T_{{\rm ad},z} |$ and the unit composition is $ \alpha d | T_{0z} - T_{{\rm ad},z} | / \beta $.  Note that we use $|T_{0z} - T_{{\rm ad},z} |$ instead of $T_{0z} - T_{{\rm ad},z}$ to ensure that all units are positive.  Letting, e.g. $ \nabla  = d^{-1} \hat \nabla$, $t = (d^2/\kappa_T) \hat t$, $(u,v,w) = (\kappa_T / d) (\hat u,\hat v, \hat w)$, $T = d |T_{0z} - T_{{\rm ad},z} | \hat T$, and so forth, we obtain the non-dimensional equations 
\begin{eqnarray}
&&\frac{1}{\rm Pr} \left( \frac{\partial {\bf \hat u}}{\partial \hat t}  + \hat {\bf u} \cdot \hat \nabla \hat { \bf u} \right) = -  \hat \nabla \hat p +  (\hat T - \hat C) {\bf e}_z + \hat \nabla^2  \hat {\bf u} \mbox{      ,      } \hat \nabla \cdot \hat {\bf u } = 0  \label{eq:momnondim} \\
&&\frac{\partial \hat T}{\partial \hat t} + \hat {\bf u} \cdot \hat \nabla \hat T \pm \hat w  =  \hat \nabla^2  \hat T ,  \label{eq:tempnondim} \\
&& \frac{\partial \hat C}{\partial \hat t} + \hat {\bf u} \cdot \hat \nabla \hat C \pm R_0^{-1}  \hat w   = \tau  \hat \nabla^2  \hat C, 
  \label{eq:compnondim} 
\end{eqnarray}
where the $+$ sign applies in (\ref{eq:tempnondim}) and (\ref{eq:compnondim}) for fingering convection, and the $-$ sign applies for ODDC. It is rather remarkable to see that the only {\it mathematical} difference between these two instabilities is the sign difference in the $\hat w$ term in these equations.
Since the notation is somewhat heavy as is, in what follows the hats on the non-dimensional independent variables $x$, $y$, $z$, $t$ and on the operator $\nabla$ will be dropped, but those on the non-dimensional dependent variables will be kept for clarity. Three nondimensional parameters have appeared in these equations, namely the Prandtl number ${\rm Pr} = \nu / \kappa_T$, the diffusivity ratio $\tau = \kappa_C / \kappa_T$, and the density ratio 
\begin{equation}
R_0 = \frac{\alpha |T_{0z} - T_{{\rm ad},z} |}{\beta |C_{0z}|} = \frac{ | N_T^2 | }{|N_C^2|} = \frac{\delta (\nabla_T - \nabla_{\rm ad} ) }{\phi \nabla_\mu} 
\end{equation}
where the last expression has been written assuming $C$ is the mean molecular $\mu$, and using standard definitions in stellar astrophysics \cite{CoxGiuli1968,kippenhahnweigert} for $\delta$, $\phi$ and $\nabla_T-\nabla_{\rm ad}$ and $\nabla_\mu$. 

The Prandtl number and diffusivity ratio depend principally on the nature of the fluid considered, and their dependence on thermodynamical quantities such as temperature, pressure, composition, etc. is neglected as part of the Boussinesq approximation. In salt water for instance ${\rm Pr} = O(10)$ and $\tau = O(0.01)$. In degenerate regions of White Dwarf (WD) stars ${\rm Pr} = O(0.01)$ and $\tau = O(0.001)$. In non-degenerate regions of Main Sequence (MS) and Red Giant Branch (RGB) stars ${\rm Pr} = O(10^{-6})$ and $\tau  = O(10^{-7})$. Note that these are just indicative order of magnitudes, and the user should compute these numbers more accurately for the application of their choice. Examples of such computations are given in \cite{Garaudal2015} for WDs, MS stars and RGB stars. 

The density ratio on the other hand measures the relative strengths of the potential temperature and compositional stratifications. With this definition, a Ledoux-neutral stratification has $R_0 = 1$. As we will now demonstrate, the density ratio is the most important parameter controlling the dynamics of both fingering convection and ODDC.  

The linear stability of a doubly-stratified system satisfying equations (\ref{eq:momnondim})--(\ref{eq:compnondim}) can be computed more-or-less as we did before for the non-diffusive case. We first assume a similar ansatz (bearing in mind everything is now non-dimensional), $\hat q(x,y,z,t) = \tilde q  \exp( i \hat l x + i \hat m y + i \hat k z + \hat \lambda t)$. Substituting that ansatz into the linearized equations, and successively eliminating $\tilde p$, $\tilde v$, $\tilde u$, $\tilde T$, $\tilde C$ and finally $\tilde w$, results in the following {\it cubic} for the growth rate $\hat \lambda$ \cite{baines1969}: 
\begin{eqnarray}
\hat \lambda^3 + ({\rm Pr} + \tau + 1) | \hat {\bf  k} |^2 \hat \lambda^2 &+& \left[ ({\rm Pr} + \tau + {\rm Pr} \tau) | \hat {\bf  k} |^4   \pm {\rm Pr}  \frac{\hat k_h^2}{| \hat {\bf  k} |^2}  (1-R_0^{-1}) \right] \hat \lambda 
 \nonumber \\ &+& {\rm Pr} \tau | \hat {\bf  k} |^6 \pm {\rm Pr} \hat k_h^2  (\tau - R_0^{-1} ) = 0,  \label{eq:DDCcubic}
\end{eqnarray}
where $\hat {\bf k} = (\hat l,\hat m,\hat k)$ is the total wavevector, $\hat k_h  = \sqrt{\hat l^2 + \hat m^2}$ is the horizontal wavenumber, and where the $+$ sign applies for fingering convection, and the $-$ sign applies for ODDC. As in the non-diffusive case, the existence of solutions of this cubic with positive real part demonstrates instability. 

Despite the strong similarity of the cubics obtained in the fingering and ODDC cases, respectively, the solutions can be quite different. Using standard properties of cubics, it can be shown that fastest-growing modes in the fingering regime (with $R_0 > 1$) are real while they are complex in the ODDC regime (with $R_0<1$). For this reason, we now discuss the two cases separately. 

\subsubsection{Linear stability properties of the fingering regime}
\label{sec:linearfing}

To study the fingering regime, we consider (\ref{eq:DDCcubic}) with the $+$ sign. Solutions are real so the condition for marginal stability can simply be written as $\hat \lambda = 0$, which implies that the critical density ratio for instability for a given mode with wavenumber $\hat {\bf k}$ is 
\begin{equation}
R_{0,c}(\hat {\bf k})  = \left[ \tau + \frac{\tau | \hat {\bf  k} |^6 }{  \hat k_h^2  } \right] ^{-1} .
\end{equation}
The largest possible $R_{0,c}(\hat {\bf k})$ is obtained when $\hat {\bf k} = 0$, and is equal to $R_{fc} = \tau^{-1}$. We have therefore demonstrated that the fingering instability exists when $ 1 \le R_0 \le \tau^{-1}$. The system is unstable to standard overturning according to the Ledoux criterion when $R_0 < 1$, and stable when $R_0 > \tau^{-1}$. These findings are not too surprising: in this regime, $R_0$ can be interpreted as the ratio of the stabilizing stratification to the destabilizing one, so the larger $R_0$ is, the more stable the system is. We also see that the range of instability depends only on the value of the diffusivity ratio $\tau = \kappa_C / \kappa_T$. If temperature and composition diffuse at the same rate, then fingering is not possible. On the other hand, the smaller $\tau$ is, the larger the range of density ratios for which instability can exist. Because $\tau$ is usually asymptotically small in stars, this implies that even a very small inverse $\mu$ gradient can destabilize a radiative zone. 

With a little work, it can be shown \cite{radko2013double} that the modes with $\hat k = 0$ are always the most rapidly-growing ones, and are often called {\it elevator modes} for obvious reasons: the flow within these elevator modes is strictly vertical, and all the quantities of interest are invariant with $z$. In the limit of low Prandtl number and low diffusivity ratio appropriate for stellar interiors, and close to marginal stability (so $R_0 \rightarrow \tau^{-1}$ and both $\hat \lambda$ and $| \hat {\bf  k} |$ are small), we can roughly estimate $\hat \lambda$ for these elevator modes by neglecting the cubic and quadratic terms in (\ref{eq:DDCcubic}) (see \cite{Ulrich1972}):
\begin{equation}
\hat \lambda  \simeq    \frac{R_0^{-1}  - \tau }{ 1-R_0^{-1}}  \hat k_h^2   .
\label{eq:lambdamargfing}
\end{equation}
In reality, however, the horizontal wavenumber $\hat k_h$ of the  {\it fastest growing} elevator mode also varies with $R_0$ close to marginal stability, so this expression does not directly tell us what $\hat \lambda$ is for that mode. To do so, one needs to perform a formal asymptotic expansion of (\ref{eq:DDCcubic}) in the limit of low Pr and $\tau$. This was done by Brown et al. \cite{Brownal2013}, who derived a number of approximate analytical solutions for the growth rate $\hat \lambda$ of the fastest-growing modes as a function of $R_0$, Pr and $\tau$, in the stellar limit (${\rm Pr} , \tau \ll 1$). The reader is referred to their Appendix B for details. Among other results, they find that for $R_0$ not too close to 1 nor too close to marginal stability (i.e. $ 1 \ll R_0 \ll \tau^{-1}$, which seems to be a reasonable limit for stars), 
\begin{equation}
\hat \lambda \simeq \sqrt{ {\rm Pr} \frac{1-\tau}{R_0 - 1} } \simeq  \sqrt{  \frac{{\rm Pr}}{R_0} } ,
\label{eq:lambdaapprox}
\end{equation}
while the horizontal wavenumber of the fastest-growing mode is always of order unity, with 
\begin{equation}
 \hat k_h^2  \simeq \sqrt{\frac{\rm Pr}{{\rm Pr} + \tau} } .
\end{equation}
Dimensionally, this implies that the wavelength of these fingers is of order $2\pi d$ (corresponding to one up-flowing and one down-flowing finger), where $d$ was estimated in (\ref{eq:d}), and that their growth rate is 
\begin{equation}
\lambda \simeq \sqrt{  \frac{{\rm Pr}}{R_0} } \frac{\kappa_T}{d^2} \simeq \frac{N_T}{\sqrt{R_0}} \simeq 10^{-4} {\rm s}^{-1} \left( \frac{ |N_T|^2}{10^{-4}{\rm s}^{-2}} \right)^{1/2} \left(\frac{R_0}{10^4} \right)^{-1/2} , 
\end{equation} 
 which is usually substantially smaller than the buoyancy frequency unless $R_0$ is close to 1.

\subsubsection{Linear stability properties of the ODDC regime}
\label{sec:linearODDC}

To determine the stability properties of ODDC, we study the solutions of (\ref{eq:DDCcubic}) with the $-$ sign. In that case, it is common to use the inverse density ratio $R_0^{-1}$ as the relevant parameter; $R_0^{-1}$ is now the ratio of the stabilizing compositional stratification to the destabilizing potential temperature stratification, with $R_0^{-1} < 1$ being Ledoux-unstable (hence convective in the standard sense), while $R_0^{-1} > 1$ is Ledoux-stable. Increasing $R_0^{-1}$ correspond to increasingly stable systems. 
The criterion for marginal stability to ODDC is a little more difficult to derive than in the fingering case, because $\hat \lambda$ is not real. So we let $\hat \lambda = \hat \lambda_R + i \hat \lambda_I$ and first rewrite (\ref{eq:DDCcubic}) in terms of $\hat \lambda_R$ only. The same argument as in the fingering case can be used to show that the elevator modes are still the fastest-growing modes, so focussing on these only, we let $\hat k = 0$. 
With a little work (see the Appendix A.1 in \cite{Mirouh2012}), we obtain
 \begin{eqnarray}
8 \hat \lambda_R^3 + 8  \hat k_h^2  \hat \lambda_R^2 ( \tau+ \Pr + 1 ) + 2 \hat \lambda_R \left[     \hat k_h^4  \left( \tau  + \Pr\tau + \Pr
  +\left( \tau+ \Pr + 1
\right)^2   \right) +  \Pr (R_0^{-1} -1 )  \right] \nonumber \\ 
+   \hat k_h^6  (\tau + \Pr)(\tau+1) ( \Pr + 1)  +   \hat k_h^2  \Pr (R_0^{-1} \left( \tau+ \Pr \right) - \left(
     \Pr + 1 \right) ) 
=0 \mbox{   .}
\label{eq:lmax1}
\end{eqnarray}
Setting $\hat \lambda_R = 0$, we then find that the critical inverse density ratio for instability is achieved for $\hat k_h \rightarrow 0$ (as in the fingering case). This shows that ODDC can only occur for 
\begin{equation}
1 \le R_0^{-1} \le  \frac{ {\rm Pr} + 1}{{\rm Pr} + \tau} .
\end{equation}
Since ${\rm Pr} \gg \tau$ in geophysical environments, $({\rm Pr} + 1)  /( {\rm Pr} + \tau) \simeq 1$. As a result, the range of $R_0^{-1}$ for which a doubly stratified system  on Earth may be linearly unstable to ODDC is very small and almost never naturally realized. By contrast, both ${\rm Pr}$ and $\tau$ are very small in stars, so the range of instability to ODDC can be very substantial.

Applying the same arguments as in the fingering case, we can roughly estimate $\hat \lambda_R$ close to marginal stability (i.e. when $R_0^{-1} \rightarrow ({\rm Pr} + 1)  /( {\rm Pr} + \tau)$ where both $\hat \lambda_R$ and $\hat k_h$ are small) by neglecting the quadratic and cubic terms in (\ref{eq:lmax1}). 
In the stellar limit where ${\rm Pr} , \tau \ll 1$, we then get \cite{Langer1983},  
\begin{equation} 
 \hat \lambda_R  
 \simeq  \frac{1}{2} \hat k_h^2 \frac{ \left(
     \Pr + 1 \right) - R_0^{-1} \left( \tau+ \Pr \right)  }{R_0^{-1} -1} \simeq \frac{  \hat k_h^2  }{2R_0^{-1}}  \label{eq:langerlambda}
\end{equation}
for sufficiently large $R_0^{-1}$ (but not too close to marginal stability).  Again, this should be viewed as a rough estimate, and does not provide information on the wavenumber $\hat k_h$ for the fastest-growing modes. A more formal asymptotic analysis needs to be carried out to get this information. Details can be found in Appendix A.2 of Mirouh et al. \cite{Mirouh2012}. The solutions have an analytical but non-trivial dependence on $R_0^{-1}$, whose expression is not particularly illuminating. At best, 
one can state that $ \hat \lambda_R $ is proportional\footnote{This is indeed consistent with (\ref{eq:langerlambda}) since that equation is only valid in the limit where $R_0^{-1} \rightarrow ({\rm Pr} + \tau)^{-1}$.} to ${\rm Pr}$, so from a dimensional point of view, we find that $\lambda_R \propto \sqrt{\rm Pr} N_T$. Finally, it is also relatively easy to show that for ${\rm Pr} , \tau \ll 1$, the imaginary part of $\hat \lambda$ is well-approximated by the local buoyancy frequency including the compositional stratification, i.e. 
\begin{equation}
\hat \lambda_I \simeq \sqrt{{\rm Pr} (R_0^{-1} - 1)}  \rightarrow  \lambda_I \simeq \sqrt{ N_C^2 + N_T^2} . 
\end{equation} 

\subsubsection{Summary for linear stability of double-diffusive systems}

\begin{figure}[h]
\centering
\includegraphics[width=0.9\textwidth]{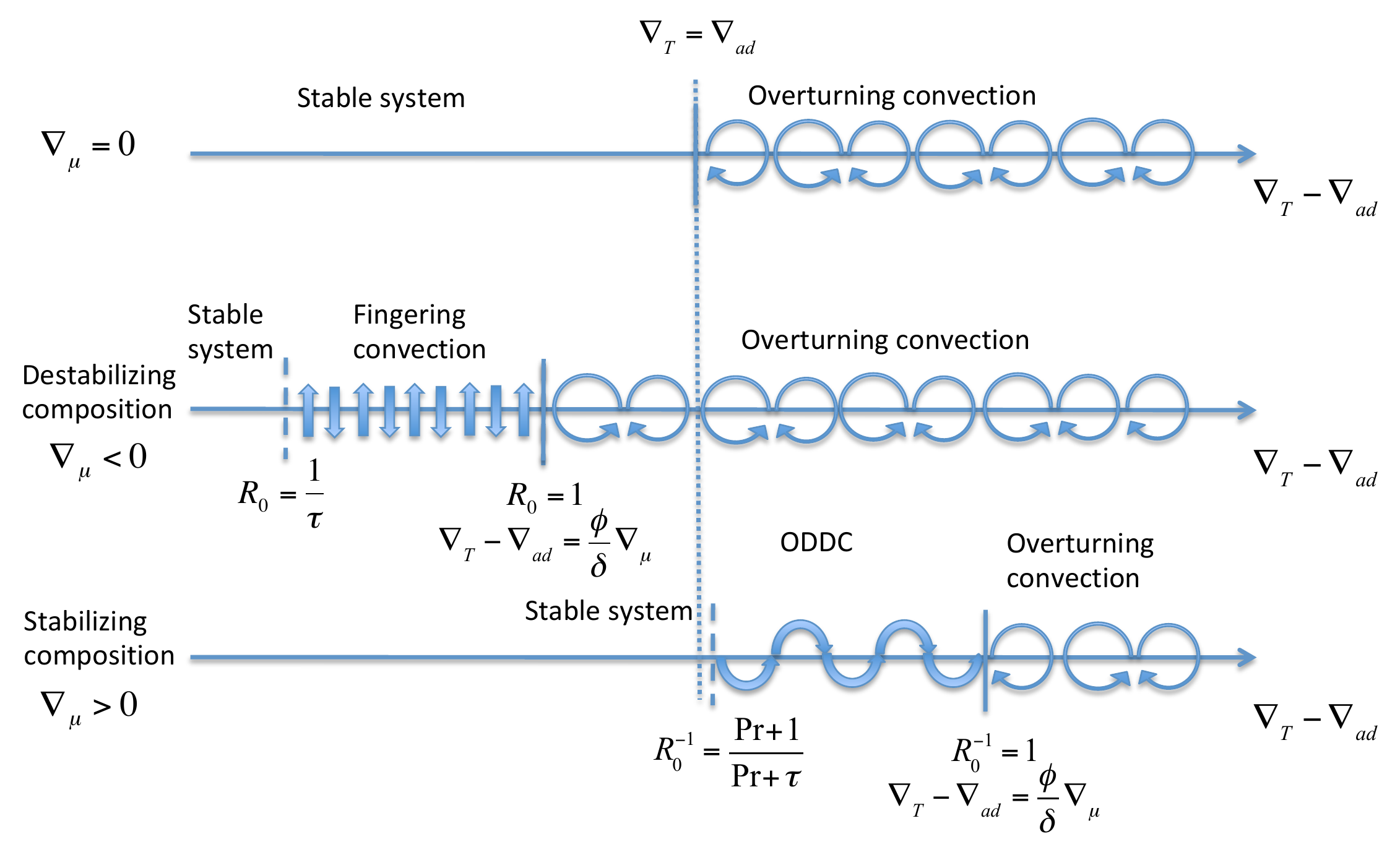}
\caption{Illustration of the various regimes of instability as a function of $R_0$ (or equivalently, $\nabla_T - \nabla_{\rm ad}$ and $\nabla_\mu$), for fingering convection and ODDC.  }
\label{fig:linearstab}       
\end{figure}

An illustrative summary of the findings from linear stability analysis is presented in Figure \ref{fig:linearstab}. In the absence of compositional stratification, the threshold between stability and instability to overturning convection is simply given by the Schwarzschild criterion, namely $R_0 = 0$ or equivalently $ \nabla_T = \nabla_{\rm ad}$. With compositional stratification, instability to overturning convection is set by the Ledoux criterion, $R_0 = 1$ or equivalently $\delta (\nabla_T - \nabla_{\rm ad}) = \phi \nabla_\mu$. We see that double-diffusive effects destabilize regions that are Ledoux-stable. In the presence of a destabilizing compositional gradient, fingering instabilities are excited in a wide region of parameter space that would normally be purely radiative. In the presence of a stabilizing compositional gradient, ODDC is excited almost everywhere in the region of parameter space bound by the Schwarzschild criterion on one side, and the Ledoux criterion on the other. In all cases, the fastest-growing modes of instability are elevator modes, i.e. vertically-invariant structures, whose horizontal size is $O(10 d)$.

While this analysis has been performed for an idealized system without any added physics, we are of course also interested in knowing whether rotation, shear, magnetic fields, and other dynamics, could affect these results. 
 Interestingly, neither rotation, magnetic fields, nor shear seem to affect the range of instability of fingering convection \cite{SenguptaGaraud2018,HarringtonGaraud2019,Garaudal2019}, because there always appears to be a way for the elevator modes to develop. Horizontal gradients, however, can extend the unstable fingering range significantly  \cite{Holyer1983,Medrano2014}. Gravitational settling may induce as similar effect if the settling velocity of individual species approaches the value $\kappa_T / d$ \cite{Alsinanal2017,Realial2017}, though this has not yet been studied in the astrophysical context. To my knowledge, there is no evidence for subcritical instabilities (i.e. instabilities that only develop given specific finite-amplitude initial conditions instead of from infinitesimal perturbations) in fingering-prone stratified fluids. 
 
 The impact of added dynamics on ODDC has not yet been studied as exhaustively as in the fingering case. Rotation does not affect the unstable range \cite{MollGaraud2017}, but shear does (at least in geophysics), through the newly discovered thermo-shear instability \cite{Radko2016}. Whether that instability is relevant at astrophysical parameters has not yet been established. The effect of magnetic fields on the linear stability of ODDC was investigated by Stevenson \cite{Stevenson1979}, who found that it does not affect the overall unstable range, but can change the nature of the unstable modes that are present. Finally, by contrast with the fingering case, ODDC is known to have subcritical branches of instability, at least in the geophysically-relevant region of parameter space \cite{Proctor1981}. This remains to be studied more extensively at low Prandtl number \cite{Mollal2017}. 
 
 Finally, note that all of these results have been stated in the context of the model setup described in Section \ref{sec:model}. The presence of physical boundaries can strongly affect the instability range, especially when the size of the domain is not very large compared with the natural double-diffusive scale $10d$. For bounded domains the linear stability problem needs to be investigated on a case-by-case basis, and depends on the domain size, shape, applied boundary conditions, and any added physics. 

\subsection{Where is fingering convection likely to occur in stars?}
\label{sec:wherefing}

The linear stability analysis presented above clearly shows that fingering instabilities can take place in radiative stellar regions which are Ledoux-stable, but subject to an unstable composition gradient, even if the latter is very weak. Such gradients can arise  in stellar interiors in a number of scenarios, all of which have important observational implications. 

A common mechanism to drive fingering convection involves the accretion of high $\mu$ material at the surface of a star, either from infalling planets or planetary debris, or from material transferred from a more evolved companion star. Planetary infall for instance can increase the apparent metallicity of the host star; this has been invoked in turn as a possible explanation to the planet-metallicity correlation \cite{FischerValenti2005}, and to the existence of metal-rich WDs \cite{Zuckerman2003,Wyatt2014}. It was rapidly realized however that this inverse $\mu$-gradient would create a region below the surface of the star that is unstable to fingering convection \cite{vauclair2004mfa}, which would then significantly shorten the residence time of high-$\mu$ chemical species near the surface, with implications for derived accretion rates and/or surface abundances of light elements in both MS stars \cite{Garaud2011,TheadoVauclair2012}  and WDs \cite{Deal2013,Wachlinal2017,BauerBildsten2018}. Similar arguments have been put forward in the context of binary mass transfer, with possible implications for carbon-enhanced metal poor stars, for instance \cite{StothersSimon1969,Stancliffe2007,StancliffeGlebbeek2008,Thompsonal2008} and nova outbursts \cite{MarksSarna1998}. 

Fingering regions can also appear deep within a star from off-center nuclear burning in RGB and AGB stars. In RGB stars, $^3$He burning at the outskirts of the H-burning layer can produce lower-$\mu$ material, causing that region and the regions above it to become fingering-unstable \cite{Ulrich1971,Ulrich1972}. Instabilities associated with this inverse $\mu$ gradient have been proposed as a possible solution to the peculiar surface abundances of RGB stars beyond the luminosity bump \citep{Eggletonal2006,CharbonnelZahn2007}. Fingering is also thought to occur below the carbon-burning shell of super-AGB stars \cite{Siess2009}, and to affect the carbon flame propagation by mixing the available fuel. It may also impact the properties of hybrid C/O/Ne WDs, whose electron-to-baryon fraction becomes unstable to convection and fingering convection as the star cools \cite{Brooks2017,SchwabGaraud2018}. 

 Finally, the microscopic segregation of various chemical species by gravitational settling and radiative levitation can produce chemically peculiar layers below the surface of intermediate-mass stars, that are prone to fingering instabilities \cite{Theadoal2009,Zemskova2014,Dealal2016}. Not accounting for mixing by fingering, the accumulation of such elements can be so strong that it leads to the formation of intermediate convection zones. Fingering instabilities, however, may develop much earlier and prevent the accumulation from reaching such extreme levels \cite{Zemskova2014}.

\subsection{Where is ODDC/semiconvection likely to occur is stars?} 
\label{sec:whereODDC}

We also learned using linear theory that ODDC develops in regions that are traditionally called {\it semiconvective}, i.e. regions which are unstable according to the Schwarzschild criterion $(\nabla_T - \nabla_{\rm ad} > 0)$, but stable according to the Ledoux criterion ($\phi \nabla_\mu > \delta(\nabla_T - \nabla_{\rm ad})$). Semiconvective regions are often found adjacent to convective cores and are caused by the development of stabilizing $\mu$ gradients. There are two stellar mass ranges in which this occurs, and the reason for the presence of the $\mu$ gradient differs in the two cases. For reviews on the topic, see for instance \cite{Spiegel1969,Langer1983,Spruit1992,Merryfield1995,Shibahashi2009}.

\begin{figure}[h]
\centering
\includegraphics[width=0.8\textwidth]{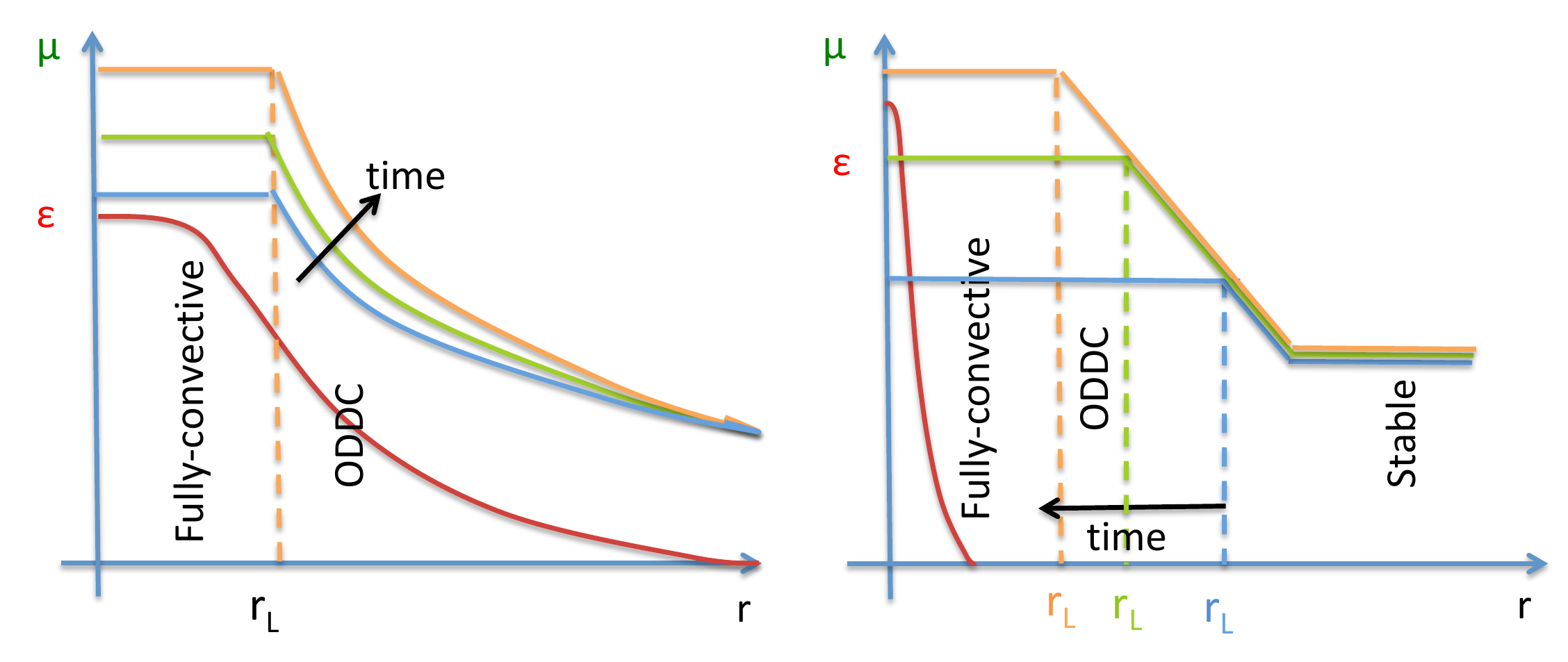}
\caption{Illustration of the two mechanisms for forming semiconvective regions at the edge of a convective core. The red line represents the nuclear energy generation rate, and the other colored lines represent the $\mu(r)$ profile, with blue at early times, then green, then orange at latest time. The vertical dashed line marks the edge of the convective core according to the Ledoux criterion. Left: In intermediate mass stars, nuclear reactions extend outside the core, and their rate depends on temperature. A $\mu$ gradient thus forms in situ. Right: In high-mass stars, nuclear generation only takes place well within the core. However the core shrinks with time (see text for detail) and leaves behind material of increasing $\mu$.  A $\mu$-gradient gradually forms as a result. In both figures $r_L$ marks the edge of the core according to the Ledoux criterion. The ODDC can exist just outside of it.}
\label{fig:semiconvectiontypes}       
\end{figure}

In intermediate mass MS stars, with masses above $1M_\odot$ and below $3M_{\odot}$, nuclear reactions can extend far beyond the radius of the convective core, owing to the relatively weak dependence of the pp-chain reaction rates on temperature. This implies that He is slowly generated outside the core, at a rate that decreases with radius. This creates a gentle $\mu$ gradient, that partially inhibits convection and causes the convective core to be smaller than what it would be according to the Schwarzschild criterion. A semiconvective region develops between this so-called Schwarzschild core radius and the Ledoux core radius, see Figure \ref{fig:semiconvectiontypes}a. 

In the core of higher mass stars (with masses greater than about 5$M_\odot$) where the CNO cycle dominates, the temperature dependence of the nuclear reaction rates is so strong that the latter only takes place deep within the convective core. As such, the mechanism described above for creating $\mu$-gradient near the core of intermediate mass stars does not work. However, the convective core shrinks with time as H is converted into He, because the opacity of the material in that region is dominated by electron scattering and is simply proportional to $1+X$, where $X$ is the hydrogen mass fraction. With a reduced opacity, more of the energy can be transported radiatively, and the convection zone retreats. This leaves behind concentric shells of increasingly high $\mu$ material, and therefore gradually builds up a substantial $\mu$ gradient outside this core (see Figure \ref{fig:semiconvectiontypes}b). That region can be unstable to ODDC\footnote{Note that the dependence of the opacity on $X$ presents an additional modeling challenge in high mass stars, which is not accounted for in the simple ODDC model presented in this review. As such whether the results presented here are applicable to semiconvection in these stars or not remains to be determined.} . 

Finally, semiconvection zones detached from nearby convective zones are sometimes found in models of high mass stars (15$M_\odot$ or higher), see for instance \cite{Langer1985}. However, whether they exist in stars or not remains to be established, because their presence or absence in models is quite sensitive to the semiconvective mixing prescription used.

\subsection{Beyond linear theory} 

Having established the conditions under which fingering convection and ODDC may develop in stars, we can now ask the more important quantitative question of how much mixing they cause, and how this affects stellar evolution. To do so requires connecting fluid dynamics, which governs the double-diffusive instabilities on the small scales / short timescales, and stellar modeling, whose equations govern the structure and evolution of stars on the large-scales / long timescales. 

In stellar evolution codes, turbulent transport by double-diffusive convection is traditionally modeled as a turbulent diffusion process. Assuming the star is spherically-symmetric,  
the conservation law for the evolution of the concentration $C$ of a particular chemical species in a given mass shell is 
\begin{equation}
\frac{D C}{Dt} = - \frac{1}{\rho} \nabla \cdot ( \rho {\bf F}_C ) +  \frac{1}{\rho} \left( \frac{D (\rho C)}{Dt}  \right)_{\rm nuc},  
\end{equation}
where the time-derivative is a Lagrangian derivative following the shell, $\rho$ is its density, $\rho C$ is the total mass of the particular chemical element considered, $\rho {\bf F}_C$ is the mass flux, and $(D (\rho C) /Dt)_{\rm nuc}$ is the rate of change due to nuclear (or chemical, if relevant) reactions. The assumption that turbulent transport takes a diffusive\footnote{In this lecture I will always use the mathematical interpretation of a diffusive process which is associated with a downgradient flux, rather than the astrophysical interpretation of element diffusion due to e.g. gravitational settling or radiative levitation.} form implies that the compositional mass flux ${\bf F}_C$ should be downgradient, with 
\begin{equation}
{\bf F}_C = - (\kappa_C + D_C) \nabla C,
\end{equation}
where $\kappa_C$ is the microscopic diffusivity and $D_C$ is the turbulent diffusivity, both of which have units of cm$^2$/s in cgs. The assumption of spherical symmetry then implies that 
\begin{equation}
\frac{D C}{Dt} = \frac{1}{\rho r^2}  \frac{\partial}{\partial r}  \left( r^2 \rho (\kappa_C + D_C) \frac{\partial C}{\partial r} \right) +  \frac{1}{\rho} \left( \frac{D (\rho C)}{Dt}  \right)_{\rm nuc} , 
\end{equation}
which can be expressed in mass coordinates $m$ as 
\begin{equation}
\frac{D C}{Dt} =  \frac{\partial}{\partial m}  \left( (4 \pi r^2 \rho)^2 (\kappa_C + D_C) \frac{\partial C}{\partial m} \right) +  \frac{1}{\rho} \left( \frac{D (\rho C)}{Dt}  \right)_{\rm nuc} .
\end{equation}
This is the formula typically implemented in stellar evolution codes for compositional transport, with a turbulent mixing coefficient $D_C$ that depends on the instability driving the turbulence. It is worth remembering, however, that turbulent transport does not always take a diffusive form, so one should ideally first question whether that assumption is correct, before attempting to model $D_C$. 

The case of turbulent heat transport by double-diffusive instabilities could in principle be treated in a similar way, but in practice this is rarely ever done, for two reasons. The first is that very few stellar evolution codes actually evolve the temperature profile in the star, preferring instead to solve for it at each timestep knowing the stellar luminosity. The second is that heat transport in fingering convection and semiconvection is usually negligible (with some notable exceptions discussed later), because temperature has to diffuse for the instability to exist in the first place.  

The rest of this lecture will be therefore be dedicated to presenting models for $D_C$ for double-diffusive instabilities, with the fingering case discussed in Section \ref{sec:fingering}, and the case of ODDC in Section \ref{sec:ODDC}. The following Table summarizes the important non-dimensional parameters that characterize the properties of double-diffusive convection. All will be extensively used in this lecture. Input parameters such as ${\rm Pr}$, $\tau$ and $R_0$ have already been discussed in this Section. Output non-dimensional parameters based on $D_C$ (and other turbulent mixing processes) will be introduced and used in Sections \ref{sec:fingering} and \ref{sec:ODDC}, to help quantify turbulent transport of heat and composition in the system, both in comparison with the respective diffusive transport rates, or in comparison with one another.

\begin{table}[h]
\centering
\caption{Non-dimensional parameters governing double-diffusive mixing, written as the ratio of dimensional quantities. The dimensional turbulent fluxes of temperature and composition are $F_T = \langle wT \rangle$ and $F_C = \langle wC \rangle$, where the notation $\langle \cdot \rangle$ denotes a volume average. Assuming turbulent transport is diffusive implies that $F_C = - D_C C_{0z}$ in this notation.}
\label{tab-parameters}       
\begin{tabular}{p{1cm}p{3cm}p{4cm}p{3cm}}
\hline
                              & Name & Symbol and Formula & Interpretation  \\ \hline
 Input params. & Prandtl number &  ${\rm Pr} = \nu / \kappa_T$ & Ratio of diffusivities  \\
                             & Diffusivity ratio & $\tau  = \kappa_C / \kappa_T$ & Ratio of diffusivities  \\
                             & Density ratio & $R_0 = \frac{\alpha |T_{0z} - T_{{\rm ad},z}|}{\beta |C_{0z}|} = \frac{|N^2_T|}{|N_C^2|}$ & Ratio of stratifications  \\\hline
 Output params.   		            & Nusselt number for $T$ & ${\rm Nu}_{T} = \frac{-\kappa_T (T_{0z} - T_{{\rm ad},z}) + \langle wT \rangle }{ -\kappa_T (T_{0z} - T_{{\rm ad},z})  }$ & Ratio of total to diffusive potential temperature flux  \\
                                & Nusselt number for $C$ & ${\rm Nu}_{C} = \frac{-\kappa_C C_{0z} + \langle wC \rangle }{ -\kappa_C C_{0z}  } = \frac{D_C + \kappa_C}{\kappa_C}$ & Ratio of total to diffusive compositional flux  \\
                                & Turbulent flux ratio & $\gamma_{\rm turb} = \frac{\alpha   \langle wT \rangle }{ \beta \langle wC \rangle } $ &    Ratio of turbulent fluxes \\                         
                                & Total  flux ratio & $\gamma_{\rm tot} = \frac{ \alpha }{\beta}  \frac{-\kappa_T (T_{0z} - T_{{\rm ad},z}) + \langle wT \rangle }{-\kappa_C C_{0z} + \langle wC \rangle}     $ &    Ratio of  total fluxes \\ \hline                        
\end{tabular}
\end{table}

\section{Mixing by fingering convection}
\label{sec:fingering}

\subsection{Traditional models of "thermohaline" mixing}
\label{sec:traditionalfing}

The first turbulent diffusion model proposed for mixing by fingering convection in astrophysics was put forward by Ulrich \cite{Ulrich1972}, and is based on a very simple dimensional analysis. Noting, as it is common to do so in astrophysics, that the turbulent diffusion coefficient $D_C$ has the units of a velocity times a length, or equivalently, of a length squared divided by time, it is reasonable to assume that the one appropriate for fingering convection should be expressed as 
\begin{equation}
D_C = D_{\rm fing} \propto \lambda d^2 \propto \hat \lambda \kappa_T  ,
\end{equation}
where $\lambda$ is the dimensional growth rate of the fastest-growing fingers, which is related to the non-dimensional growth rate discussed in Section \ref{sec:linear} via $\lambda = \kappa_T \hat \lambda / d^2$. Using the estimate from (\ref{eq:lambdamargfing}) for the growth rate of fingers close to marginal stability, we get
\begin{equation}
D_{\rm fing} \propto  \frac{R_0^{-1}  - \tau }{ 1-R_0^{-1}}  \hat k_h^2    \kappa_T.
\end{equation}
Assuming that $\hat k_h^2$ remains close to unity, then we can write 
\begin{equation}
D_{\rm fing} = C_{\rm fing} \frac{1  - R_0 \tau }{ R_0-1}     \kappa_T.
\label{eq:propercoeff}
\end{equation}
In the limit where $R_0$ is not too close to marginal stability (which is somewhat inconsistent with the assumption made above, but let's ignore this for now) then $ 1 - R_0 \tau \simeq 1$ and we recover the formula proposed by Ulrich \cite{Ulrich1972}, with a proportionality constant he argues should be of order 700: 
\begin{equation}
D_{\rm fing} = C_{\rm U} \frac{1 }{ R_0-1}     \kappa_T.
\label{eq:DfingU}
\end{equation}

A very similar expression was later obtained from rather different arguments by Kippenhahn et al. \cite{kippenhahn80}, who argued that
\begin{equation}
D_{\rm fing} =  \frac{C_{\rm KRT} }{ R_0}     \kappa_T.
\end{equation}
We see by comparison with (\ref{eq:propercoeff}) that this expression should only be valid in the limit where $1 \ll R_0 \ll \tau^{-1}$. The coefficient $C_{\rm KRT}$ in this model is argued to be much smaller than $C_{\rm U}$, taking a proposed value of 12. 

We can already see, however, that both models as they are presented fail to account for the stabilization of the system to fingering instabilities beyond the threshold $R_0 = \tau^{-1}$. As such, we expect that they should largely overestimate the true mixing coefficient as $R_0$ approaches and/or exceeds that threshold. A better model, that would at least be consistent with linear stability theory, is given in (\ref{eq:propercoeff}) and was (to my knowledge) first derived by Denissenkov \cite{Denissenkov2010} (see his equation 15). Written in terms of the more standard astrophysical notations, the model becomes
\begin{equation}
D_{\rm fing} = C_{\rm D} \frac{\phi \nabla_\mu - \delta (\nabla_T - \nabla_{\rm ad}) \tau }{ \delta (\nabla_T - \nabla_{\rm ad}) - \phi \nabla_\mu}     \kappa_T, 
\label{eq:propercoeff2}
\end{equation}
assuming the compositional field $C$ is simply the mean molecular weight. That model, however, is ill-posed (with $D_{\rm fing} \rightarrow \infty$) as $R_0$ tends to unity, or in other words, as we approach the Ledoux criterion for convective instability. This is obviously not physically plausible, so the model should not be used when $R_0 \rightarrow 1$. This ill-posedness is not surprising since the model was derived in the first place assuming that $R_0$ is close to marginal stability, i.e. very large; better models (see below) have since been proposed to address the problem. 

\subsection{Numerical simulations of small-scale fingering convection}
\label{sec:numfing}

\begin{figure}[h]
\centering
\sidecaption
\includegraphics[width=0.6\textwidth]{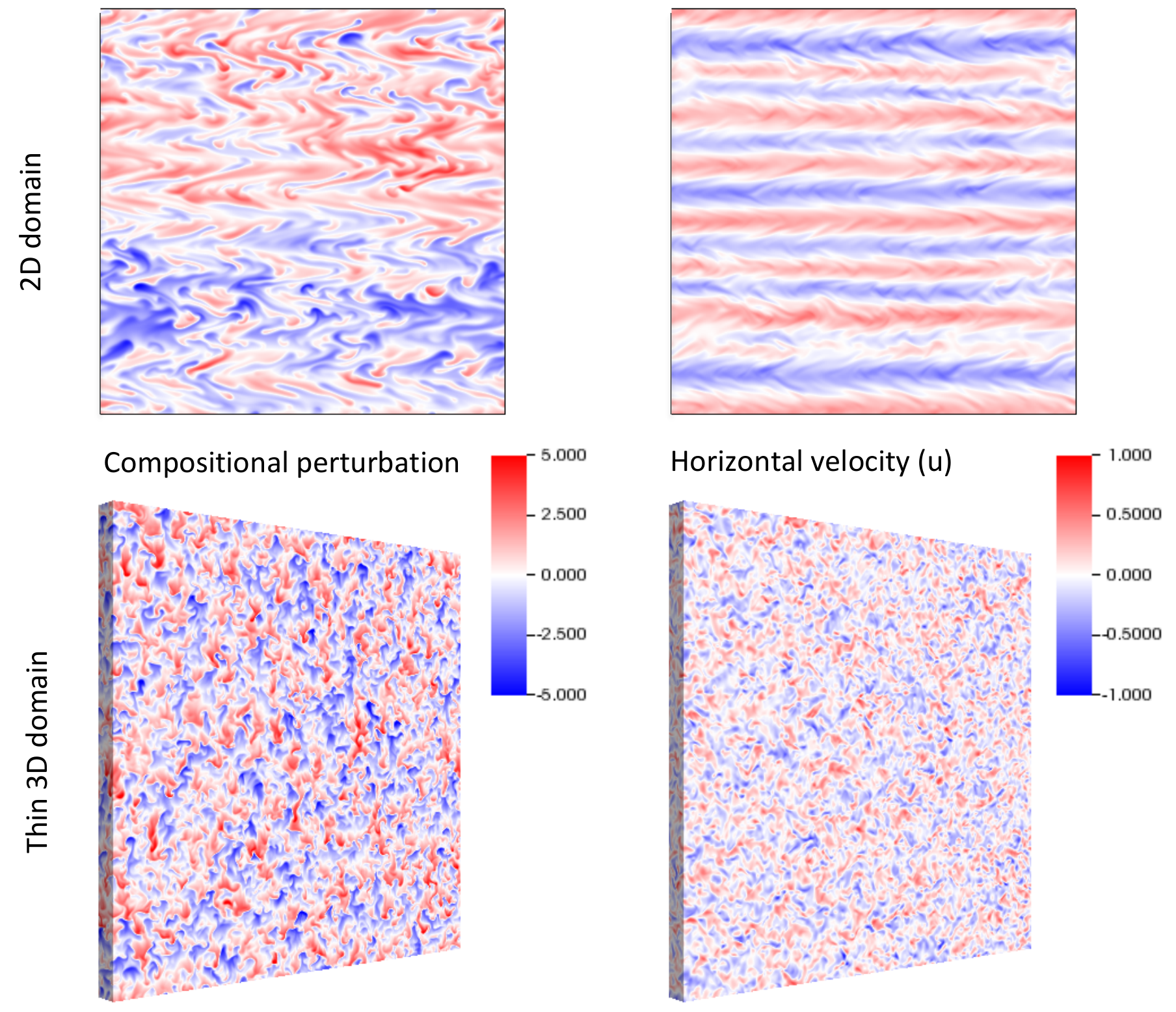}
\caption{Comparison between 2D simulations and thin-domain 3D simulations at the same parameter values (${\rm Pr} = \tau = 0.01$, and $R_0 = 5$). In the 2D case, horizontal shear layers spontaneously develop and tilt the fingers, which in turn reduces the turbulent fluxes. This is artificial, however, and these shear layers are not present in 3D as long as the domain is thick enough. Figure adapted from \cite{GaraudBrummell2015}.  }
\label{fig:2D3D}       
\end{figure}

Until recently, no robust evidence for or against the adequacy of the various expressions for $D_{\rm fing}$ listed above existed. As mentioned in the introduction, laboratory experiments at low Prandtl numbers are almost impossible, because so few low Prandtl number fluids exist on Earth, and those that do are very expensive and notoriously difficult to manipulate (e.g. liquid mercury, liquid lithium, liquid potassium). Numerical experiments are also challenging. Indeed, the computational domain needs to contain at least 5-10 fingers in each direction for good statistics. Furthermore, we saw that the scale of the fingers is related to the thermal diffusion scale, while the scale of boundary layers between the fingers is dictated by the viscous and compositional diffusion scales, which are asymptotically small under stellar conditions (since ${\rm Pr} = \nu / \kappa_T \ll 1$ and $\tau = \kappa_C / \kappa_T \ll 1$). 
All of these scales need to be resolved to adequately model fingering convection, which poses a hard computational challenge. Finally, it has recently been demonstrated that fingering convection at low Prandtl number cannot be modeled in 2D, as it develops unphysical pathological behavior  \cite{GaraudBrummell2015}. In 2D simulations, artificial horizontal shear layers appear spontaneously from the fingering instability, and in turn affect the fingering structures. These shear layers are not present in 3D domains\footnote{at least not as ubiquitously and with such strong amplitude.}, as long as the third dimension is wide enough (i.e at least 2 finger widths), see Figure \ref{fig:2D3D}.

Within the last decade, however, numerical simulations with a physical grid resolution of $\sim 300^3$ have become routine, and state-of-the-art ones are now exceeding $\sim 3000^3$ or more. With the smaller-sized runs, it is possible to 
systematically explore parameter space (i.e. cover the entire range of density ratios) for fingering convection at diffusivity ratios ${\rm Pr}$ and $\tau$ down to $\sim 0.01$. The higher resolution simulations can be used to test models down to, e.g. ${\rm Pr} , \tau \sim 0.001$, but only for a few values of the density ratio. 

\begin{figure}[h]
\centering
\sidecaption
\includegraphics[width=0.6\textwidth]{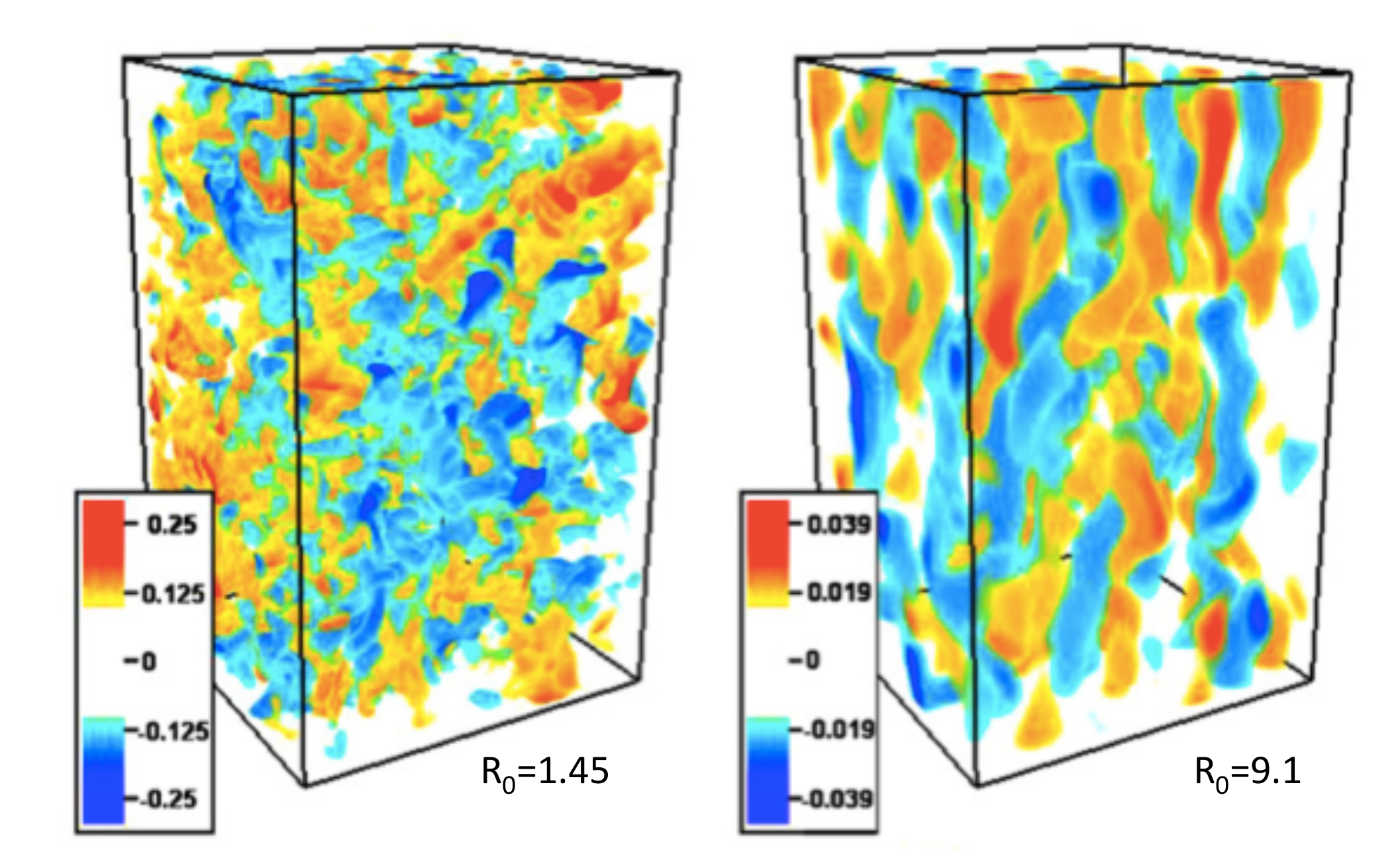}
\caption{Volume renderings of the compositional field in homogeneous fingering convection at ${\rm Pr} = \tau = 0.1$, for $R_0 = 1.45$ (left), which is strongly unstable, and $R_0 = 9.1$ (right), which is close to marginal stability. Compositionally rich (red) fingers are flowing downward, and compositionally poor (blue) fingers are flowing upward. Figure from \cite{Traxler2011b}.}
\label{fig:Traxler}       
\end{figure}

A first series of 3D DNSs of fingering convection at low Prandtl number was presented by Traxler et al. \cite{Traxler2011b}, using the PADDI code developed by S. Stellmach. This code is a pseudo-spectral code that solves equations (\ref{eq:momnondim})--(\ref{eq:compnondim}) in a triply periodic domain, ensuring incompressibility is maintained using a standard projection method \cite{Traxler2011a}. 
Traxler et al. limited their study to moderate ${\rm Pr}, \tau$, from 1/3 down to 1/30. 
Figure \ref{fig:Traxler} shows representative snapshots of the compositional field in their simulations, for two density ratios (one low, close to the Ledoux threshold, and one large, close to the marginal stability threshold). Simulations with lower ${\rm Pr}$ and $\tau$ down to 1/300 made with the PADDI code are now also available \cite{Brownal2013,Garaud18} and show qualitatively similar features.  

Each simulation can easily be used to measure a turbulent compositional diffusivity. Indeed, in a triply-periodic domain, the horizontal average (marked with an overbar) of the dimensional composition evolution equation (\ref{eq:compdim}) is  
\begin{equation}
\frac{\partial \bar C}{\partial t} + \frac{\partial}{\partial z} \overline{w C}  = \kappa_C \frac{\partial^2 \bar C}{\partial z^2}, 
\label{eq:cbardim}
\end{equation}
which reveals $\overline{w C}$ to be the vertical turbulent compositional flux (i.e. the transport of $C$ due to advection by vertical fluid motions). While this quantity would generally depend on both $z$ and $t$, in a statistically steady state and assuming that $\bar C$ remains small (which we can verify is true in these simulations), $\overline{w C}$ is approximately constant. We can then obtain a good estimate of the vertical turbulent flux by taking both a time-average and a volume average of $w C$ in the simulation, and define this as $F_C = \langle wC \rangle$. The same can be done for the temperature. Incidentally, it is easy to show \cite{Malkus1954} that $F_C$ and $F_T$ {\it must} be negative for statistically stationary fingering convection. Indeed, multiplying equation (\ref{eq:compdim}) by $C$, integrating over the volume, and using incompressibility and the periodicity of the boundary conditions to eliminate the nonlinear advection terms and the boundary terms, we get
\begin{equation}
\frac{1}{2} \frac{\partial \langle C^2 \rangle }{\partial t} + \langle wC \rangle C_{0z} = \kappa_C \langle C \nabla^2 C \rangle = - \kappa_C \langle | \nabla  C  |^2 \rangle , 
\end{equation}
using one of Green's identities. Since $C_{0z} > 0$ in fingering convection, and since $| \nabla  C  |^2 > 0$, then $\langle wC \rangle$ must be negative if the turbulence is statistically stationary (i.e. if we can neglect the time-derivative). The same is true for $\langle wT \rangle$, showing that both $T$ and $C$ are transported downward (which is not surprising, given the physical mechanism described in Section \ref{sec:intro}). 

The volume-averaged compositional flux can finally be used to derive $D_{\rm fing}$ from each simulation as 
\begin{equation}
D_{\rm fing} = - \frac{\langle wC \rangle }{C_{0z}} =  - R_0 \kappa_T \langle \hat w \hat C \rangle , 
\label{eq:numDdef}
\end{equation}
where the hats denote non-dimensional quantities (see Section \ref{sec:lineargeneral}). 

\begin{figure}[h]
\centering
\sidecaption
\includegraphics[width=0.6\textwidth]{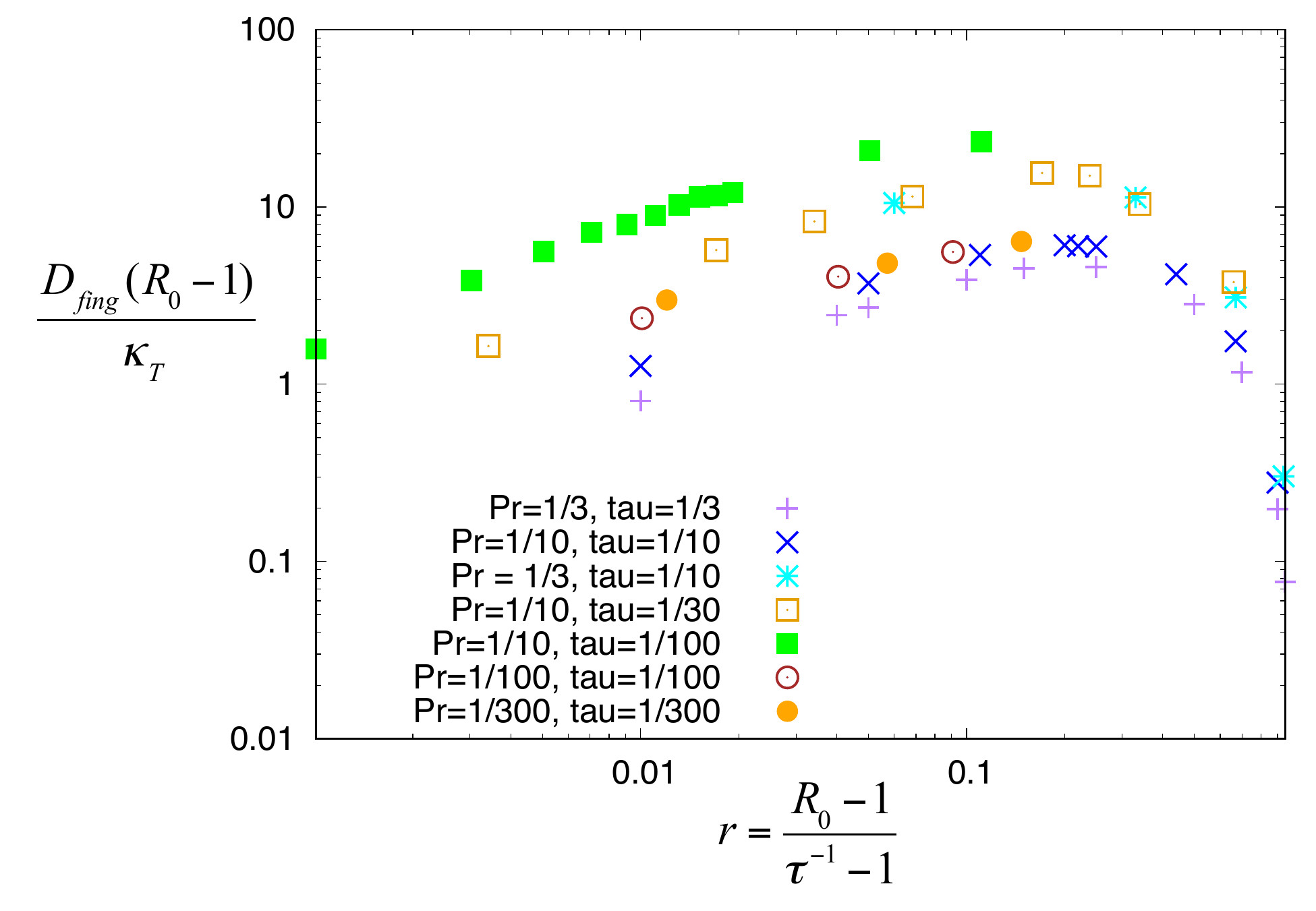}
\caption{Non dimensional ratio $D_{\rm fing} (R_0-1)/\kappa_T$, against the reduced density ratio $r$, for a wide range of simulations. This quantity should be constant according to the traditional models of fingering convection \cite{Ulrich1972,kippenhahn80} but is not in reality.}
\label{fig:testUKRT}       
\end{figure}

The numerical results can be used to test the traditional models of Ulrich and Kippenhahn et al. \cite{Ulrich1972,kippenhahn80} described in Section \ref{sec:traditionalfing}. Writing both models as $D_{\rm fing} = C_{\rm U,KRT} \frac{\kappa_T}{R_0-1}$, with only the constant differing between them, the theory would predict that $D_{\rm fing} (R_0-1)/\kappa_T$ should be constant. The results are shown in Figure \ref{fig:testUKRT}, for a wide range of simulations \cite{Traxler2011b, Brownal2013,Garaud18}. For ease of visualization the results are shown, not as a function of $R_0$, but as a function of the reduced density ratio \cite{Traxler2011b}
\begin{equation}
r = \frac{R_0 - 1}{\tau^{-1} - 1} , 
\label{eq:reducedrfing}
\end{equation}
which maps the entire fingering range into the interval $[0,1]$ regardless of $\tau$, with $r = 0$ corresponding to the Ledoux criterion, and $r = 1$ corresponding to marginal stability. We can immediately make several important conclusions: that $D_{\rm fing} (R_0-1)/\kappa_T$ is not constant across the entire range, that it seems to depend on the ratio $\nu / \kappa_C = {\rm Pr} / \tau$ (called the Schmidt number) and finally that the numerical results do not support the large value of $C_{\rm U} \simeq 700$ \cite{Ulrich1972} but are more consistent (within the list of caveats listed) with the smaller value $C_{\rm KRT} \simeq 12$ \cite{kippenhahn80}.

As discussed earlier, the fact that $D_{\rm fing} (R_0-1)/\kappa_T$  is not constant across the fingering range is not surprising, since $D_{\rm fing} \propto \frac{\kappa_T}{R_0-1}$ is both unphysically singular as $R_0 \rightarrow 1$, and does not account for the stabilization of the system as $R_0 \rightarrow \tau^{-1}$ (which is clearly visible in the data). The fact that the results depend principally on the Schmidt number is also expected, both on physical and mathematical grounds. Physically speaking \cite{Traxler2011b}, when $\kappa_T \ll (\kappa_C, \nu)$, the role of the temperature fluctuations essentially becomes negligible, and the instability is driven by the compositional field. It is therefore not surprising to find that the dynamics solely depend on $\nu/\kappa_C$, rather than on any parameter that depends on $\kappa_T$. From a mathematical point of view, it can be shown \cite{Pratal2015,Xieal2019} that in the limit of $\kappa_T \rightarrow 0$ (and for sufficiently weak fingering)  the non-dimensional governing equations (\ref{eq:momnondim})--(\ref{eq:compnondim}) reduce to a new set of asymptotic equations that only depend on two parameters instead of the usual three, namely the Schmidt number, and the so-called Rayleigh ratio ${\rm R} = (R_0 \tau)^{-1}$. As such, the data dependence on the Schmidt number, rather than ${\rm Pr}$ and $\tau$, is expected. 

Based on their first set of numerical experiments, Traxler et al. \cite{Traxler2011b} proposed a simple empirical formula for mixing by fingering convection, namely
\begin{equation}
D_{\rm fing} =  101 \sqrt{\kappa_C \nu} \exp(-3.6r) (1-r)^{1.1},
\label{eq:Dfingtraxler}
\end{equation}
where the numbers 101, 3.6 and 1.1 were fitted to the data. This model is easy to use, and fits the results adequately except for very low $r$ \cite{Brownal2013}, where the system is only weakly stratified and close to the Ledoux threshold for overturning convection. In that limit (\ref{eq:Dfingtraxler}) dramatically underestimates the mixing coefficient \cite{Brownal2013} .

\subsection{The Brown et al. 2013 model for small-scale fingering convection}
\label{sec:brown}

\begin{figure}[h]
\centering
\sidecaption
\includegraphics[width=0.6\textwidth]{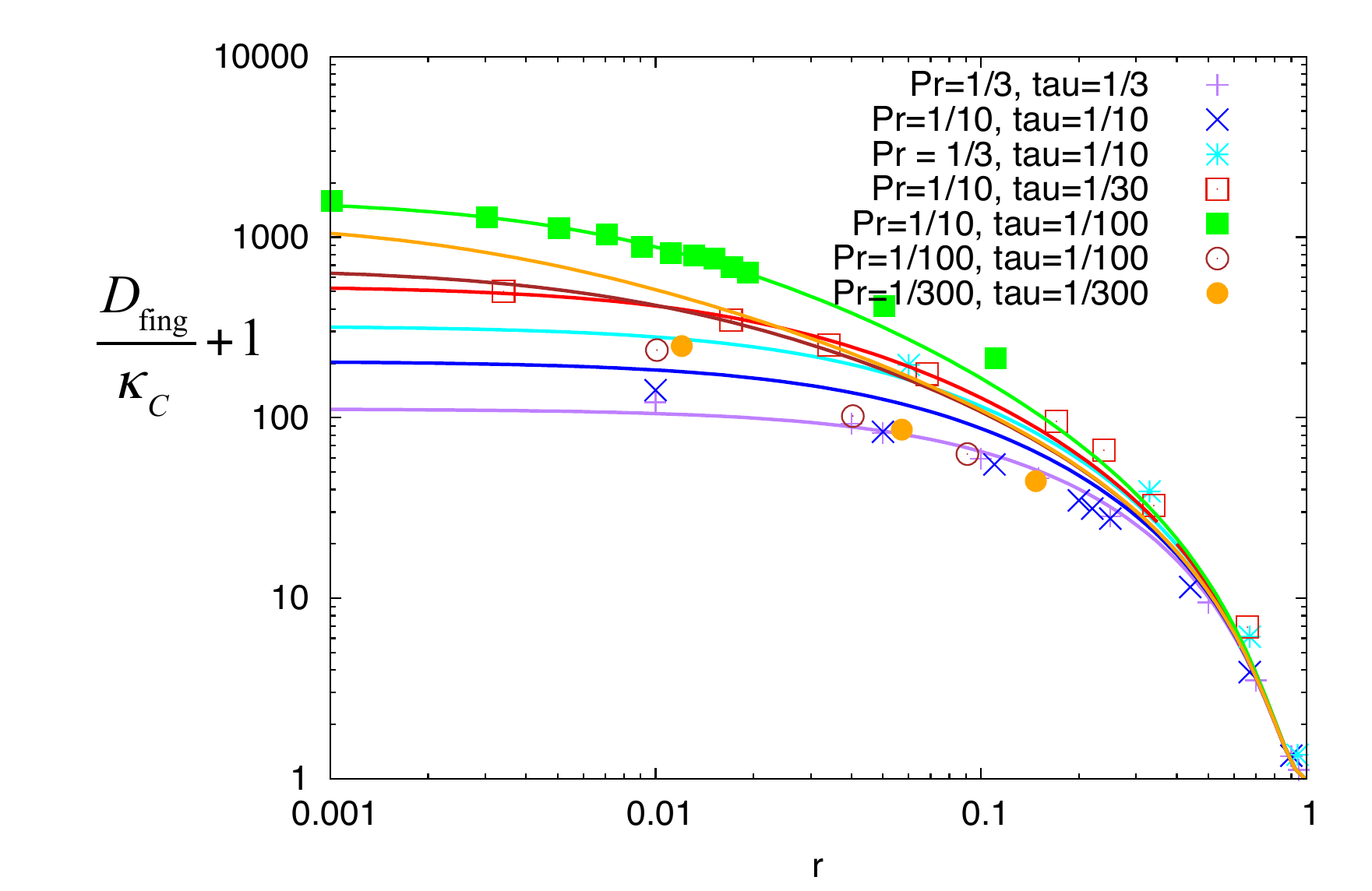}
\caption{Comparison between data (symbols) and theory (equation \ref{eq:Dfingbrown}) (lines) for the Nusselt number $D_{\rm fing}/\kappa_C + 1$, against the reduced density ratio $r$, for a wide range of parameters. Figure from \cite{Garaud18}. }
\label{fig:Browncompare}       
\end{figure}

In order to address the shortfalls of the Traxler et al. model for fingering convection, Brown et al. \cite{Brownal2013} revisited the problem and proposed a new theory for the mixing coefficient. 
Starting with the definition of $D_{\rm fing}$ given in (\ref{eq:numDdef}), we see that the key to creating a model for this coefficient is to estimate both the typical non-dimensional vertical velocity of the fingers, $\hat w_f$, and their typical compositional perturbation, $\hat C_f$, so 
\begin{equation}
D_{\rm fing} = - R_0 \kappa_T \langle \hat w \hat C \rangle = - R_0 \kappa_T K_B \hat w_f \hat C_f 
\end{equation}
where $K_B$ is a constant that depends on the geometry of the fingers, and the typical correlation between $\hat w$ and $\hat C$. Assuming that transport is controlled by the most rapidly growing fingers (which are elevator modes) and that these fingers are controlled by the linearized equations (at least until they nonlinearly saturate) we can relate $\hat w_f$ and $\hat C_f$ using the linearized version of (\ref{eq:compnondim}), in which $\partial/\partial t$ is replaced by the finger growth rate $\hat \lambda$, and horizontal derivatives are replaced by the horizontal wavenumber $\hat k_h$, so
\begin{equation}
\hat \lambda \hat C_f  + R_0^{-1} \hat w_f = - \tau \hat k_h^2 \hat C_f  \rightarrow \hat C_f = - \frac{ R_0^{-1}  \hat w_f }{\hat \lambda + \tau \hat k_h^2 } .
\end{equation}
Hence 
\begin{equation}
D_{\rm fing} =  K_B \frac{  \hat w_f^2  }{\hat \lambda + \tau \hat k_h^2 } \kappa_T, 
\label{eq:Dfingw}
\end{equation}
and the only problem remaining is to estimate $\hat w_f$. To do so requires specifying the mechanism by which the linear fingering instability saturates. Following Radko \& Smith \cite{RadkoSmith2012}, Brown et al. \cite{Brownal2013} assumed that saturation occurs by parasitic shear instabilities that develop between up-flowing and down-flowing fingers, and that the fingers stop growing when the parasitic instability growth rate $\hat \sigma$ approaches 
$\hat \lambda$. By dimensional analysis (or by solving the problem exactly, see Appendix A of Brown et al.), it can be shown that $\hat \sigma = K_\sigma \hat w_f \hat k_h$ where $K_\sigma$ is a universal constant of order unity (whose value is irrelevant for reasons explained below) so the saturation condition reads
\begin{equation}
K_\sigma \hat w_f \hat k_h = K_w \hat \lambda \rightarrow \hat w_f = \frac{K_w}{K_\sigma} \frac{\hat \lambda}{\hat k_h} , 
\end{equation} 
where $K_w$ is another universal constant of order unity. We see that these two constants end up folded into a single one, which is the reason why their individual values are irrelevant. The relationship between $\hat w_f$ at saturation of the fingering instability and $\hat \lambda/\hat k_h$ was recently verified \cite{SenguptaGaraud2018} against a re-analysis of the existing numerical data, and revealed that $K_w / K_\sigma \simeq 2\pi$. With this, we finally have the prediction that 
\begin{equation}
D_{\rm fing} =  K_B \left( \frac{K_w}{K_\sigma} \right)^2 \frac{   \hat \lambda^2  }{\hat \lambda \hat k_h^2+ \tau \hat k_h^4 } \kappa_T. 
\label{eq:Dfingbrown}
\end{equation}
This can easily be tested against available data, and the combination of constants $K_B (K_w/K_\sigma)^2$ is found to be around 49 (implying that $K_B \simeq 1.24$). The comparison between the model and the data is very good (though not perfect) for all cases where $\nu/\kappa_C \le 1$, which is indeed the case in stellar interiors, for all $r$, ${\rm Pr}$ and $\tau$ tested. For decreasing ${\rm Pr}$ and $\tau$ the good fit deteriorates somewhat, though remains reasonable (within a factor of order unity), see Figure \ref{fig:Browncompare}. 

In summary, to compute $D_{\rm fing}$ one should first find the growth rate and wavenumber of the fastest-growing modes of the fingering instability at the desired parameter values $(R_0, {\rm Pr}, \tau)$, which can be done numerically using a cubic root-finding algorithm. Once $\hat \lambda$ and $\hat k_h$ are known, they are used in expression (\ref{eq:Dfingbrown}) to predict $D_{\rm fing}$. This method has already been implemented for instance in MESA. Alternatively, analytical approximations for $\hat \lambda$ and $\hat k_h$ (found in Appendix B of Brown et al. \cite{Brownal2013}) can be used instead  to speed up the process. Routines providing such estimates are also available in MESA.

\subsection{Large-scale instabilities?}
\label{sec:largescaleinstab}

While the process of small-scale fingering convection in stellar environment is now arguably well-understood, this may not be the full story. Indeed, fingering-unstable regions of the ocean on Earth are often associated with thermohaline staircases, which are horizontally-invariant stepped structures in the vertical profiles of temperature and salinity (see \cite{radko2013double} for a review). These staircases are formed of stacked {\it layers} and {\it interfaces}. Within a layer, the density is very slightly unstably stratified and subject to large-scale convective overturning. Both temperature and salinity are almost constant within each layer as a result of the strong mixing. In between the layers are stably stratified interfaces undergoing fingering convection. While the layers can be tens or even hundreds of meters deep, the interfaces are much shallower (tens of centimeters), so the temperature and salinity gradients across the interface are very large (hence the characteristic appearance of the staircase). When thermohaline staircases are present, vertical mixing can be enhanced by two orders of magnitude \cite{schmitt2005enhanced} compared to a similar overall stratification without staircases. It is therefore crucial to understand why and how such staircases form in the ocean, and whether similar processes may be taking place in stars. 

\begin{figure}[h]
\centering
\sidecaption
\includegraphics[width=0.6\textwidth]{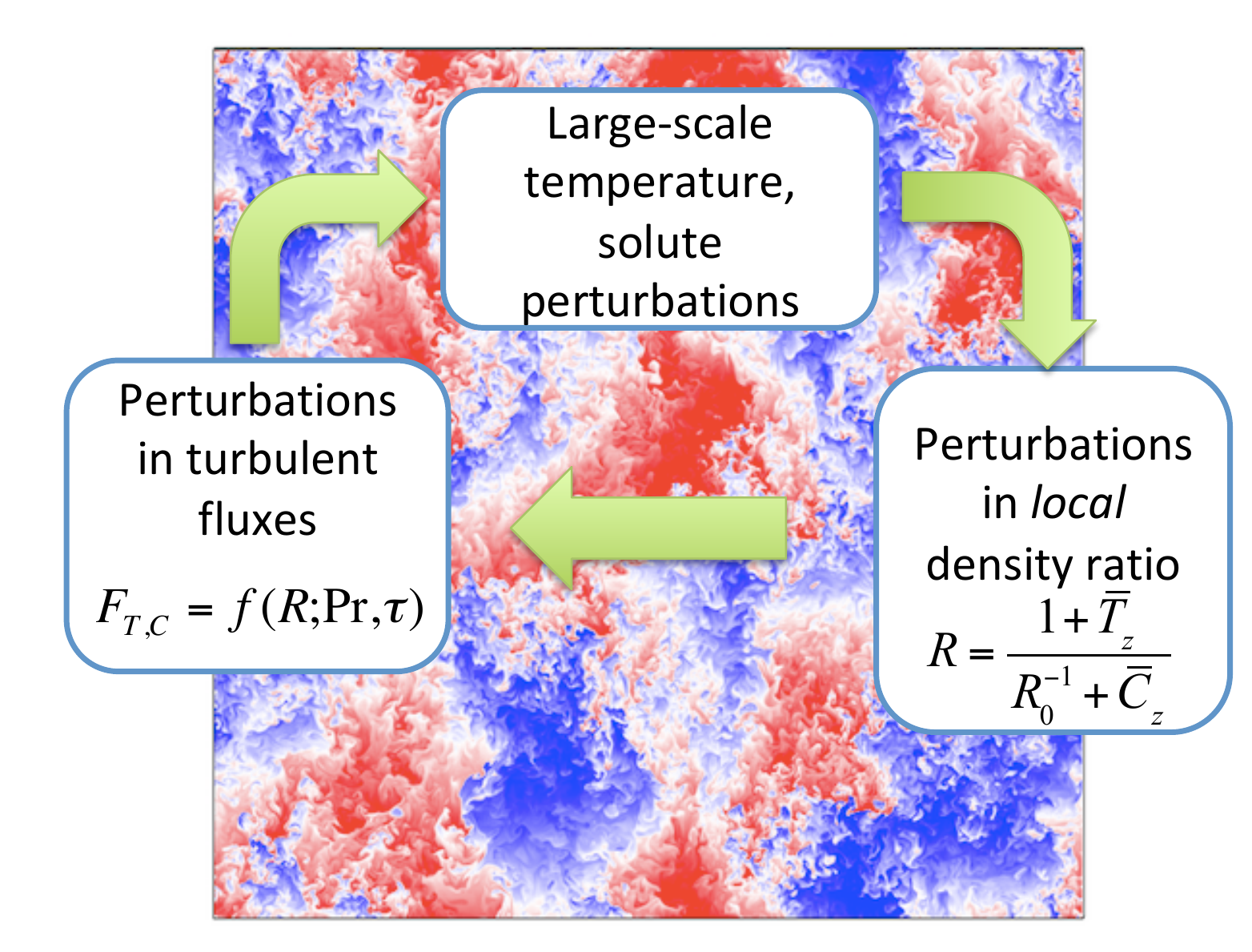}
\caption{Illustration of the possible positive feedback loop between large-scale perturbations in $\bar T$ and $\bar C$, and turbulent fluxes. See text for detail. Figure adapted from \cite{Garaud18}. The background image is a snapshot from a simulation exhibiting the spontaneous emergence of large-scale gravity waves from \cite{GaraudBrummell2015} (see also \cite{Garaudal2015}).}
\label{fig:Feedback}       
\end{figure}

Thankfully, significant progress in modeling the formation of oceanic thermohaline staircases (and the emergence of other large-scale dynamics) from fingering convection has been made in the past 20 years. Given the wide separation between the finger scale and the scale of the staircases, mean field hydrodynamics turns out to be a fruitful approach to the problem. In this type of theory, the large scales (such as the layer scale) are modeled exactly, while the effect of small scales (i.e. the basic fingering convection) is parameterized using some form of turbulence closure. As we saw earlier,  for a given fluid (i.e. given Pr and $\tau$), the turbulent fluxes in fingering convection only appear to depend on the local density ratio. With this in mind, it is easy to see how large-scale instabilities might develop (see Figure \ref{fig:Feedback}). Indeed, any large-scale perturbation in the temperature and compositional fields causes large-scale modulations in the local density ratio (which depends on their gradients). This in turn modulates the temperature and compositional fluxes due to fingering, and the convergence or divergence of these fluxes can in some cases enhance the original perturbation, and in some cases suppress it. Positive feedback loops, when they exist, thus drive the growth of large-scale instabilities. As we have discovered \cite{Traxler2011a,Garaudal2015}, several distinct types of positive feedback loops can in fact exist, each leading to the amplification of different kinds of perturbations.

To model this idea mathematically and determine whether thermocompositional layers may form in stars, we essentially follow the theory of Radko \cite{radko2003mechanism}, and expand it to account for the diffusive contribution to the fluxes \cite{Traxler2011b} (which, as we shall demonstrate, is a crucial element of the stellar problem). 
Since layers are horizontally invariant, we begin by taking the horizontal average of the non-dimensional equations for temperature and composition, namely 
\begin{equation}
\frac{\partial \bar T}{\partial t} + \frac{\partial }{\partial z}  \overline{wT}   = \frac{\partial^2 \bar T }{\partial z^2}  \mbox{   ,    } 
\frac{\partial \bar C}{\partial t} + \frac{\partial }{\partial z}  \overline{wC}   = \tau \frac{\partial^2 \bar C }{\partial z^2} ,  \label{eq:horizav}
\end{equation}
where we expect $\bar w = 0$ otherwise mass is not conserved. Note that the hats have been dropped from the horizontally averaged quantities to avoid crowding the notations, but everything in this section is implicitly non-dimensional. If we then define the total fluxes
\begin{equation}
\hat F_{T,{\rm tot}} = \overline{ wT} - \left(1+\frac{\partial \bar T}{\partial z}\right)  \mbox{   and  } \hat F_{C,{\rm tot}} = \overline{ wC} - \tau \left(R_0^{-1} + \frac{\partial \bar C}{\partial z} \right), 
\end{equation}
as the sum of the turbulent flux plus the diffusive flux of each quantity, then these equations simply become the conservation laws 
\begin{equation}
\frac{\partial \bar T}{\partial t} + \frac{\partial \hat F_{T,{\rm tot}} }{\partial z}   =0   \mbox{   and  }   \frac{\partial \bar C}{\partial t} + \frac{\partial \hat F_{C,{\rm tot}} }{\partial z}   =0  .
\label{eq:fluxevol} 
\end{equation}
We also define two non-dimensional quantities: the Nusselt number ${\rm Nu}_T$, and the flux ratio $\gamma_{\rm tot}$, as
\begin{equation}
{\rm Nu}_T = - \frac{ \hat F_{T,{\rm tot}}}{1 + \partial \bar T/ \partial z} \mbox{   and } \gamma_{\rm tot} = \frac{\hat F_{T,{\rm tot}}}{\hat F_{C,{\rm tot}}} ,
\label{eq:gammanudef}
\end{equation}
which are similar to those defined in Table \ref{tab-parameters}, but expressed as the ratio of non-dimensional quantities, and including large-scale perturbations $\bar T$ and $\bar C$. 
The Nusselt number is the ratio of the total temperature flux to the diffusive (potential) temperature flux, the latter being simply equal to $-(1 + \partial \bar T/ \partial z)$ in these units. 
The key assumption made by the Radko \cite{radko2003mechanism} is that ${\rm Nu}_T$ and $\gamma_{\rm tot}$ can only depend on other local non-dimensional properties of the fluid. Aside from Pr and $\tau$, which are fixed once the fluid is specified, the only other relevant non-dimensional quantity is the local density ratio. The latter is given by 
\begin{equation}
R = \frac{1+ \partial \bar T / \partial z}{R_0^{-1} +  \partial \bar C/ \partial z } , 
\label{eq:Rdef}
\end{equation}
since 1 is the non-dimensional background potential temperature gradient, and $R_0^{-1}$ is the non-dimensional background compositional gradient. To determine the evolution of large-scale horizontally-invariant perturbations, we therefore simply evolve the equations in (\ref{eq:fluxevol}) together with 
equation (\ref{eq:gammanudef}), where the functions ${\rm Nu}_T(R;{\rm Pr},\tau)$ and $\gamma_{\rm tot}(R;{\rm Pr},\tau)$ are assumed to be known (they can be derived for instance from the Brown et al. model or from experimental data), and $R$ given in (\ref{eq:Rdef}). 

A trivial solution of these equations exists: when $\bar T \equiv \bar C \equiv 0$, then $R = R_0$, ${\rm Nu}_T = {\rm Nu}_T(R_0;{\rm Pr},\tau)$ and $\gamma_{\rm tot} = \gamma_{\rm tot}(R_0;{\rm Pr},\tau)$ are all constant, so the fluxes $\hat F_{T,{\rm tot}}$ and $\hat F_{C,{\rm tot}}$ are also constant. This defines a turbulent state that is spatially homogeneous and statistically steady -- this is the basic state of small-scale fingering convection. 

When $\bar T$ and $\bar C$ vary with $z$, solutions cannot be found analytically in general because the functional form of  ${\rm Nu}_T$ and $\gamma_{\rm tot}$ can be quite complex, and (\ref{eq:Rdef}) is nonlinear. Instead, we proceed by linearizing the large-scale equations around the homogeneous fingering solution described above to study its stability to the development of large-scale perturbations. Let us therefore assume that $\bar T$ and $\bar C$ are small, so that $\partial \bar T / \partial z \ll 1$ and $\partial \bar C / \partial z \ll R_0^{-1}$. Then 
\begin{equation}
R \simeq R_0 \left(1+ \frac{\partial \bar T }{ \partial z  } -  R_0 \frac{\partial \bar C}{ \partial z}  \right) \equiv R_0 + R'  \rightarrow R ' = R_0 \frac{\partial \bar T }{ \partial z  } -  R_0^2 \frac{\partial \bar C}{ \partial z} .
\label{eq:Rdef}
\end{equation}
We continue by linearizing ${\rm Nu}_T (R)$ (at fixed ${\rm Pr}$ and $\tau$) in the vicinity of $R_0$, as
\begin{equation}
{\rm Nu}_T (R) = {\rm Nu}_T (R_0) + R'  \left. \frac{\partial {\rm Nu}_T}{\partial R}  \right|_{R=R_0} .
\end{equation}
Using this with (\ref{eq:fluxevol}) and (\ref{eq:gammanudef}), we get after successive simplifications (from linearization)
\begin{eqnarray}
\frac{\partial \bar T}{\partial t} =  \frac{\partial}{\partial z} \left[  \left(1 +  \frac{\partial \bar T}{\partial z} \right)  \left( {\rm Nu}_T (R_0) + R'  \left. \frac{\partial {\rm Nu}_T}{\partial R}  \right|_{R=R_0}  \right) \right] 
\nonumber \\
=  \frac{\partial^2 \bar T}{\partial z^2}  {\rm Nu}_T (R_0)  +  \left. \frac{\partial {\rm Nu}_T}{\partial R}  \right|_{R=R_0}  \frac{\partial R'}{\partial z}  \nonumber \\
=  \frac{\partial^2 \bar T}{\partial z^2}  {\rm Nu}_T (R_0)  +  A_{\rm Nu} \left( \frac{\partial^2 \bar T }{ \partial z^2  } -  R_0 \frac{\partial^2 \bar C}{ \partial z^2}  \right), \label{eq:Tlayereq}
\end{eqnarray}
where $A_{\rm Nu} \equiv R_0 \left. \frac{\partial {\rm Nu}_T}{\partial R}  \right|_{R=R_0}$. 
Similarly, it is easy to show that 
\begin{align}
\frac{\partial \bar C}{\partial t} = -  \frac{\partial}{\partial z} \left( \gamma_{\rm tot}^{-1}  \hat F_{T,{\rm tot}}  \right) =  \gamma_{\rm tot}^{-1}(R_0) \frac{\partial \bar T}{\partial t}  -  \hat F_{T,{\rm tot}} \frac{\partial  \gamma_{\rm tot}^{-1} }{\partial z} \nonumber \\
= \gamma_{\rm tot}^{-1}(R_0) \frac{\partial \bar T}{\partial t}   -  \frac{\partial  \gamma_{\rm tot}^{-1} }{\partial R}  \frac{\partial R'}{\partial z}  \hat F_{T,{\rm tot}}(R_0)   \nonumber \\ = \gamma_{\rm tot}^{-1}(R_0) \left[  \frac{\partial^2 \bar T}{\partial z^2}  {\rm Nu}_T (R_0)  +  A_{\rm Nu} \left( \frac{\partial^2 \bar T }{ \partial z^2  } -  R_0 \frac{\partial^2 \bar C}{ \partial z^2}  \right)  \right] + A_{\gamma} \left( \frac{\partial^2 \bar T }{ \partial z^2  } -  R_0 \frac{\partial^2 \bar C}{ \partial z^2}  \right) {\rm Nu}_T (R_0) ,   \label{eq:Clayereq}
\end{align} 
where $A_\gamma \equiv R_0 \left. \frac{\partial \gamma_{\rm tot}^{-1}}{\partial R}  \right|_{R=R_0}$, and where we used the fact that in the absence of any large-scale perturbations, $ \hat F_{T,{\rm tot}}(R_0)  = - {\rm Nu}_T (R_0) $. 

\begin{figure}[h]
\centering
\sidecaption
\includegraphics[width=0.6\textwidth]{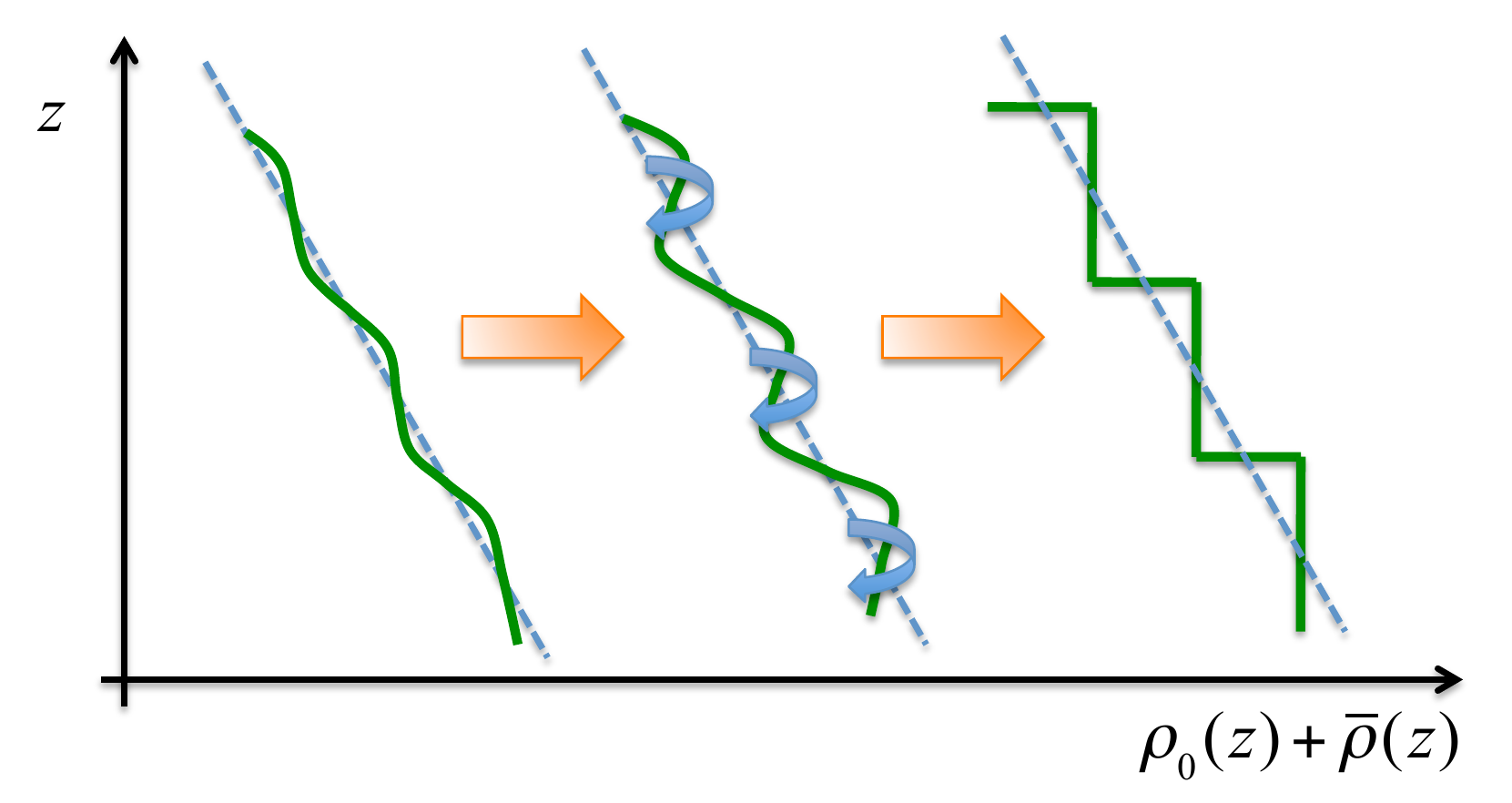}
\caption{Eigenmodes of the layering instability are sinusoidal perturbations in the density profile $\bar \rho(z)$. They grow exponentially, and eventually cause the total density profile $\rho_0(z) + \bar \rho(z)$ to increase with height at specific positions in the domain. Being unstable to overturning convection these regions are then fully mixed and become the layers of the staircase. Credit: S. Stellmach.}
\label{fig:LayeringInstab}       
\end{figure}

Finally, assuming that $\bar T  \propto \exp(i \hat Kz + \hat \Lambda t)$ and similarly for $\bar C$, we can substitute these solutions into (\ref{eq:Tlayereq}) and (\ref{eq:Clayereq}), and with a little algebra, obtain a quadratic equation for the non-dimensional growth rate $\hat \Lambda$ of these horizontally-invariant perturbations: 
\begin{equation}
\hat \Lambda^2 + \hat \Lambda \hat K^2 \left[ A_{\rm Nu} (1-R_0 \gamma_0^{-1} ) + {\rm Nu}_0 (1-A_\gamma R_0) \right] -\hat  K^4  A_{\gamma} {\rm Nu}_0^2 R_0  = 0 ,
\label{eq:gammaquadratic}
\end{equation}
where for simplicity of notation I have used ${\rm Nu}_0 = {\rm Nu}_T (R_0) $ and $\gamma_0= \gamma_{\rm tot} (R_0)$. 
Solutions for $\hat \Lambda$ that have a positive real part denote instability. When unstable, perturbations of the form $\bar T  \propto \cos(\hat Kz) $ or $\sin(\hat Kz)$ (and similarly for $\bar C$) grow exponentially with time, until the density profile itself has regions that are dynamically unstable (i.e. where density increases upwards). As soon as this happens, overturning convection sets in, and stacked convective layers appear separated by sharp interfaces (see Figure \ref{fig:LayeringInstab}). 

From the form of (\ref{eq:gammaquadratic}), we can immediately deduce a few important properties of the solutions. First, for unstable modes to exist (i.e. solutions with positive real $\hat \Lambda$), it is sufficient to require that $A_\gamma > 0$, or equivalently, that $\gamma_{\rm tot}$ be a {\it decreasing} function of $R$. This is called the $\gamma$-instability criterion, first derived by Radko \cite{radko2003mechanism}.

Second, we see that there is no stabilization of the system for large $\hat K$. Instead, it is easy to show that $\hat \Lambda$ is always proportional to $\hat K^2$, so the smallest scale  modes grow the most rapidly, which is not particularly physical. At the root of this so-called {\it ultraviolet} catastrophe, with $\hat \Lambda \propto \hat K^2$, is the anti-diffusive nature of the fingering fluxes when $\gamma_{\rm tot}$ decreases with $R$. To see this, let us construct the evolution equation for the horizontally-averaged density perturbation (using as before a linearization of the equations around $R = R_0$) 
\begin{eqnarray}
\frac{\partial \bar \rho}{\partial t} =  - \frac{\partial \bar T}{\partial t} + \frac{\partial \bar C}{\partial t} = \frac{\partial }{\partial z} \left[ \hat F_{T,{\rm tot}} ( 1 - \gamma_{\rm tot}^{-1} ) \right]  \nonumber \\
 = ( 1 - \gamma_{\rm tot}^{-1} )  \frac{\partial \hat F_{T,{\rm tot}}}{\partial z} - \frac{\partial  \gamma_{\rm tot}^{-1} }{\partial R}  \frac{\partial R}{\partial z}  \hat F_{T,{\rm tot}}  \nonumber \\
= ( 1 - \gamma_{\rm tot}^{-1} )  \frac{\partial \hat F_{T,{\rm tot}}}{\partial z} - A_\gamma \left( \frac{\partial^2 \bar T }{ \partial z^2  } -  R_0 \frac{\partial^2 \bar C}{ \partial z^2}  \right)    \hat F_{T,{\rm tot}}  \nonumber \\
 = - A_\gamma {\rm Nu}_0  \frac{\partial^2 \bar \rho }{ \partial z^2  } + ( 1 - \gamma_0^{-1} )  \frac{\partial \hat F_{T,{\rm tot}}}{\partial z} - A_\gamma  (R_0-1)  \frac{\partial^2 \bar C}{ \partial z^2}   {\rm Nu}_0  , 
\end{eqnarray}
where I have linearized $R$ to go from the second to the third line, and again used $\hat F_{T,{\rm tot}}(R_0)  = - {\rm Nu}_0$ in the last line.  Since ${\rm Nu}_0$ is always positive in fingering systems, we see that the first term behaves anti-diffusively if $A_\gamma > 0$ and diffusively if $A_\gamma < 0$, which confirms the statement made above. Note that in reality the mean-field equations stop being valid on scales approaching the finger scale, and should not be used in that limit. As such, this ultraviolet catastrophe is only a feature of the mean field equations but would not occur in a real system. 

The $\gamma$-instability model has been quantitatively  validated in the geophysical context for fingering convection at high Prandtl number in both 2D \cite{radko2003mechanism} and 3D \cite{Stellmach2011}, confirming its ability to  predict not only when layers should form, but also at what rate (as long as the wavenumber of the layering mode is not too large). As a result, we can use it to determine whether layers are expected to form in stars or not. To do so, we simply need to compute the function $\gamma_{\rm tot}(R)$ and apply the $\gamma$-instability criterion. 

\begin{figure}[h]
\centering
\sidecaption
\includegraphics[width=0.6\textwidth]{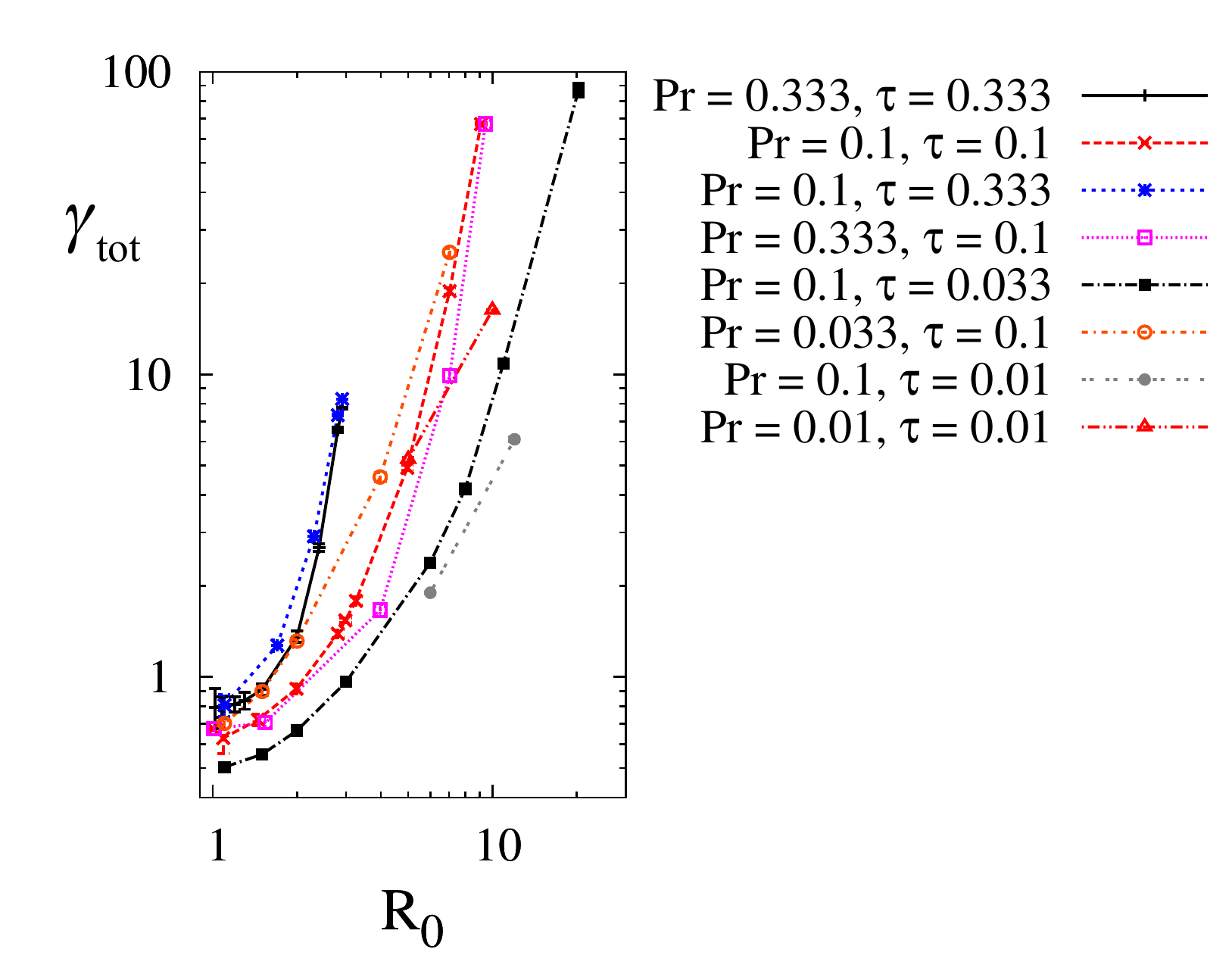}
\caption{The variation of $\gamma_{\rm tot}$ with $R_0$ for a range of fingering simulations at varying ${\rm Pr}$ and $\tau$ smaller than one, showing that this function is always increasing. Figure from \cite{Brownal2013}.}
\label{fig:gammatot}       
\end{figure}

Nondimensionally, in a homogeneous fingering simulation at density ratio $R_0$, 
\begin{equation}
\gamma_{\rm tot}(R_0) = \frac{ \hat F_{T,{\rm tot}}}{\hat F_{C,{\rm tot}}} = \frac{\langle \hat w \hat T \rangle -1  }{\langle \hat w \hat C\rangle - \tau R_0^{-1} } ,
\end{equation}
so we can use the numerically-determined fluxes $\langle \hat w \hat T \rangle$ and $\langle \hat w \hat C \rangle $ from e.g. Traxler et al. and Brown et al. \cite{Traxler2011b,Brownal2013} to compute $\gamma_{\rm tot}(R_0)$ at moderate vales of ${\rm Pr}$ and $\tau$. The results are shown in Figure \ref{fig:gammatot}. In all cases, we see that $\gamma_{\rm tot}$ is a strictly increasing function of the density ratio. The same can be shown to be true at stellar values of ${\rm Pr}$ and $\tau$, with estimates for $\langle \hat w \hat C \rangle$  and $\langle \hat w \hat T \rangle$  obtained using the Brown et al. model. To understand why this is the case, note that the turbulent fluxes decrease significantly when ${\rm Pr}$ and $\tau$ do, and become small compared with the diffusive fluxes \cite{Brownal2013}. As a result, for sufficiently low ${\rm Pr}$ and $\tau$, we have $\gamma_{\rm tot} \simeq R_0 \tau^{-1}$ which clearly increases with density ratio. This effectively demonstrates that {\it layering cannot be produced from the $\gamma-$instability in the stellar regime}.
 
Curiously, Brown et al. \cite{Brownal2013} did report that one of their simulations at very low density ratio spontaneously developed layers. However, these layers were not formed by the $\gamma-$instability, but instead, by the nonlinear development of large-scale internal gravity waves, that were themselves excited by the small-scale fingering. An example of such gravity waves can be seen for instance in the background snapshot of Figure \ref{fig:Feedback}. The spontaneous emergence of internal waves from fingering convection is another well-known large-scale instability, sometimes called the {\it collective instability}, that can also be modeled using mean field hydrodynamics \cite{Stern2001sfu,Traxler2011a,Garaudal2015}. The algebra associated with that calculation is however much more involved than the one outlined above to model the layering instability, so I will not derive it here. The reader is referred to the work of Traxler et al. \cite{Traxler2011a} for details of the calculation (see also \cite{Garaudal2015}). 

Using the turbulent flux laws from Brown et al. \cite{Brownal2013} to close the mean field equations, we were able to determine that fingering convection is expected to excite internal gravity waves down to ${\rm Pr} , \tau \sim 10^{-3}$ but not lower \cite{Garaudal2015,Garaud18}. Furthermore, they can only develop when the density ratio is close to one (i.e. close to the Ledoux limit). This finding rules out the excitation of gravity waves from basic fingering convection in non-degenerate regions of stellar interiors (where ${\rm Pr}, \tau \sim 10^{-6}$), but not in degenerate regions of e.g. WDs and evolved stars. Whether a fingering region would extend all the way into the degenerate region remains to be determined, however. Furthermore, even if gravity waves are indeed excited by this mechanism, they would remain relatively small scale, so would probably not be observable anyway \cite{Garaudal2015}. As such, this effect seems to be more of an interesting curiosity rather than something that is likely to impact stellar evolution and observations\footnote{Of course I would love to be proved wrong.}.
 
 \subsection{Conclusions for now}

Having established that fingering convection is not likely to spontaneously excite larger-scale dynamics (layers or internal gravity waves) at low Prandtl number and diffusivity ratio, we can conclude that in the absence of any other dynamical process (rotation, magnetic fields, shear, etc.), fingering-induced mixing in stars remains small scale, and is well-described by the model of Brown et al. \cite{Brownal2013} (see Section \ref{sec:brown}). The role of rotation, magnetic fields, and shear on fingering-induced mixing is the subject of ongoing work and will be briefly discussed in Section \ref{sec:ccl}.

 \subsection{Applications to stellar astrophysics}
 \label{sec:fingapp}

In Section \ref{sec:wherefing}, we saw a number of examples where fingering convection may occur in stars. Armed with a better quantitative understanding of the process, we are now equipped to answer the question of how much mixing it causes and what its observable effect on stellar evolution may be. Here, it is almost impossible to give a comprehensive review of the topic. Instead, I will focus on a few examples where the newly established transport laws have enabled us to make somewhat definitive statements about the role of fingering in explaining (or not explaining, in some cases) observations. 

\begin{figure}[h]
\centering
\includegraphics[width=0.9\textwidth]{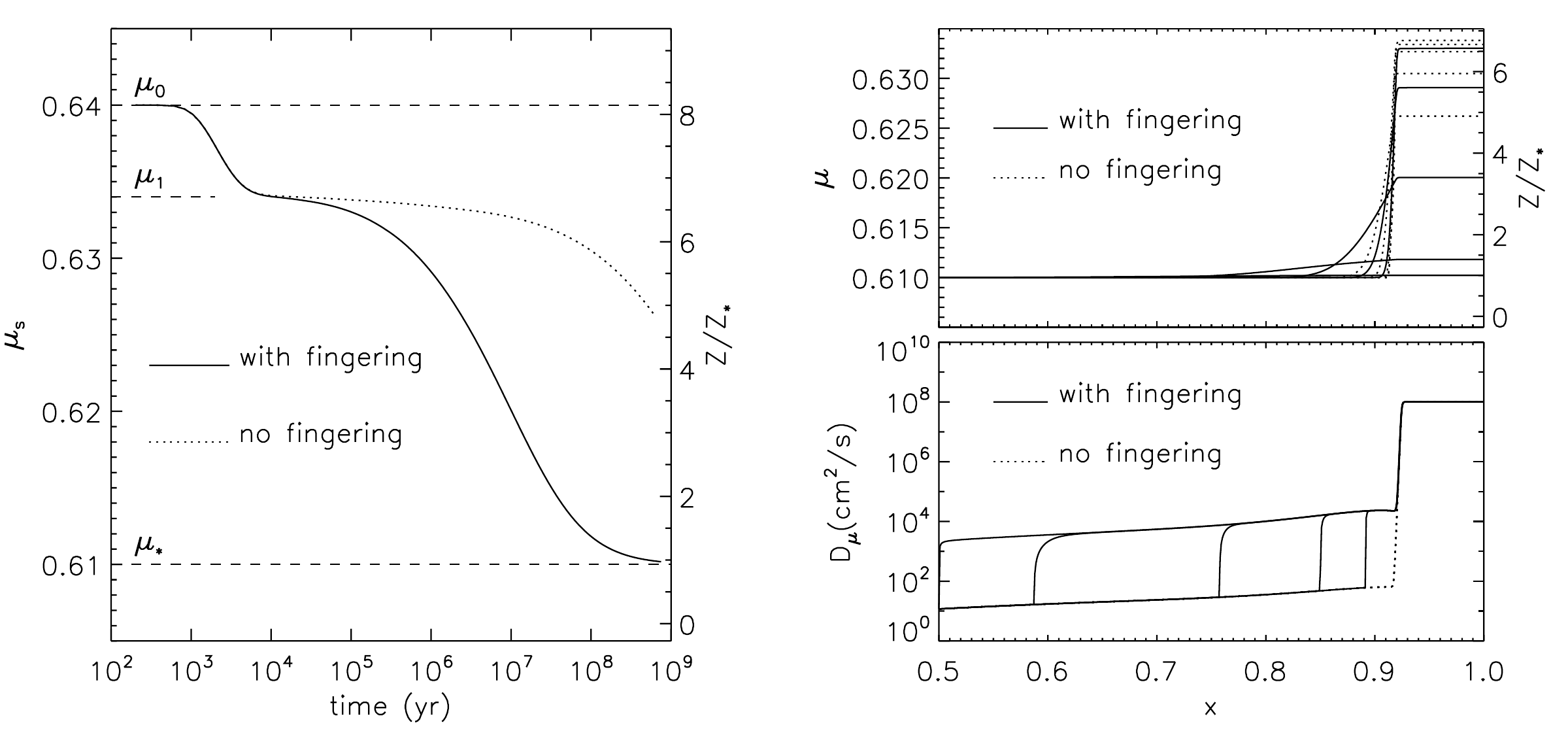}
\caption{Evolution of the metallicity profile and compositional mixing coefficient $D_\mu$ (calculated using equation \ref{eq:Dfingtraxler}), in a 1.4$M_\odot$ model impacted by a 1 Jupiter mass planet. Left: Evolution of the surface metallicity of the star $\mu_s$ with and without taking fingering into account. The early-time drop in $\mu_s$ from the initial value $\mu_0$ of the polluted material to the intermediate value $\mu_1$ is caused by mixing within the surface convection zone, while the later, slower decay in $\mu_s$ corresponds to the effect of mixing below the convection zone (the decay in the case without fingering is due to microscopic molecular diffusion). Right: Evolution of the subsurface metallicity profile (top) and compositional mixing coefficient (bottom) after $10^5$yrs, $10^6$yrs, $10^7$yrs, $10^8$yrs, and 7$\times 10^8$ years. Figure adapted from \cite{Garaud2011}.}
\label{fig:planet}       
\end{figure}

Understanding the surface metallicity of stars undergoing accretion of high-$\mu$ material from infalling planets is one such example. It is thought that, through the combined effect of type I migration in a protostellar disk (which can bring terrestrial planets close to the central star, \cite{LinPap1986}) and tidal interactions (which causes orbital decay of sufficiently close-in planets even in the absence of disk), planets could regularly fall into their host star even after the disk has disappeared \cite{Jacksonal2009}. The event is not expected to have a noticeable effect on the surface metallicity if the star has a deep outer convective zone, but could be important for stars with sufficiently shallow outer convection zones (or none at all).  Planetary infall has thus been proposed as a possible mechanism that could explain the observed planet-metallicity correlation \cite{LaughlinAdams1997,FischerValenti2005}. 
Vauclair \cite{vauclair2004mfa} (see also \cite{TheadoVauclair2012}) correctly argued that this scenario would not work if mixing induced by fingering convection drains the metals into the interior on a short timescale compared with the time since the last infall event (for a given star). Using the mixing coefficient proposed by Traxler et al. \cite{Traxler2011b} (see equation \ref{eq:Dfingtraxler}), which is valid in the limit where the density ratio is not too small (which is the case in these systems), I confirmed Vauclair's idea and demonstrated that any evidence for an infall event would disappear on a timescale of around 100Myr (see Figure \ref{fig:planet}). Since this is relatively short compared with the typical age of planet-bearing stars, we are left to conclude that the planet-metallicity correlation must be of primordial origin \cite{Garaud2011}. 

Fingering convection similarly affects the surface metallicity of WDs undergoing accretion from a planetary debris disk, and this effect should be taken account if one wishes to use the observed metallicities to derive the debris accretion rates \cite{Deal2013,Wachlinal2017,BauerBildsten2018}.

\begin{figure}[h]
\centering
\includegraphics[width=0.9\textwidth]{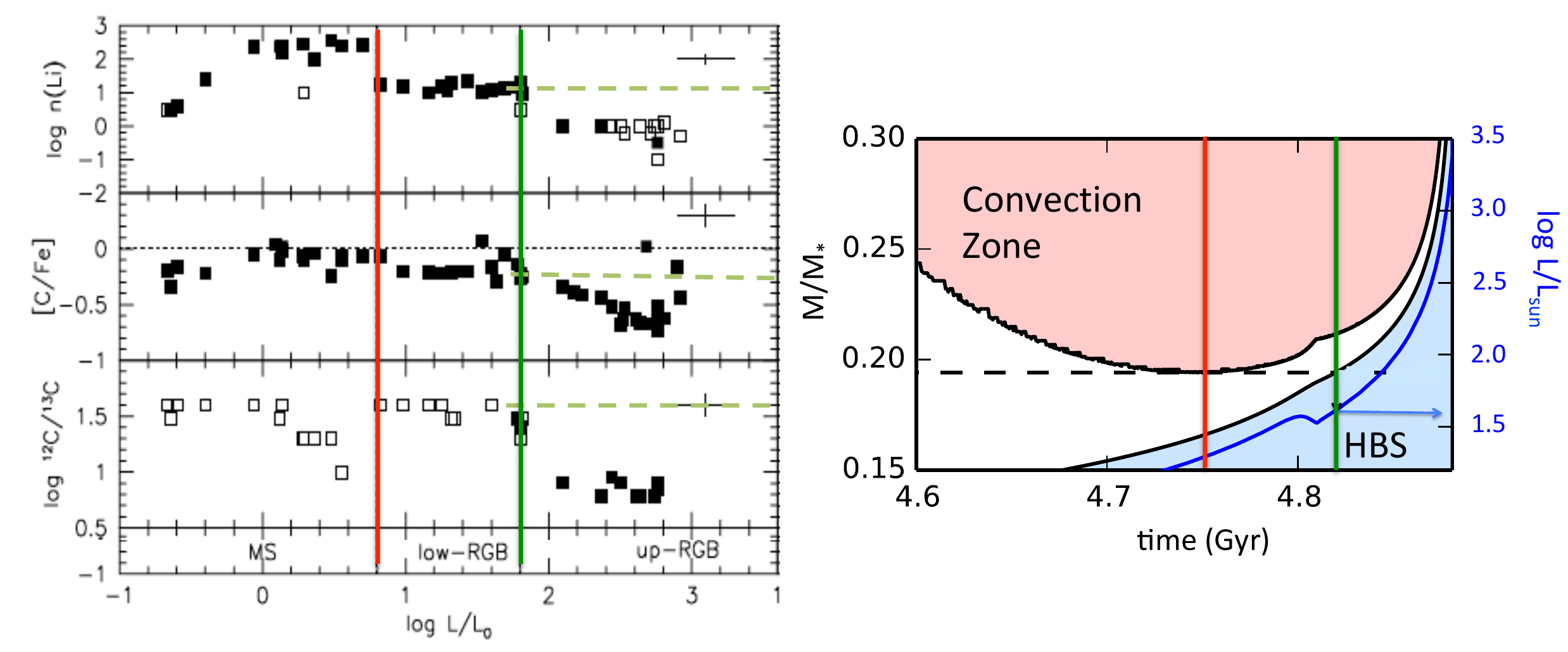}
\caption{Left: Surface abundances of metal poor field stars as a function of luminosity, adapted from Gratton et al. \cite{Gratton2000}. The red line marks the approximate location of the first dredge up, and the green line marks the approximate location of the luminosity bump. The dashed horizontal green line marks the abundances a star would have in the absence of post-dredge-up mixing. Right: Kippenhahn diagram for a 1.3$M_\odot$, $Z = 0.2$ star, illustrating how the luminosity bump occurs when the hydrogen burning shell reaches the radius of lowest descent of the convection zone during prior dredge-up. Figure produced by C. Cadiou using MESA.}
\label{fig:dredgeup}       
\end{figure}

Meanwhile, RGB stars are a good example of objects where a better understanding of fingering convection has made it {\it harder} to explain observations. Detailed spectroscopy of metal-poor stars by Gratton et al. \cite{Gratton2000}
has revealed that the surface abundances of lithium and of elements participating in the CNO cycle changes noticeably along the RGB. A first sudden change occurs as expected during the first dredge-up event (i.e. when the outer convection zone penetrates most deeply into the star), see \cite{SmithTout1992} for early work on the topic. A second much more unexpected change occurs once the convection zone has retreated, around the so-called luminosity bump that corresponds to the time where the hydrogen-burning shell enters the region that was previously mixed in the first dredge up, see Figure \ref{fig:dredgeup}. 

\begin{figure}[h]
\centering
\sidecaption
\includegraphics[width=0.7\textwidth]{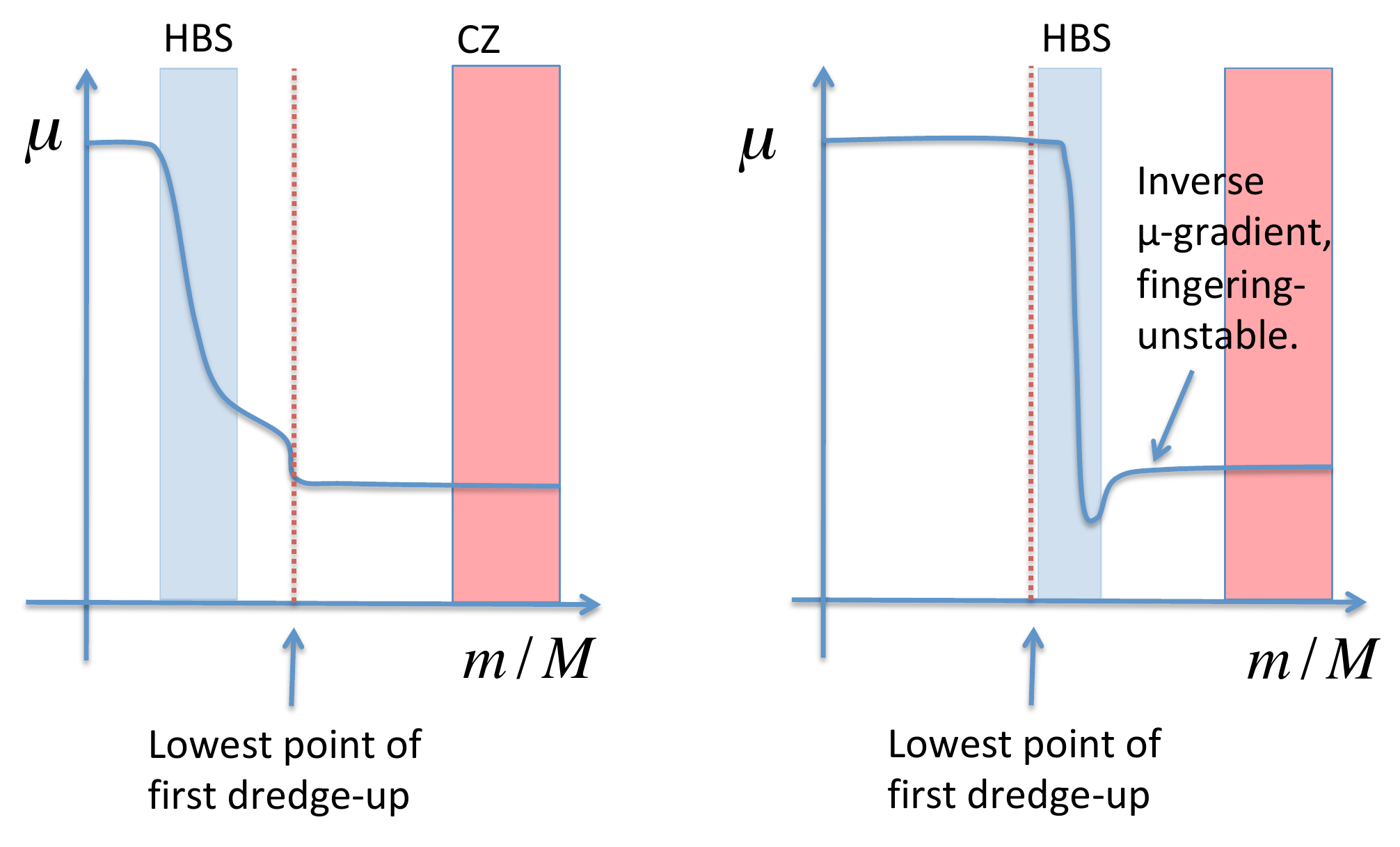}
\caption{Left: Illustrative sketch of the mean molecular weight profile before the luminosity bump. Right: Same but after the luminosity bump. In that case, an inverse $\mu$ gradient forms that drives fingering convection between the hydrogen burning shell (HBS) and the convection zone (CZ).}
\label{fig:charbonnel}       
\end{figure}

Eggleton et al. \cite{Eggletonal2006} noted that $^3$He burning ($^3$He + $^3$He $\rightarrow$ $^4$He +  p + p) is the dominant reaction in the cooler outer edge of the hydrogen burning shell, and locally lowers the mean molecular weight slightly. This can cause an inversion of the mean molecular weight gradient, but only once the shell has moved into a region that was previously homogenized by the dredge-up (see Figure \ref{fig:charbonnel}). In the 3D simulations of Eggleton et al. \cite{Eggletonal2006}, this inversion caused the development of a Rayleigh-Taylor instability, which mixed material between the hydrogen burning shell and the convection zone above. Charbonnel \& Zahn \cite{CharbonnelZahn2007} however pointed out that this inverse $\mu$-gradient would first become unstable to fingering convection, rather than the Rayleigh-Taylor instability. By including the effects of fingering convection in their stellar evolution code, modeled using the traditional formula where $D_{\rm fing} = C_f\kappa_T / (R_0-1)$ (see equation, e.g. \ref{eq:DfingU}), they were able to explain the Gratton et al. data provided $C_f = O(1000)$, but not if $C_f = O(10)$. At the time, this was viewed as evidence in favor of the Ulrich prescription for fingering convection \cite{Ulrich1972}, but we now know that such a high constant $C_f$ is not supported by the numerical experiments presented in this lecture (see Figure \ref{fig:testUKRT} and also \cite{Denissenkov2010,DenissenkovMerryfield2011}). 

This is a very frustrating state of affairs, since in all other respects the Charbonnel \& Zahn model \cite{CharbonnelZahn2007} provides a simple and elegant explanation for the data: fingering convection is triggered at exactly the right time and in the right place to cause the extra mixing needed -- but its efficiency is too low. With this in mind, we dedicated the last 5 years (since 2013) trying to determine if another process, in combination with fingering convection, could increase the induced turbulent transport. This was at the heart of our attempts to establish whether thermo-compositional staircases could form through fingering in stars, but it really seems that they cannot (see Section \ref{sec:largescaleinstab} and \cite{Garaudal2015}), at least spontaneously. It was also what prompted us to study added physics, such as rotation \cite{SenguptaGaraud2018}, magnetic fields \cite{HarringtonGaraud2019} and shear \cite{Garaudal2019}, all of which are present in RGB stars. Of all these processes, magnetic fields appear to be the most promising. Preliminary results for fingering convection in a medium permeated by a vertical background magnetic field of amplitude $B_0$ obtained by Harrington and Garaud \cite{HarringtonGaraud2019} suggest that
\begin{equation}
D_{\rm fing,mag}(B_0) \simeq \frac{2.5 B_0^2}{\rho_m \mu_0} \sqrt{ \frac{N_T^2 + N_C^2}{- N_T^2 N_C^2} }, 
\label{eq:DfingB}
\end{equation}
for sufficiently large magnetic field strengths, where $\mu_0$ is the permeability of the vacuum, and all the other quantities were defined earlier. This coefficient can be two orders of magnitude larger than one appropriate for non-magnetic fingering for $B_0$ as low as a few hundred Gauss, which are plausibly present in these stars \cite{HarringtonGaraud2019}. More work is needed, however, to establish whether this result holds for arbitrarily aligned magnetic fields.

 \section{Oscillatory double-diffusive convection and layered convection}
\label{sec:ODDC}

\subsection{Traditional models of mixing by ODDC}
\label{sec:traditionalODDC}

As discussed in Section \ref{sec:intro}, the existence of regions that are both Schwarzschild-unstable but Ledoux-stable predates the discovery of ODDC by several years, and ODDC did not really become widely known in the astrophysical literature until much later. As such, there are many different prescriptions for mixing in semiconvective regions that have very little to do with the process of ODDC. These prescriptions are usually quite simplistic, either assuming that the region is adiabatically stratified, or purely radiative, or some interpolation between the two regimes, with some associated mixing model for chemical species. 

It was only later that the first attempts to create models of semiconvective mixing based on the physics of ODDC were put forward \cite{Stevenson1977,StevensonSalpeter1977,Stevenson1979,Langer1983,Spruit1992}. Owing to its simplicity, the model of Langer et al. \cite{Langer1983} is probably the most commonly-used one in stellar evolution codes today. It can be derived using the same arguments as the ones I presented in Section \ref{sec:traditionalfing} to recover the Ulrich \cite{Ulrich1972} and Kippenhahn et al. \cite{kippenhahn80} models for fingering convection. If we simply assume that semiconvection acts as a turbulent diffusivity with coefficient $D_C = D_{\rm semi}$, then by dimensional arguments 
\begin{equation}
D_{\rm semi} \propto \lambda d^2  = \hat \lambda \kappa_T 
\end{equation}
As discussed in Section \ref{sec:linearODDC}, for sufficiently large $R_0^{-1}$, $\hat \lambda  \propto \hat k_h^2  / (R_0^{-1} - 1)$, where $\hat k_h$ is close to one. As a result, we can estimate $D_{\rm semi}$ as 
\begin{equation}
D_{\rm semi} = C_{\rm L} \frac{\kappa_T}{R_0^{-1} - 1} , 
\end{equation}
where $C_{\rm L}$ is a constant factor that can only be determined by comparison with experimental data. This expression is very reminiscent of the traditional fingering coefficient $D_{\rm fing}$ (see equation \ref{eq:DfingU}), with $R_0$ replaced by $R_0^{-1}$. Using astrophysical notations that might be more familiar to this audience, this becomes
\begin{equation}
D_{\rm semi} = C_{\rm L} \frac{\kappa_T  \delta ( \nabla_T - \nabla_{\rm ad} ) }{\phi \nabla_\mu -  \delta ( \nabla_T - \nabla_{\rm ad} ) }  , 
\end{equation}
which recovers the prescription of Langer et al. \cite{Langer1983}. The coefficient $C_{\rm L}$ is usually argued to be of order unity. Note that heat transport in this model is assumed to be negligible, so the background temperature gradient is the radiative one. 

Less well know perhaps is the model of Stevenson \cite{Stevenson1979}, who argued that the saturation of the unstable oscillatory modes of ODDC arises from a parametric subharmonic instability, in which smaller scale modes  rapidly grow once the parent mode has reached a certain amplitude. He argues, using analogies with the geophysical literature, that the mixing coefficient should take the form\footnote{See equation 31 of \cite{Stevenson1979}.}
\begin{equation}
D_{\rm semi} = C_{\rm S} R_0^{2} \kappa_T , 
\end{equation}
which, for $R_0^{-1} \gg 1$, only differs from the Langer et al. proposal in the exponent applied to the inverse density ratio $R_0^{-1}$. Stevenson and his colleagues \cite{StevensonSalpeter1977,Stevenson1979} were also the first to clearly argue for the existence of a distinct regime where {\it layered} double-diffusive convection takes place (though Spiegel \cite{Spiegel1969} hinted at its possibility) instead of the small-scale turbulence considered so far, and to discuss where in parameter space it could take place. 

Indeed, as discussed in Section \ref{sec:linearODDC}, the range of linear instability for ODDC in geophysics (and in laboratory experiments) is very small, so ODDC almost never derives from it. Instead, it is excited from a subcritical branch of instability. As a result, none of the existing laboratory experiments on ODDC show the presence of unstable gravity waves, but instead, all take the form of layered double-diffusive convection (layered convection, for simplicity) \cite{TURNER1965,LindenShirtcliffe1978,turner1985mc,radko2013double}.  Layered convection is also ubiquitously found in nature on Earth in the combined presence of unstable temperature gradients and stable salt gradients, such as in volcanic lakes \cite{Wuest2012}, in the arctic ocean \cite{Timmermans2008}, and under the ice shelf in the antarctic \cite{Kimuraal2015}. The temperature and salinity profiles associated with layered convection take the form of a thermohaline staircase similar to fingering staircases discussed in Section \ref{sec:largescaleinstab}, although the overall gradients now have the opposite signs. 

An entirely different class of astrophysical semiconvection models therefore exists that assumes the presence of layered convection \cite{StevensonSalpeter1977,Spruit1992,LeconteChabrier2012,Spruit2013,LeconteChabrier2013}. Despite their differences, these models generally start from the same assumption, namely that the thermocompositional staircase is in equilibrium. This implies that the fluxes of $T$ and $C$ through a layer must be equal to the fluxes of $T$ and $C$ through the adjacent interfaces on either side, otherwise the interfaces and/or the layers would have to evolve on a rapid timescale\footnote{Evolution on the slower stellar evolution timescale is of course allowed.}. As a result of this assumption, one can easily build a transport model from two ingredients only: the first is a prescription for the temperature flux within a convective layer (which is not negligible in layered convection), and the second is a prescription for the {\it ratio} of temperature and compositional fluxes across an interface. Once these two quantities are known, the fluxes of heat and composition through the entire staircase can easily be computed. 
 
 The turbulent temperature flux through individual layers can be put forward on dimensional grounds, to be 
\begin{equation}
F_{T} \equiv \langle wT \rangle  = -  ({\rm Nu}_T - 1) \kappa_T \left( T_{0z} - T_{{\rm ad},z} \right) = \frac{ T}{H_p}  \kappa_T ({\rm Nu}_T - 1)\delta(\nabla_T - \nabla_{\rm ad}) ,
\end{equation}
where ${\rm Nu}_T$ is a potential temperature Nusselt number, defined in Table \ref{tab-parameters}; going from the second to the third expression above uses that definition. Traditional geophysical models of overturning convection between solid boundaries separated by a distance $H$ usually argue \cite{priestley1954convection,Malkus1954,kraichnan1962turbulent,howard1963heat,spiegel1963generalization} that for very turbulent flows the Nusselt number should be equal to some power of the so-called Rayleigh number ${\rm Ra} = |N_T^2| H^4 / \kappa_T \nu$, which is the ratio of the buoyancy force driving convection, to the viscous force that damps fluid motions. In layered convection in stars, however, the viscous force is thought to be negligible, and the relevant Rayleigh number is
\begin{equation}
{\rm Ra}_\star \equiv \frac{|N_T^2| H_L^4}{\kappa_T^2} = {\rm Ra} {\rm Pr} .
\label{eq:Rastar}
\end{equation}
instead, where $H_L$ is the layer height. As such, models of layered convection in stars traditionally have
 \begin{equation}
{\rm Nu}_T - 1  = C_{\rm semi} {\rm Ra}_\star^a ,
\label{eq:NuTsemi}
\end{equation}
where the pre-factor $C_{\rm semi}$, and the power $a$, vary between models (for instance $a = 1/4$ in \cite{Spruit1992} and $a = 1/3$ in \cite{LeconteChabrier2012,Spruit2013}). We see that $F_T$ 
depends rather sensitively on the layer height, which must also be specified by the model. Without further justification, $H_L$ is usually taken to be some fraction of the pressure scaleheight, to be specified by the user. Note that unless $H_L$ is {\it very} small, ${\rm Nu}_T$ is much larger than one, so the diffusive contribution to the heat transport is in this case negligible. 

The second ingredient of these models is a parametrization for the non-dimensional ratio of the temperature flux to the compositional flux, $\gamma_{\rm tot}^{-1}$ (see Table \ref{tab-parameters}). In this case, all traditional astrophysical models of layered convection \cite{StevensonSalpeter1977,Spruit1992,LeconteChabrier2012,Spruit2013} agree with one another and are based on laboratory experiments \cite{TURNER1965,Shirtcliffe1973} and theory \cite{LindenShirtcliffe1978} of layered convection in salt water. In this theory, the interface is assumed to be entirely diffusive, so the ratio of the interfacial fluxes is equal to the ratio of the diffusive fluxes, which in turn depends on the respective temperature and salinity gradients across the interface. With further assumptions, Linden \& Shirtcliffe \cite{LindenShirtcliffe1978} arrive at the conclusion that 
\begin{equation} 
\gamma_{\rm tot}^{-1} = \sqrt{\frac{\kappa_C}{\kappa_T}}  = \tau^{1/2} ,
\end{equation}
(with no proportionality constant between the two expression). This prediction is consistent with many of results obtained in laboratory experiments, and is generally considered to be correct in the geophysical literature. 

This flux ratio can then be used to compute the compositional flux across the staircase given the temperature flux, as in  
\begin{equation}
\frac{\beta F_C}{\alpha F_T} = \tau^{1/2} \rightarrow F_C   = - \frac{\alpha}{\beta} \tau^{1/2}  ({\rm Nu}_T - 1) \kappa_T \left( T_{0z} - T_{{\rm ad},z} \right), 
\end{equation}
which in turn implies that the compositional mixing coefficient for layered semiconvection can be written as 
\begin{equation}
D_{\rm semi} = - \frac{F_C }{ C_{0z} } =  \gamma_{\rm tot}^{-1}  ({\rm Nu}_T - 1) \frac{ \kappa_T }{R_0^{-1}} ,
\label{eq:Dsemi}
\end{equation}
where the Nusselt number is given in (\ref{eq:NuTsemi}) and $\gamma_{\rm tot}^{-1} = \tau^{1/2}$. Again, unless the layer height is very small, this quantity is usually much larger than the microscopic diffusion coefficient $\kappa_C$, which can be neglected. 

\subsection{Numerical simulations of ODDC}

\begin{figure}[h]
\centering
\includegraphics[width=\textwidth]{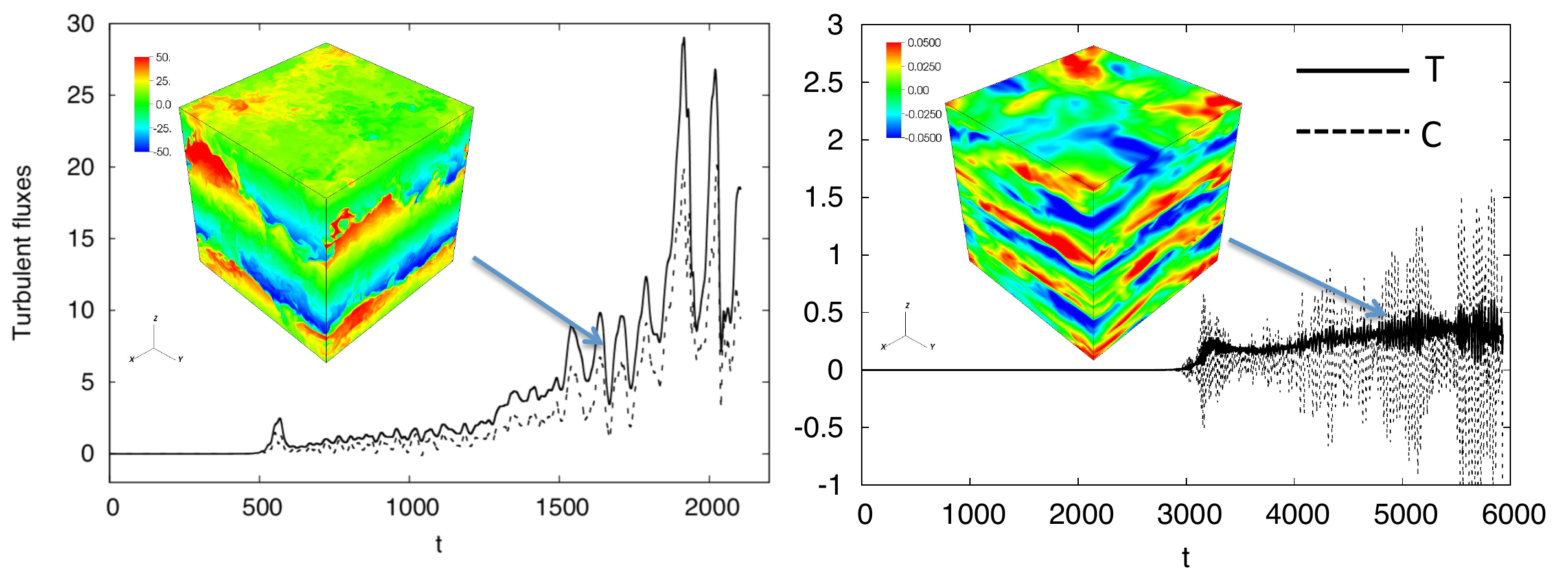}
\caption{Snapshots of the compositional perturbations in the two forms of ODDC, extracted from simulations at ${\rm Pr} = \tau = 0.03$, see text for detail. Left: Layering at low inverse density ratio ($R_0^{-1} = 1.5$) causes substantial increase in the turbulent temperature (solid line) and compositional (dotted line) fluxes. Right: No layering at high inverse density ratio ($R_0^{-1} = 5$). The turbulent fluxes remain weak. Figure from \cite{Garaud18}, with data from \cite{Mirouh2012}.}
\label{fig:ODDCtypes}       
\end{figure}

As in the fingering case, none of these models had ever been tested in the low Prandtl number regime more appropriate of stellar astrophysics until recently. In particular, whether ODDC takes the form of small-scale wave-like turbulence or layered convection remained unknown. Early simulations of ODDC/semiconvection in a two-dimensional vertically bounded domain were presented by Merryfield \cite{Merryfield1995} (see also \cite{ZaussingerSpruit2013}), but the results were limited in scope by the resolution affordable at the time. The first three-dimensional direct numerical simulations of ODDC at low Prandtl number were presented by Rosenblum et al. \cite{rosenblumal2011}, using the PADDI code \cite{Traxler2011a} with the model setup described in Section \ref{sec:model}, and solving equations (\ref{eq:momnondim})--(\ref{eq:compnondim}) with the $-$ sign in the temperature and composition equations. In this paper, we demonstrated that {\it both} layered and non-layered outcomes are possible, depending on the local thermocompositional stratification (as measured by the inverse density ratio $R_0^{-1}$). This was later confirmed by Mirouh et al. \cite{Mirouh2012}, who performed a more comprehensive exploration of parameter space. These two possible outcomes are illustrated in Figure \ref{fig:ODDCtypes}. 

For high inverse density ratios, i.e. for systems that are more strongly stratified, we have found that ODDC is excited as expected from the linear theory described in Section \ref{sec:linearODDC}, and saturates into a state of weak wave turbulence. The fastest-growing modes are, as expected, elevator modes. However, once they reach a certain amplitude, nonlinear interactions causes a transfer of energy to modes that have a higher vertical wavenumber (see Figure \ref{fig:ODDCtypes}b). This is at least qualitatively consistent with the idea put forward by Stevenson \cite{Stevenson1979}. The turbulent fluxes in that regime are fairly weak, and decrease with increasing $R_0^{-1}$, and decreasing ${\rm Pr}$ and $\tau$ (see below). 

For low inverse density ratios, on the other hand, the initial state of wave-like turbulence always transitions to a layered state. The initial height of the convective layers is fairly small, of the order of a few tens of $d$, but the layers then always merge until a single one is left in the computational domain (the mergers cannot proceed beyond that owing to the periodicity of the boundary conditions). The initial formation of the layers, and each subsequent merger, is accompanied by a substantial increase in the turbulent fluxes, suggesting that the latter indeed depend on the layer height.  

In view of these results, it is clear that any model of ODDC/semiconvection in stars should include a way to determine which regime is expected (layered or non-layered), a prescription for transport in the layered regime and a prescription for transport in the non-layered regime \cite{Mirouh2012}. 

\subsection{Transport in non-layered ODDC}
\label{sec:nonlayeredtransport}

Starting with the latter, we can use the simulations of Mirouh et al. \cite{Mirouh2012} to extract turbulent fluxes in non-layered convection\footnote{For simulations that become layered, we extract the fluxes prior to layer formation}, and compare them to the predictions of Stevenson \cite{Stevenson1979} and Langer et al. \cite{Langer1983} (which are for the non-layered regime). The turbulent diffusion coefficient $D_{\rm semi}$ is extracted from the simulations as usual by computing $- \langle wC \rangle / C_{0z} = R_0 \kappa_T  \langle \hat w \hat C \rangle$. As a side note, it is easy to show using arguments very similar to those put forward in the fingering regime (see Section \ref{sec:numfing}), that both $\langle \hat w \hat C \rangle$ and $\langle \hat w \hat T \rangle$ have to be positive in a statistically stationary state. As such, $D_{\rm semi}$ is indeed positive. 

\begin{figure}[h]
\centering
\includegraphics[width=\textwidth]{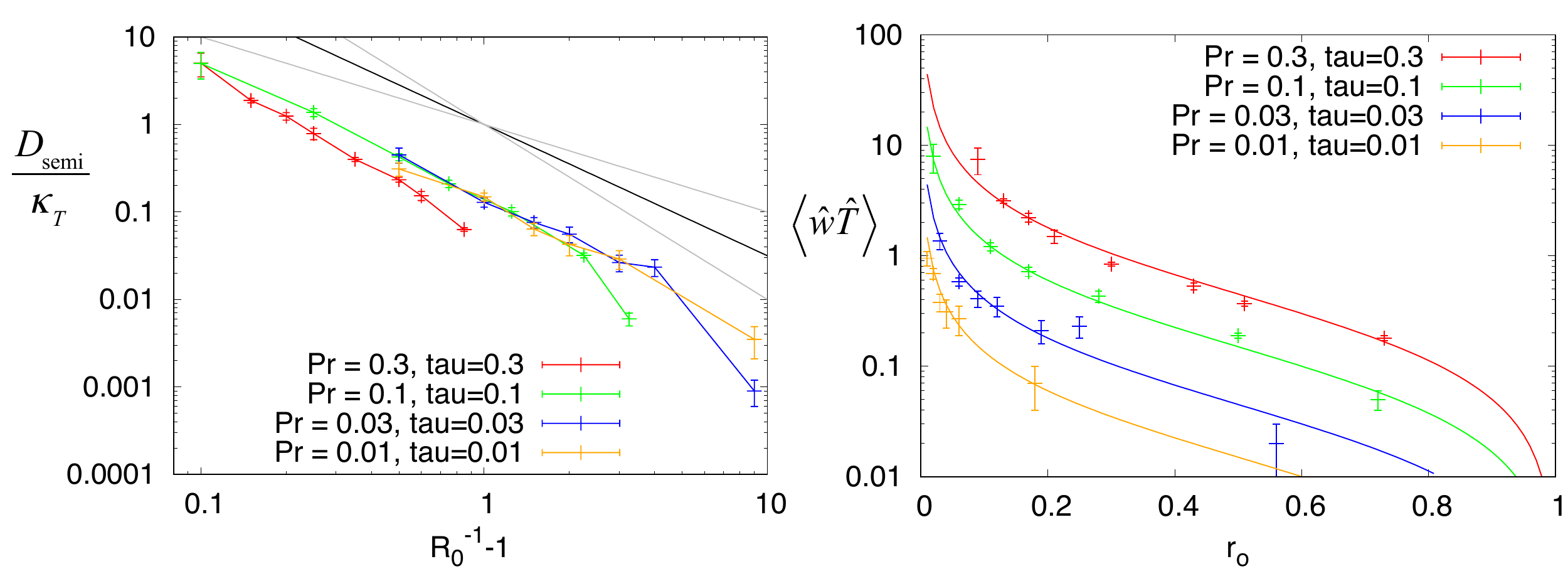}
\caption{Left: Variation of the ratio $D_{\rm semi}/\kappa_T$ with $R_0^{-1}-1$, in the data, and comparison with models where $D_{\rm semi}/\kappa_T \propto (R_0^{-1}-1)^{-2}$ and $(R_0^{-1}-1)^{-1}$ respectively (grey lines). A better fit to the data suggests that $D_{\rm semi}/\kappa_T \propto (R_0^{-1}-1)^{-1.5}$ (black line), except when $R_0^{-1}$ approaches marginal stability. Right: Non-dimensional turbulent temperature flux. The symbols are from the data \cite{Mirouh2012}, and the lines are from the model given in equation (\ref{eq:MirouhNut}).}
\label{fig:Mirouhfluxes}       
\end{figure}

The results are shown in Figure \ref{fig:Mirouhfluxes}a.  In the model of Stevenson \cite{Stevenson1979}, $D_{\rm semi}/\kappa_T \propto R_0^2$ while in Langer et al. \cite{Langer1983} $D_{\rm semi}/\kappa_T \propto (R_0^{-1}-1)^{-1}$. 
The data clearly favors a model where $D_{\rm semi}/\kappa_T$ is a power law function of $R_0^{-1}-1$ and the slope is roughly equal to $-1.5$, which is somewhere in between these existing models, at least for large enough $R_0^{-1}$. We also see that these models fail to account for the stabilization of the system (with $D_{\rm semi}\rightarrow 0$) that occurs when $R_0^{-1}$ approaches the marginal stability threshold $({\rm Pr}+1)/({\rm Pr} + \tau)$. 

In an attempt to create a new model for mixing by ODDC in the non-layered regime, we (i.e. me and my students) have tried various approaches, including an extension of the Brown et al. model to the ODDC case, and various attempts at weakly nonlinear theory to capture the nonlinear evolution of the unstable gravity waves. None of these approaches were able to model the variation of $D_{\rm semi}$ with all input parameters (${\rm Pr}$, $\tau$, and $R_0^{-1}$) satisfactorily. This does not mean that such a theory cannot be developed -- only that we have been unable to do so to date. For lack of a more theoretically-grounded model, Mirouh et al. \cite{Mirouh2012} put forward an {\it empirical} parametrization of the existing data that seems to capture adequately the variation of both compositional and temperature fluxes across parameter space. In this parametrization, the non-dimensional turbulent temperature flux $\hat F_T = \langle \hat w \hat T \rangle$ is proposed to be 
\begin{equation}
\hat F_T = 0.75 \left( \frac{\rm Pr}{\tau} \right)^{1/4} \frac{1-\tau}{R_0^{-1} - 1 } (1-r_o),
\label{eq:MirouhNut}
\end{equation}
where $R_{oc}^{-1} = ({\rm Pr}+1)/({\rm Pr} + \tau)$ is the marginal stability threshold for ODDC, and 
\begin{equation}
r_o = \frac{R_0^{-1} - 1}{R_{oc}^{-1} - 1} ,
\end{equation}
is the reduced inverse density ratio associated with ODDC. This parameter, as in the fingering case, maps the instability range $R_0^{-1} \in [1,R_{oc}^{-1}]$ to the interval $r_o \in [0,1]$. A comparison of (\ref{eq:MirouhNut}) with data from \cite{Mirouh2012} is shown in Figure \ref{fig:Mirouhfluxes}b, suggesting that this models is adequate. We also note that $\hat F_T $ rapidly decreases with ${\rm Pr}$ and $\tau$, to the extent that for any reasonable stellar value $\hat F_T $ is either smaller than or negligible compared to the diffusive flux (which is one in these units). We therefore confirm that the turbulent heat flux is negligible in comparison with the diffusive flux for non-layered ODDC in stars. 

To model the compositional flux, we could use a similar empirical parametrization, starting for instance from the scaling with $R_0^{-1}-1$ discovered in Figure \ref{fig:Mirouhfluxes}a; this may be an interesting idea to pursue in the future. 
Generally speaking, the data suggests that $D_{\rm semi}$ rapidly decreases with increasing stratification. Moll et al. \cite{Moll2016} in fact suggested that compositional mixing in the non-layered regime (which takes place at large  $R_0^{-1}$) is negligible compared with pure diffusive transport at stellar parameters. This remains to be confirmed. 

As an alternative approach, in Appendix A.3. of Mirouh et al. \cite{Mirouh2012} we extended the work of Schmitt \cite{schmitt1979fgm,schmitt1983} (also used by Brown et al. \cite{Brownal2013}), who noted that it is possible to estimate the {\it ratio} of the turbulent temperature flux to the turbulent compositional flux, named $\gamma_{\rm turb}$ (see Table \ref{tab-parameters}) in double-diffusive systems simply from linear theory, even though nonlinear arguments are needed to estimate each of them individually. Indeed, using non-dimensional fluxes, we can write $\gamma_{\rm turb} = \langle \hat w \hat T \rangle / \langle \hat w \hat C \rangle$. Assuming that $\hat w$, $\hat T$ and $\hat C$ are dominated by the fastest-growing linearly unstable modes, we can replace them by their proposed ansatz from linear theory (see Section \ref{sec:linear}), to get 
\begin{equation}
\gamma_{\rm turb} = \frac{\langle  {\cal R}e \left[ \tilde w  \exp(i \hat k_h x + \hat \lambda t)  \right]   {\cal R}e \left[ \tilde T  \exp(i \hat k_h x + \hat \lambda t)  \right]   \rangle }{\langle  {\cal R}e \left[ \tilde w  \exp(i \hat k_h x + \hat \lambda t)  \right]   {\cal R}e \left[ \tilde C  \exp(i \hat k_h x + \hat \lambda t)  \right]  \rangle } , 
\label{eq:gammaturb1}
\end{equation}
 In this expression $\hat \lambda$ and $\hat k_h$ are the growth rate and wavenumber of the fastest-growing elevator modes for the desired set of parameters ${\rm Pr}$, $\tau$ and $R_0^{-1}$, computed from linear theory (see Section \ref{sec:linear}). From the linearized versions of equations (\ref{eq:momnondim})--(\ref{eq:compnondim}), we have the following system of equations relating the amplitudes $\tilde w$, $\tilde T$ and $\tilde C$ of each mode. 
\begin{equation}
\hat \lambda \tilde  T \pm \tilde w = - \hat k_h^2  \tilde  T , \quad \hat \lambda \tilde  C \pm R_0^{-1}  \tilde w = - \tau \hat k_h^2  \tilde  C, 
\end{equation}
where as usual $+$ refers to the fingering regime and $-$ refers to ODDC. Solving for $\tilde T$ and $\tilde C$, we get
\begin{equation}
\tilde T   = \mp  \frac{\tilde w}{\hat \lambda +  \hat k_h^2 } , \quad  \tilde  C = \mp  \frac{ R_0^{-1}  \tilde w }{ \hat \lambda +  \tau \hat k_h^2  } .
\label{eq:gammaturb2}
\end{equation} 
If we finally assume, without loss of generality, that the amplitude $\tilde w$ is real, then we combine equations (\ref{eq:gammaturb1}) and (\ref{eq:gammaturb2}) to get the following prediction for the turbulent flux ratio $\gamma_{\rm turb}$ \cite{Mirouh2012}:
\begin{equation}
\gamma_{\rm turb} = R_0 \frac{ ( \hat \lambda_R + \tau \hat k_h^2)^2 + \hat \lambda_I^2}{  ( \hat \lambda_R +  \hat k_h^2)^2 + \hat \lambda_I^2  }  \frac{ \hat \lambda_R +  \hat k_h^2}{  \hat \lambda_R + \tau  \hat k_h^2  }  , 
\label{eq:gammaturbpred}
\end{equation}
where we have written $\hat \lambda = \hat \lambda_T + i \hat \lambda_I$. Conveniently, the $\pm$ signs cancel out, so this expression can be used for {\it both} fingering convection and ODDC. Even more conveniently, the unknown amplitude of the mode $\tilde w$ cancels out entirely from that expression, so $\gamma_{\rm turb}$ does not contain any unknown parameter. 

A comparison of this theory with the simulation data (in terms of $\gamma_{\rm tot}$, rather than $\gamma_{\rm turb}$) is presented in the next section, and shows reasonably good (but not perfect) agreement between the two.  This expression can finally be used to compute the dimensional mixing coefficient 
\begin{equation}
D_{\rm semi} = R_0\kappa_T  \langle \hat w \hat C \rangle  = R_0  \kappa_T \gamma_{\rm turb}^{-1} \langle \hat w \hat T \rangle, 
\end{equation}
with $\langle \hat w \hat T \rangle$ given by (\ref{eq:MirouhNut}). 

One may rightfully argue that this method is much more complicated than proposing a simple empirical formula for $D_{\rm semi}$ based on the data shown in Figure \ref{fig:Mirouhfluxes}, which would be more practical for stellar evolution purposes. I do not disagree, especially since we find that in most instances $D_{\rm semi} \ll \kappa_C$ at stellar parameters, making the additional effort of computing $\gamma_{\rm turb}$ (which involves solving a cubic equation) somewhat pointless. It does however have the advantage (in my opinion) of being based on solid theoretical arguments for {\it why} the turbulent flux ratio $\gamma_{\rm turb}$ should be of a particular form, and can in addition be used to establish a criterion for layer formation, as shown below. 

\subsection{Criterion for layer formation}
\label{sec:layercrit}

As discovered by Rosenblum et al. \cite{rosenblumal2011} and Mirouh et al. \cite{Mirouh2012}, simulations of ODDC at a low inverse density ratio always spontaneously evolve into a state of layered convection, suggesting that the mechanism leading to layer formation must be very robust. In Section \ref{sec:largescaleinstab}, we learned about the $\gamma$-instability as a mechanism for layer formation in oceanographic fingering convection, and found that it should occur whenever the total flux ratio $\gamma_{\rm tot}$ is a decreasing function of the density ratio $R_0$. It is therefore natural to wonder whether a similar mechanism may be at play here. 

As it turns out, the mechanism is not just similar, it is indeed exactly the same. To see this, simply note that the horizontally-averaged temperature and composition equations which were used as the starting point for the development of the $\gamma-$instability theory in Section \ref{sec:largescaleinstab} (see equation \ref{eq:horizav}) are exactly the same in fingering convection and in ODDC, since the horizontal average of the vertical advection terms containing $\pm w$ are identically zero (for mass conservation). As a result, Rosenblum et al. \cite{rosenblumal2011} demonstrated that the criterion for layer formation in ODDC is the same as in fingering convection: layers are expected to form when $\gamma_{\rm tot}$ is a decreasing function of $R_0$, or equivalently, when $\gamma_{\rm tot}^{-1}$ is a decreasing function of $R_0^{-1}$. 

\begin{figure}[h]
\centering
\includegraphics[width=\textwidth]{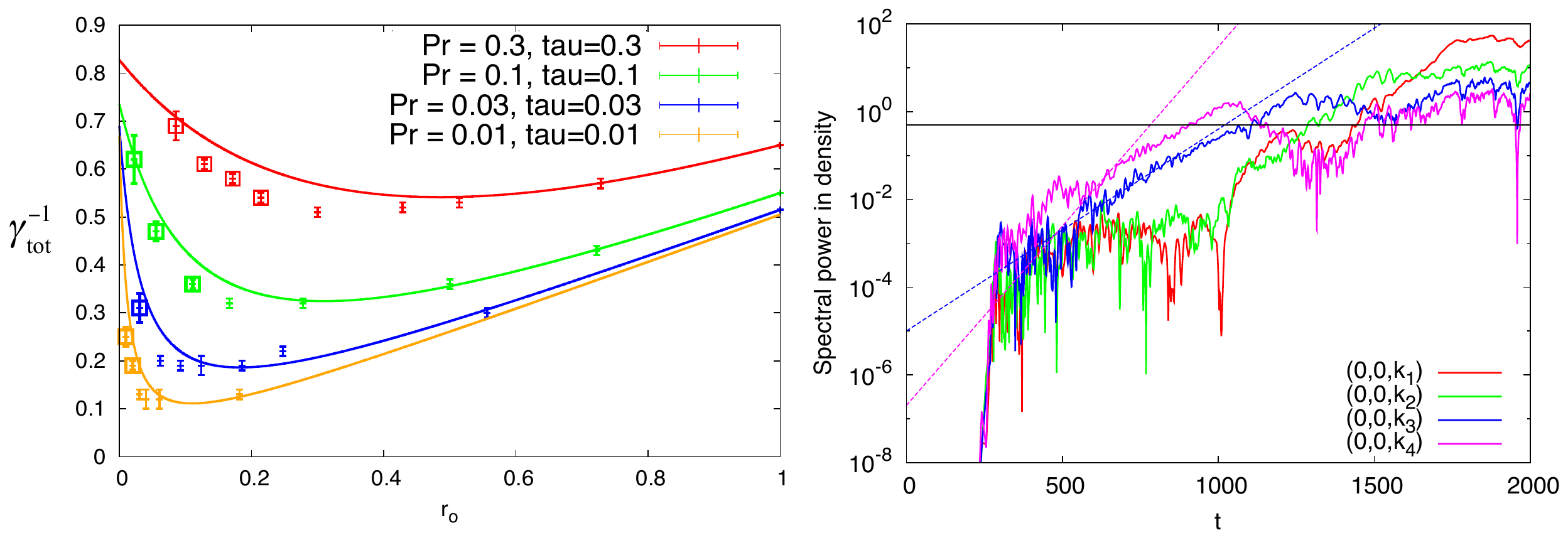}
\caption{Left: The symbols show the total inverse flux ratio $\gamma_{\rm tot}^{-1}$ measured from simulation data in non-layered ODDC for various ${\rm Pr}$ and $\tau$ (see \cite{Mirouh2012} for a more extended dataset). Larger symbols denote simulations that ultimately become layered. The solid lines are predictions for $\gamma_{\rm tot}^{-1}$ from the model of Mirouh et al.  \cite{Mirouh2012}. Figure adapted from \cite{Garaud18}. Right: The solid lines show the growth of the spectral power in the horizontally-average density field for various layering modes with wavenumber $k_n$ (where $n$ is the number of layers in the domain) for a simulation with ${\rm Pr} = \tau  = 0.3$ and $R_0^{-1} = 1.2$ \cite{rosenblumal2011}. The 3- and 4-layered modes grow until they reach an amplitude that causes inversions in the density profile (black line), at which point layered convection begins. The 1- and 2-layered modes do not initially grow but emerge out of subsequent layer mergers. The dotted lines show the corresponding prediction from the $\gamma-$instability theory for the growth of the 3-layered and 4-layered mode.}\label{fig:GammaMirouh}       
\end{figure}

This theory can be tested from the extended dataset of Mirouh et al. \cite{Mirouh2012}, and some of the results are presented in Figure \ref{fig:GammaMirouh}a. All symbols in this figure show the measured values of $\gamma_{\rm tot}^{-1}$ extracted from the data, by computing 
\begin{equation}
\gamma_{\rm tot}^{-1} =  \frac{ \tau R_0^{-1} +  \langle \hat w \hat C \rangle }{ 1 + \langle \hat w \hat T \rangle}
\label{eq:oddcgammatot}
\end{equation}
in the non-layered phase (this expression is the non-dimensional version of the one given in Table \ref{tab-parameters}). {\it Large} symbols in this figure denote simulations that eventually transitioned into layered convection. We see that this indeed happens whenever $\gamma_{\rm tot}^{-1}$ is a decreasing function of $R_0^{-1}$, confirming the role of the $\gamma$-instability as suspected. 

A more quantitative way of testing the theory is to compare the predicted and observed growth rates of the layering modes in simulations. Let's take for instance a simulation presented by \cite{rosenblumal2011}, with ${\rm Pr} = \tau  = 0.3$ and $R_0^{-1} = 1.2$. In this simulation, 4 layers were first observed to form, though by inspection of the spectral power in the horizontally-averaged density field (see Figure \ref{fig:GammaMirouh}b) we see that the 3-layer mode is also growing. In order to compute the theoretical growth rate $\hat \Lambda$ of the layering modes, we first need to estimate the parameters of the quadratic dispersion relation (\ref{eq:gammaquadratic}). We can easily measure the potential temperature Nusselt number ${\rm Nu}_0$ and the total flux ratio $\gamma_0$ in the non-layered phase preceeding layer formation in the simulation. We can also, by running simulations at slightly different $R_0^{-1}$, compute $A_{\rm Nu} = R_0 \left. \frac{\partial {\rm Nu}_T}{\partial R}  \right|_{R=R_0}$ and $A_\gamma = R_0 \left. \frac{\partial \gamma_{\rm tot}^{-1}}{\partial R}  \right|_{R=R_0}$. We find that 
\begin{equation}
{\rm Nu}_0 \simeq 3.4, \gamma_0^{-1} \simeq 0.55, A_{\rm Nu} \simeq 12.9  \mbox{  and } A_{\gamma} \simeq 0.45, 
\end{equation}
for ${\rm Pr} = \tau  = 0.3$ and $R_0^{-1} = 1.2$.
Using these parameters and coefficients in the quadratic equation (\ref{eq:gammaquadratic}), we then predict that 
\begin{equation}
\hat \Lambda(\hat K) \simeq 0.47 \hat K^2
\end{equation}
for this simulation. 

We can compare the temporal evolution of the spectral power in density for each growing mode with this prediction, knowing their wavenumber is $\hat K = k_n$, where $k_n$ is the wavenumber with $n$ layers in the domain. 
We see in Figure \ref{fig:GammaMirouh}b that the growth rate of the 4-layer mode is over-predicted by the model, but that of the 3-layer mode is adequately captured. As discussed in Section \ref{sec:largescaleinstab}, the fact that the mean-field theory fails when the mode wavenumber is large is not surprising given the ultraviolet catastrophe inherent to the mean field model, but it is very encouraging to see that it works well for the 3-layer mode. Similar comparisons were later made by Mirouh et al. \cite{Mirouh2012}, with equally satisfactory success. This completes the validation of the $\gamma-$instability theory for layer formation in ODDC. 

If one believes the Mirouh et al. \cite{Mirouh2012} prescriptions for the turbulent heat flux and turbulent flux ratio described above, it is easy to predict {\it when} layers are expected to form in stellar ODDC. Indeed, from (\ref{eq:oddcgammatot}) and the definition of $\gamma_{\rm turb} = \langle \hat w \hat T \rangle / \langle \hat w \hat C \rangle$ we have 
\begin{equation}
\gamma_{\rm tot}^{-1} =  \frac{ \tau R_0^{-1} +  \gamma^{-1}_{\rm turb} \langle \hat w \hat T \rangle }{ 1 + \langle \hat w \hat T \rangle }.
\end{equation}
We can then use (\ref{eq:MirouhNut}) and (\ref{eq:gammaturbpred}) to predict $\gamma_{\rm tot}^{-1}$ as a function of $R_0^{-1}$ for any parameter set $({\rm Pr},\tau)$. A comparison of this prediction with existing data is shown in Figure \ref{fig:GammaMirouh}a, and reveals very good agreement (see \cite{Mirouh2012} for a comparison with a more extended dataset), giving us some confidence in the model. Using this method, we can determine when $\gamma_{\rm tot}^{-1}$ is a decreasing function of $R_0^{-1}$ for any ${\rm Pr}$ and $\tau$ in the stellar parameter regime. 

\begin{figure}[h]
\centering
\sidecaption
\includegraphics[width=0.6\textwidth]{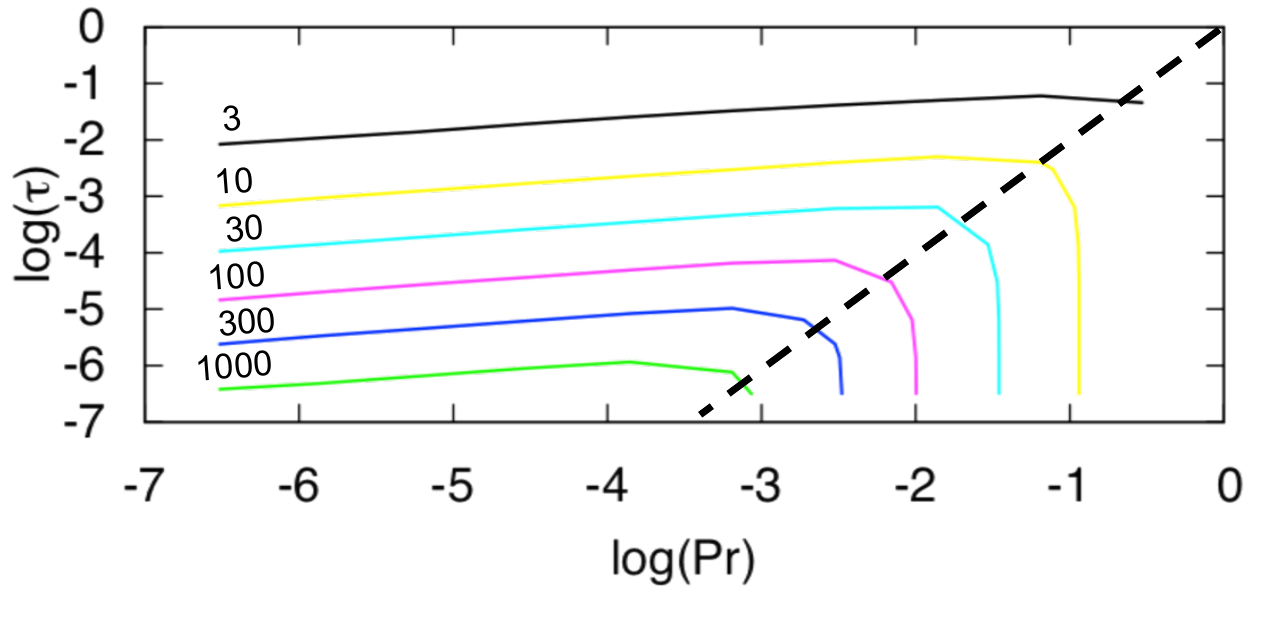}
\caption{Contour lines of value of the inverse density ratio $R_0^{-1}$ below which layered convection is expected, as a function of Pr and $\tau$, according to the model of Mirouh et al. \cite{Mirouh2012}. The dashed line is the ${\rm Pr} = \tau^{1/2}$ line. Figure adapted from \cite{Mirouh2012}.}\label{fig:layercrit}       
\end{figure}

The results were first presented by Mirouh et al. \cite{Mirouh2012} and are shown in Figure \ref{fig:layercrit}, which is a contour plot of the critical value of $R_0^{-1}$ below which layering can occur, as a function of both ${\rm Pr}$ and $\tau$. Generally speaking, we see that when ${\rm Pr} > \tau^{1/2}$ the layering threshold is more-or-less independent of $\tau$ and is roughly equal to ${\rm Pr}^{-1}$. This is not a particularly relevant situation in stellar astrophysics, however. On the other hand, when ${\rm Pr} < \tau^{1/2}$, then the layering threshold is roughly equal to $\tau^{-1/2}$ (with a weak dependence on Pr). For stellar interiors with Pr and $\tau$ of order $10^{-6}$, we therefore expect layers to spontaneously form whenever the inverse density ratio $R_0^{-1} < O(10^3)$, implying that all cases of semiconvective zones near a convective core should lie in the layered regime. For planetary interiors or degenerate regions of stellar interiors with Pr and $\tau$ of order $10^{-2}$, we expect layers to spontaneously form whenever the inverse density ratio $R_0^{-1} < O(10)$ \cite{Mirouh2012}.

\subsection{Mixing in layered convection}

Having established {\it when} layered convection is naturally expected, we now turn to the question of modeling turbulent transport induced by the process. Figure \ref{fig:ODDCtypes}a revealed that the total heat and compositional flux through layered convection depends on the layer height, and increases every time a merger occurs. To study this more quantitatively, Wood et al. \cite{Woodal13} (see also earlier preliminary work by \cite{rosenblumal2011}) re-analyzed the simulations of Mirouh et al. \cite{Mirouh2012} that became layered, and also presented new simulations that were produced using the same model setup and code, but in much larger domains. They measured the average turbulent temperature and compositional fluxes in the domain as a function the average layer height. With this information, the Rayleigh number ${\rm Ra}_\star$, the potential temperature Nusselt number ${\rm Nu}_T$ and the total flux ratio $\gamma_{\rm tot}^{-1}$ can be computed and compared with the models introduced in equation (\ref{eq:Rastar}) for layered convection. 

\begin{figure}[h]
\centering
\includegraphics[width=\textwidth]{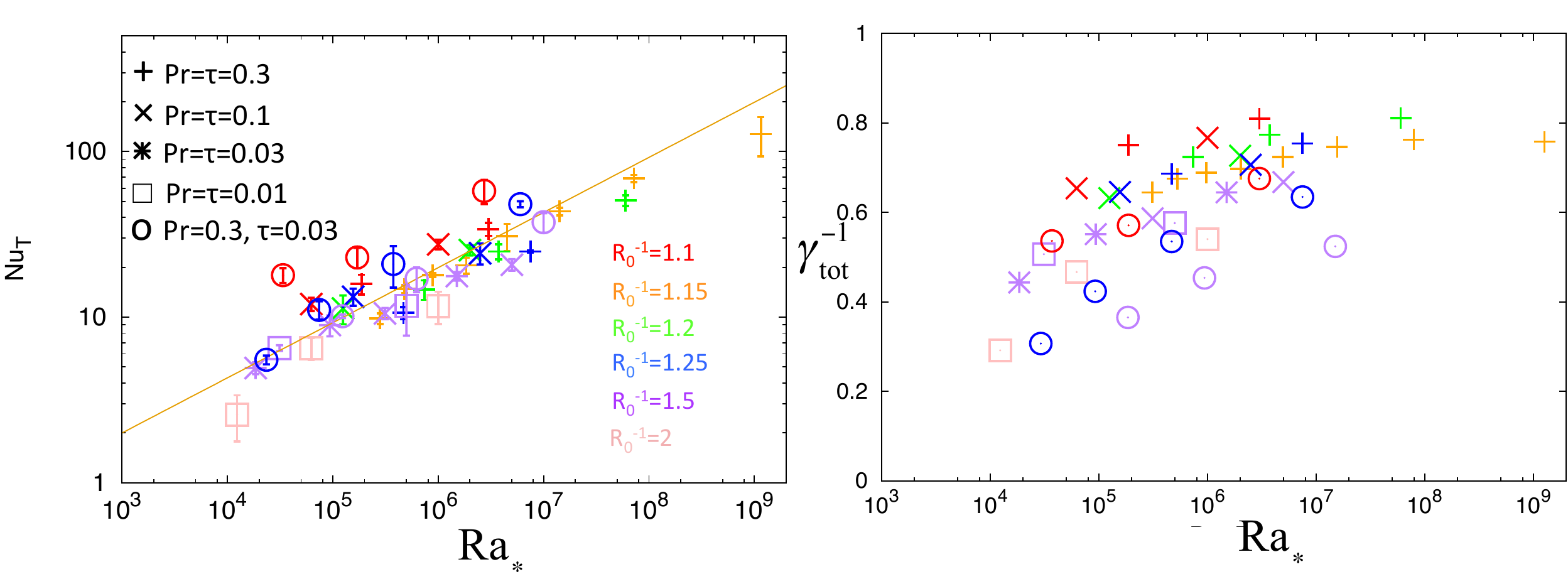}
\caption{Left: Nusselt number as a function of Rayleigh number Ra$_\star$ for simulations at various Pr, $\tau$ (see symbols legend) and $R_0^{-1}$ (see color legend). The solid line shows the relationship ${\rm Nu}_T = 0.2 {\rm Ra}_\star^{1/3}$. Right: Total inverse flux ratio $\gamma_{\rm tot}^{-1}$ as a function of Ra$_\star$  for the same simulations. Figure from \cite{Garaud18} based on data from \cite{Woodal13}.}\label{fig:layerflux}       
\end{figure}

The results for existing simulations with varying ${\rm Pr}$, $\tau$ and $R_0^{-1}$ are shown in Figure \ref{fig:layerflux}. Figure \ref{fig:layerflux}a reveals that ${\rm Nu}_T$ is indeed a power law function of Ra$_{\star}$, with:  
\begin{equation}
{\rm Nu}_T \simeq 1+ C_{\rm semi}(\tau,R_0^{-1}) {\rm Ra}_\star^{1/3}, 
\label{eq:NutWood}
\end{equation}
with $C_{\rm semi} \sim O(0.1)$ when fitted to the existing data. 
The exponent of that power law therefore rules out \cite{Spruit1992} but is consistent with \cite{LeconteChabrier2012,Spruit2013}. The pre-factor $C_{\rm semi}$ seems to have an additional weak dependence on both $R_0^{-1}$ and $\tau$, which could not be determined without much more demanding numerical simulations \cite{Woodal13}. Indeed, we just saw that layered convection only spontaneously emerges when $R_0^{-1}$ is smaller than the layering threshold (which is $\sim \tau^{-1/2}$). As such, running simulations of layered convection for a wider range of $R_0^{-1}$ requires using much smaller values of $\tau$, which has so far been computationally prohibitive.

Figure \ref{fig:layerflux}b tests the second assumption of traditional models \cite{StevensonSalpeter1977,Spruit1992,LeconteChabrier2012,Spruit2013}, namely that $\gamma_{\rm tot}^{-1} = \tau^{1/2}$ \cite{LindenShirtcliffe1978}. We see that this prescription does not hold: $\gamma_{\rm tot}^{-1}$ is generally larger that $\tau^{1/2}$ (see also the results of Moll et al. \cite{Mollal2017}), having instead a complicated dependence on Pr, $\tau$ and $R_0^{-1}$, and perhaps even on the layer height (though there are hints that $\gamma_{\rm tot}^{-1}$ may become independent of $H_L$ for sufficiently tall layers). The discrepancy between these numerical results and the Linden and Shirtcliffe \cite{LindenShirtcliffe1978} model for the interfacial flux ratio is easy to understand. The interfaces in their model (and more generally in the high Prandtl number laboratory experiments of layered convection for which the model was created) are quiescent so the interfacial transport is purely diffusive. However, the interfaces in our low Prandtl number simulations are quite turbulent (see Figure \ref{fig:ODDCtypes}). This example nicely illustrates the risk of applying geophysically-derived parametrizations to stellar astrophysics without verifying them first with low Prandtl number numerical experiments.

Based on their limited data, Wood et al. \cite{Woodal13} proposed the following empirical parametrization for the compositional mixing coefficient:
\begin{equation}
D_{\rm semi} = ({\rm Nu}_C-1) \kappa_C \simeq 0.03 {\rm Pr}^{-0.12} {\rm Ra_\star}^{0.37} \kappa_T. 
\label{eq:dsemiwood}
\end{equation}
Note, however, that combining this with (\ref{eq:NutWood}) would imply that  
\begin{equation}
\gamma_{\rm tot}^{-1} = \tau R_0^{-1} \frac{{\rm Nu}_C}{{\rm Nu}_T} \simeq 0.3 R_0^{-1} , 
\label{eq:gammawood}
\end{equation}
(neglecting very weak powers of ${\rm Pr}$ and ${\rm Ra}_\star$ and assuming that ${\rm Nu}_C, {\rm Nu}_T \gg 1$). This is not really consistent with Figure \ref{fig:layerflux}b, where it appears that $\gamma_{\rm tot}^{-1}$ may be {\it decreasing} instead of increasing with $R_0^{-1}$. This discrepancy most probably comes from the fact that neither (\ref{eq:NutWood}) nor (\ref{eq:dsemiwood}) correctly account for the weak dependence of ${\rm Nu}_T$ and ${\rm Nu}_C$ (and in turn $D_{\rm semi}$) on $R_0^{-1}$. This problem can only be resolved by running simulations for a wider range of $R_0^{-1}$, which is computationally very challenging as discussed above.

Ultimately, however, the main uncertainty in modeling both heat and compositional fluxes in layered convection is their dependence on the layer height (i.e. with $F_{T,C} \sim H_L^{4/3}$). In all the simulations performed to date, the layers were ultimately seen to merge until a single one remains in the domain. This poses two problems. First, it has prevented us from proposing a model for what the layer height may be in a star, since it is not clear what mechanism (if any) eventually stops the merger process. Secondly, these results seem to challenge the original assumption of an equilibrium staircase, suggesting instead that layered convection may be a short temporary phase in the evolution of a star, whose duration depends on the layer merger rate.

\subsection{Conclusions for now}

Our numerical results and theoretical investigations have confirmed that there are two regimes of ODDC / semiconvection: one that is layered and one that is not layered. The layered form arises spontaneously from the $\gamma-$instability \cite{radko2003mechanism,rosenblumal2011,Mirouh2012} at sufficiently low inverse density ratio. The threshold for the layering instability to occur can be computed in the manner described in Section \ref{sec:layercrit} \cite{Mirouh2012}, and is approximately given by $R_0^{-1} \simeq \tau^{-1/2}$ at stellar (and planetary) parameters. Note that layered convection can also exist for $R_0^{-1}$ beyond that threshold, but only if layers are {\it seeded}, i.e. if they are present in the initial conditions applied to the problem (see \cite{Merryfield1995}, \cite{ZaussingerSpruit2013}, \cite{Mollal2017}). Whether this regime is relevant for stars or not remains to be determined, since one would have to specify why these layers should be present in the first place. 

Both turbulent heat and compositional transport in layered convection are very efficient, and can be computed as follows \cite{Woodal13}:
\begin{equation}
F_{{\rm heat,semi}} = \rho c_p F_T = \frac{ \rho c_p T}{H_p}  \kappa_T ({\rm Nu}_T - 1)\delta(\nabla_T - \nabla_{\rm ad}) , 
\label{eq:heatsemi}
\end{equation}
with 
\begin{equation}
{\rm Nu}_T - 1  \simeq C_{\rm semi} {\rm Ra}_\star^{1/3}  , 
\label{eq:NuTsemi}
\end{equation}
where ${\rm Ra}_\star$ depends on the layer height as $H_L^4$ (see equation \ref{eq:Rastar}), and $C_{\rm semi}$ is of the order of 0.1, with some possible weak dependence on $\tau$ and $R_0^{-1}$ that remains to be determined. What the layer height would be in a star, or whether the concept of an equilibrium staircase even applies in the first place, remains to be determined. 
Meanwhile, the turbulent mixing coefficient $D_C = D_{\rm semi}$ is 
\begin{equation}
D_{\rm semi}=  \gamma_{\rm tot}^{-1}  ({\rm Nu}_T - 1) \frac{ \kappa_T }{R_0^{-1}} , 
\label{eq:Csemi}
\end{equation}
where $\gamma_{\rm tot}^{-1}$ is of order one (but smaller than one), with some possible weak dependence on Pr, $\tau$ and $R_0^{-1}$ that also remains to be determined. 

In the non-layered regime, by contrast, we have demonstrated that the turbulent heat flux is negligible, and the turbulent compositional flux also appears to be negligible, at least for non-degenerate regions of stellar interiors where ${\rm Pr}$ and $\tau$ are asymptotically small. At more moderate ${\rm Pr}$ and $\tau$, the turbulent compositional flux could be significant, and can be determined from the model of Mirouh et al. \cite{Mirouh2012} presented in Section \ref{sec:nonlayeredtransport}, or by fitting their data empirically (which remains to be done).

 \subsection{Applications to stellar astrophysics}
 
The impact of these newly developed theories of mixing by ODDC (layered or non-layered) on stellar evolution have not really been studied in detail yet, so this section will be very limited in scope. 

As discussed in Section \ref{sec:whereODDC}, semiconvective regions are most commonly found just outside the convective cores of MS stars, in two mass ranges: intermediate-mass stars with $M_\star$ between about 1$M_\odot$ and 3$M_\odot$, and higher-mass stars with $M_\star > 10M_\odot$. In both cases, the typical inverse density ratios are well within the layered regime, with $R_0^{-1} \ll \tau^{-1/2}$. The turbulent transport of heat and composition can be substantial and depends on the unknown heights of the layers (which need not be constant in space nor time). With all this uncertainty, the best one can do (for now) is to fix the layer height to some fraction of a pressure scaleheight, and qualitatively ascertain what the effects of layered convection on the star's evolution are {\it as a function of the assumed layer height} .

It has long been known that one of the main effects of semiconvection in these stars is to prolong their lifetime on the MS by increasing the amount of hydrogen available for fusion reactions. Consequently, it also increases the size of the helium core at the end of the MS. Moore and Garaud \cite{MooreGaraud2016} attempted to quantify the effect, to determine more specifically how the helium core size of intermediate-mass stars varies with the assumed semiconvective layer height. To do so, Moore implemented both heat and compositional transport prescription by layered ODDC in MESA. The compositional transport is easy to implement because it merely requires computing the mixing coefficient $D_{\rm semi}$ with equation (\ref{eq:Dsemi}). Including a model for the turbulent heat transport on the other hand is much more complicated, since MESA computes the temperature profile within the star given the heat flux, instead of computing the heat flux given the temperature gradient (which is the way in which equation \ref{eq:heatsemi} is written). As such, including heat transport by layered semiconvection involves the numerical solution of a fourth-order polynomial in the quantity $\nabla$. Details of the implementation are given in Section 3.1.2 of \cite{MooreGaraud2016}. 

\begin{figure}[h]
\centering
\sidecaption
\includegraphics[width=0.5\textwidth]{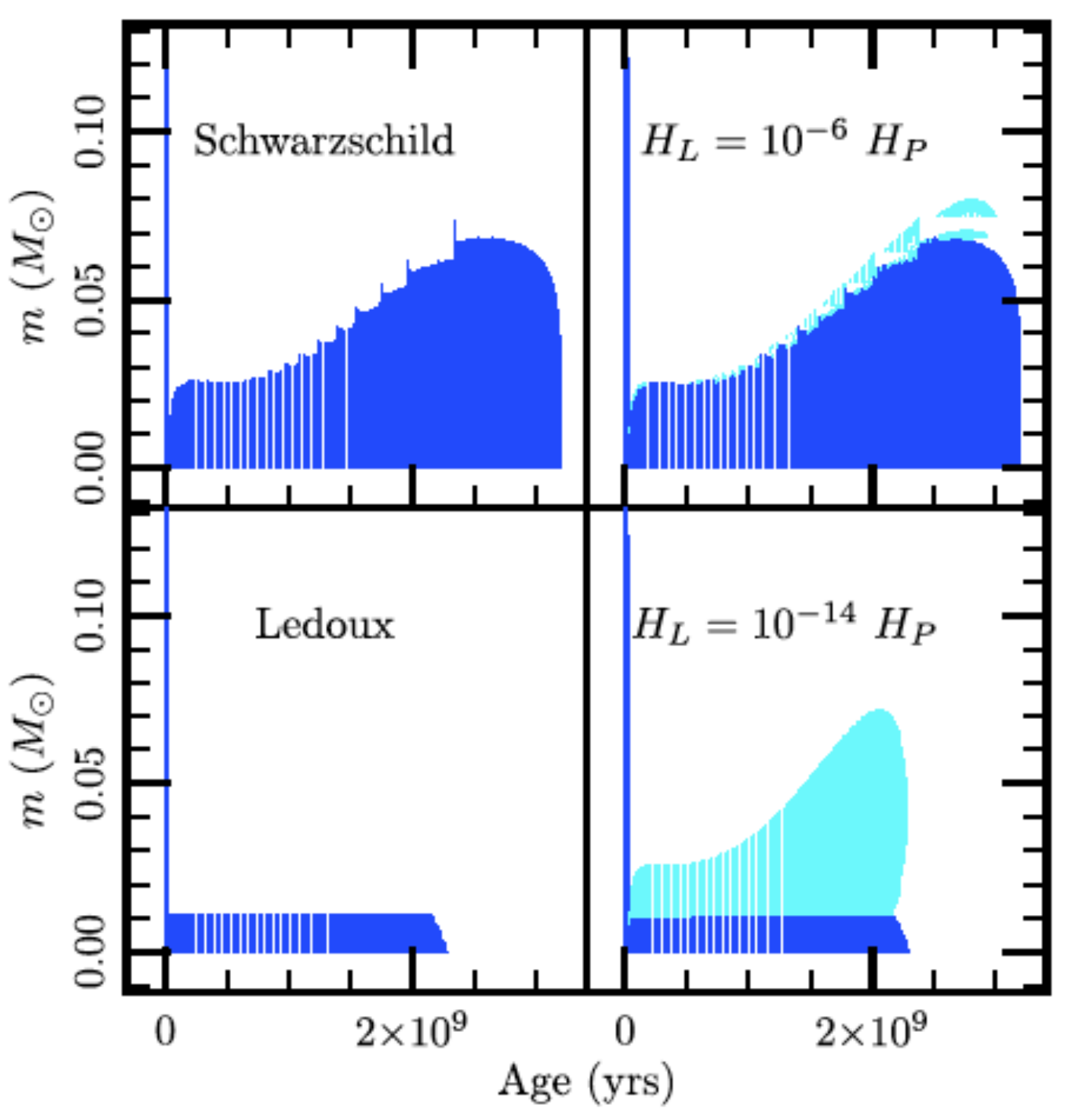}
\caption{Kippenhahn diagram for the evolution of a 1.3$M_\odot$ star under different scenarios. Left column: Models that ignore semiconvection, and use either the Schwarzschild (top) or Ledoux (bottom) criterion to compute the convective core size. Right column: Models that include layered semiconvection, with varying layer height, as shown in the figure. The dark blue regions are fully convective, while the light blue regions are semiconvective. Figure from \cite{MooreGaraud2016}.}\label{fig:moore}       
\end{figure}

The conclusions of our investigation on intermediate-mass stars are remarkably simple. We found that unless the height of the layers was assumed to be unphysically small, compositional mixing by layered convection was so efficient that it would erase any existing compositional gradient on a short timescale compared with the stellar evolution timescale. As a result, the semiconvective region rapidly becomes fully convective, and the star's core grows to be the size one would have predicted much more simply using the Schwarzschild criterion and ignoring semiconvection altogether. This is illustrated in Figure \ref{fig:moore}, which shows the predicted (fully) convective core size of a 1.3$M_\odot$ star, under four different scenarios: ignoring semiconvection and with either the Schwarzschild or Ledoux criteria to compute the core sizes, or, including layered convection and assuming that the layer height $H_L$ is either $10^{-6}$ or $10^{-14}$ times the pressure scaleheight $H_p$. We see that  with $H_L = 10^{-6} H_p$ the model already behaves almost as if the Schwarzschild criterion had been used instead, and it is only for vanishingly small layer heights ($H_L = 10^{-14} H_p$) that a substantial semiconvective region survives above the core throughout the star's evolution. 

These results are consistent with recent asteroseismic determinations of the convective core sizes in MS stars \cite{Silva2013,Deheuvels2016}, which reveal them to be even larger than what stellar evolution predicts using the Schwarzschild criterion. This implies not only that semiconvection is irrelevant (as we predicted), but also that substantial overshoot must be present. We therefore conclude that ignoring semiconvection and using the Schwarzschild criterion in these stars is a good approximation for stellar evolution purposes (for intermediate mass stars), which is particularly convenient given how technically difficult the implementation of layered convection is (for heat transport) in standard stellar evolution codes. 

Finally, I would like to add that MESA, to my knowledge, is the only stellar evolution code to date that includes the complete Wood et al. \cite{Woodal13} prescription for transport by layered convection for both heat and composition in semiconvective zones. I encourage readers to use it to see what other effects layered convection may have, in particular for more massive stars ($M_\star > 10 M_\odot$), or for stars in more evolved stages of evolution. 

\section{What next?}
\label{sec:ccl}

In the previous Sections, I have attempted to summarize what is known so far about double-diffusive instabilities and double-diffusive convection at low Prandtl number, which is the relevant limit for astrophysical fluids. However, there are a number of limitations of the model we have used to arrive at these results, which I would be remiss not to mention. Much also remains to be done in terms of modeling double-diffusive convection more realistically, including for instance the effects of rotation, magnetic fields, shear, and other relevant processes. I now discuss both of these in turn.

\subsection{Model limitations}

The numerical model used in the simulations presented in this lecture uses the Spiegel-Veronis-Boussinesq \cite{SpiegelVeronis1960} approximation, and triply-periodic boundary conditions, so it is important to determine when the limitations of this setup may begin to invalidate the results obtained. 

The Spiegel-Veronis-Boussinesq approximation is valid a long as two conditions are satisfied: (1) that the typical fluid velocities are much smaller than the sound speed $c_s$ and (2) that the region modeled is much smaller than a pressure scaleheight $H_p$. In Section \ref{sec:lineargeneral}, we learned that the typical size of basic double-diffusive structures (prior to the formation of layers, if they form) is of order $10 d = 10 (\kappa_T \nu/|N_T^2|)^{1/4}$ and provided some estimates for that quantity. We also saw that the structures remain small even in the nonlinear regime (again, prior to layer formation), so $10d$ is a good estimate for the typical eddy size in small-scale homogeneous double-diffusive convection (of both fingering and ODDC type). In terms of the typical velocity of fingering convection and ODDC, we saw that it can be estimated using $\lambda d$, where the growth rate $\lambda$ is at most equal to $N_T$, but usually much smaller (depending on the regime considered). It is clear that well within the interior of a star, we always have $10d \ll H_p$ and $N_T d \ll c_s$. Closer to the stellar surface, however, especially for intermediate and high-mass stars, $\kappa_T$ increases rapidly, and so will $d$. Meanwhile, the pressure scaleheight $H_p$ and the sound speed $c_s$ both decrease, so it is conceivable that fingering convection or ODDC in regions sufficiently close to the surface of these  stars may need to be modeled more realistically including the effects of compressibility. 

Once layers form in ODDC, the typical size of turbulent structures increases to be the entire layer height, and the turbulent velocities also increase to take typical convective (rather than double-diffusive) values. If the layers grow to a substantial fraction of a pressure scaleheight, the effects of compressibility should therefore also be taken into account.  

Finally, note that the Boussinesq approximation also assumes that the diffusivities are constant within the domain. This could be a limitation of the model when $\nu$, $\kappa_T$ or $\kappa_C$ depend sensitively on temperature or composition. 

Another set of caveats of the results comes more specifically from the use of the triply-periodic boundary conditions for the perturbations in our DNSs. The first is relatively well understood, and pertains to the relative size of the double-diffusive structures modeled compared with the computational domain size. If the structures are much smaller than the domain size (in all directions), then the latter has no effect on the double-diffusive dynamics. This is usually the case in the homogeneous regime of fingering convection and ODDC. On the other hand when the double-diffusive structures approach the domain size, it is quite likely that the results become dependent on the boundary conditions. This is the case for instance in layered convection, when there is only a single layer left. This is also the case of rapidly rotating double-diffusive convection and magnetized fingering in the very strong field regime, where the dynamics become invariant along the rotation axis and the magnetic field, respectively. When this happens, the measured fluxes can be artificially enhanced \cite{MollGaraud2017,SenguptaGaraud2018,HarringtonGaraud2019}. To control this problem, it is important to run additional DNSs in increasingly large domains until a point where the results become independent of the domain size. In all the simulations presented in these lectures, this was always verified. 

A second problem associated with triply-periodic boundary conditions is more subtle, and requires care in interpreting the turbulent flux data. In our computational model setup, the background potential temperature and composition gradients are fixed, and the developing instabilities drive turbulent heat and composition fluxes. Crucially, there is nothing in the system that limits these fluxes beyond what drives them. However, in a real star the situation is somewhat different because the total heat flux that needs to be transported through a given shell of material is fixed by the nuclear reaction rates taking place within or below. In that case, we anticipate that (on long timescales) the double-diffusive dynamics may adjust themselves to carry the required flux, rather than the flux being solely controlled by the double-diffusive dynamics. The problem is not significant when the double-diffusive dynamics transport a negligible amount of heat (which is the case of homogenous fingering convection and ODDC), but needs to be born in mind for layered convection and magnetized fingering convection with strong fields, where heat transport can be substantial (see below). These types of problems are then perhaps best studied in a model setup that can impose a fixed heat flux through the domain.

\subsection{More realism}

In the past few years, our group has begun to investigate the effects of other physical processes on the dynamics of double-diffusive convection, including those of lateral gradients of potential temperature and composition \cite{Medrano2014}, rotation \cite{MollGaraud2017,SenguptaGaraud2018} magnetic fields \cite{HarringtonGaraud2019}, and shear \cite{Garaudal2019}. Given the dramatic increase in the dimensionality of parameter space that needs to be explored, our results are in most cases preliminary, and much remains to be done. This section briefly summarizes the state of the field as of May 2019. 

\subsubsection{Horizontal gradients} 

The effects of horizontal gradients have been investigated in the context of fingering-like stratifications only \cite{Medrano2014} (the ODDC-like case has never been looked at, to my knowledge), and affect the results described in Section \ref{sec:fingering} only in the limit of very strong stratification, i.e. for density ratios approaching marginal stability or exceeding it. Horizontal gradients greatly increase the region of parameter space that is unstable \cite{Holyer1983,Medrano2014}, to the extent that almost {\it any} inverse $\mu$ gradient can cause vertical mixing. However that instability is slowly growing, and does not appear to cause very strong enhancements in the turbulent mixing coefficient above the purely diffusive value. This conclusion was obtained from simulations at very moderate Prandtl number and diffusivity ratio, and it remains to be seen whether it holds for more stellar-like parameter values. 

\subsubsection{Rotation}

The effects of rotation have been investigated in preliminary work by Moll and Garaud \cite{MollGaraud2017} for ODDC, and Sengupta and Garaud \cite{SenguptaGaraud2018} for fingering convection. In both cases, rotation does not affect the overall instability range, nor the growth rate of the fastest-growing modes of instability. Whether rotation has a weak or strong effect on the nonlinear saturation of double-diffusive instabilities can be determined by computing the turbulent Rossby number, Ro $= u_{\rm rms} / \Omega d$ where $u_{\rm rms}$ is the typical velocity of turbulent eddies, and $\Omega$ is the rotation rate. Rotation is important for Rossby numbers of order one or much smaller than one, but is negligible when Ro $\gg 1$. At moderate ${\rm Pr}$ and $\tau$ (appropriate for giant planets, or degenerate regions of stellar interiors), the turbulent velocities can be estimated roughly using $\kappa_T / d$, so the Rossby number is Ro = $\kappa_T / \Omega d^2 = {\rm Pr}^{-1/2} |N_T| / \Omega$ \cite{MollGaraud2017}. With that assumption and the fact that ${\rm Pr} \sim 0.01$ in these objects, we see that the effects of rotation will only be negligible if $\Omega \ll 10 |N_T|$, roughly speaking. In stellar interiors, the turbulent velocities associated with the basic instability are typically much smaller, of the order of $(\kappa_T / d)({\rm Pr}/R_0)^{1/2}$ for fingering convection and  $(\kappa_T / d)({\rm Pr}R_0)^{1/2}$ for non-layered ODDC. As a result, the Rossby number is of the order of Ro = $R_0^{-1/2}  |N_T| / \Omega$ for fingering convection and Ro = $(R_0^{-1})^{-1/2}  |N_T| / \Omega$ for non-layered ODDC. In other words, at very low ${\rm Pr}$ and $\tau$ and for sufficiently strongly stratified systems (i.e. large $R_0$ for fingering, and large $R_0^{-1}$ for ODDC), the Rossby number can be of order one or even smaller and rotation could be important. For layered ODDC, the situation is a little different since the estimate for the turbulent velocities $u_{\rm rms}$ must be replaced by an estimate for convective velocities, that depends on the assumed layer height. 

\begin{figure}[h]
\centering
\includegraphics[width=0.7\textwidth]{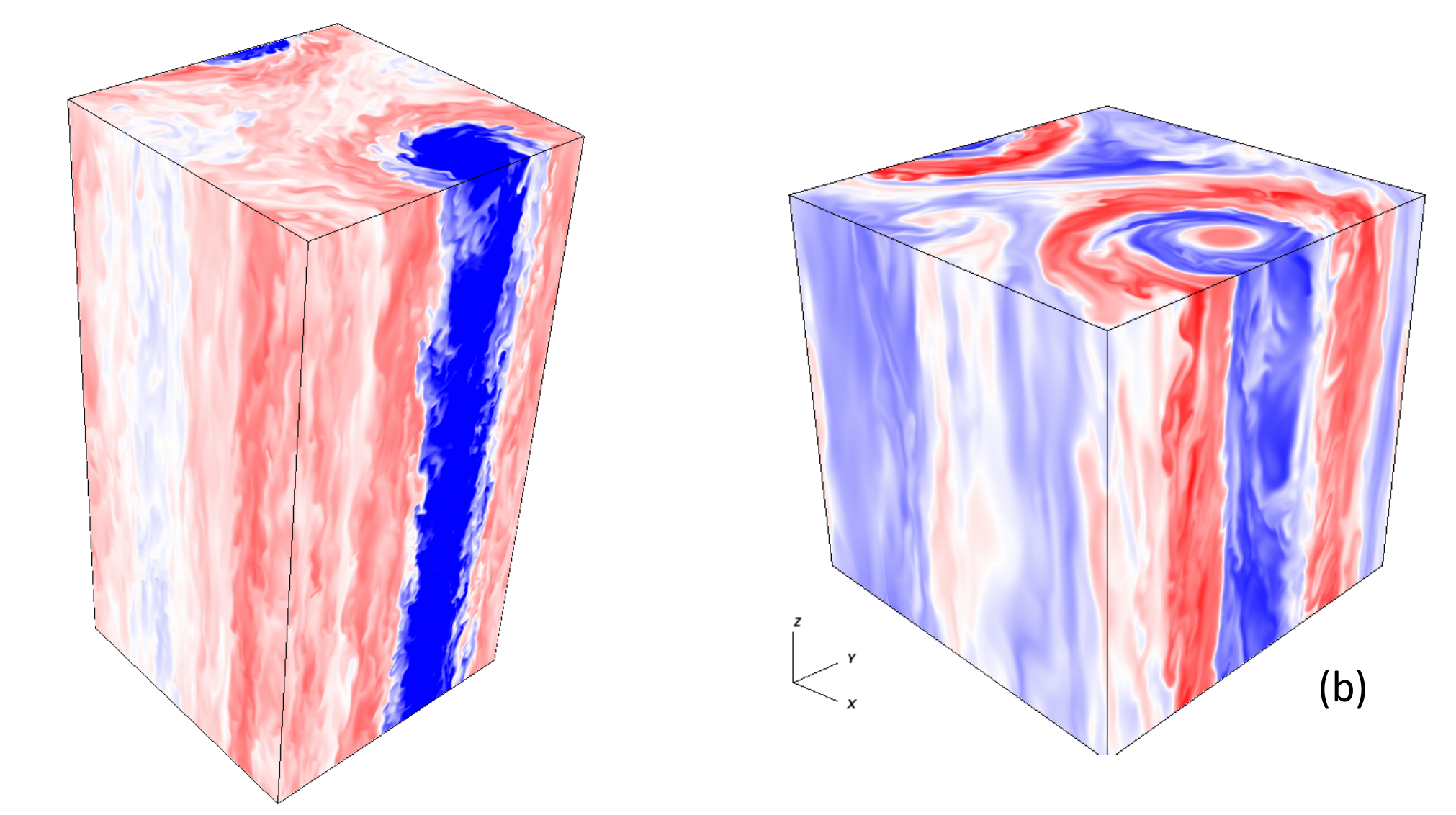}
\caption{Left: Illustrative snapshot of a large-scale vortex formed in a fingering simulation at ${\rm Pr} = \tau = 0.1$, and $R_0 = 1.45$, with ${\rm Ta}_* = 4 \Omega^2 d^2 / \kappa_T = 10$. This is the compositional field perturbation, which shows that the vortex core is low $C$, and flows upwards. Figure adapted from \cite{SenguptaGaraud2018}. Right: Illustrative snapshot of a large-scale vortex formed in an ODDC simulation at ${\rm Pr} = \tau = 0.1$, and $R_0^{-1} = 1.25$, with ${\rm Ta}_* = 10$. This time the vortex core is high $C$, and flows downwards, but has a more complex radial structure that remains to be explained. Figure adapted from \cite{MollGaraud2017}, }\label{fig:LSV}       
\end{figure}

Most of our numerical simulations of rotating double-diffusive convection to date have been performed at the poles (where the rotation axis is aligned with gravity), with a few cases only looking at mid-latitudes. We have found that weak rotation (with Ro $\gg 1$) tends to reduce the vertical compositional flux slightly compared with non-rotating simulations, in both fingering and ODDC. By contrast, when the Rossby number is of order unity, we found that rotation causes the formation of large-scale vortices that span the entire domain in both vertical and horizontal directions\footnote{Note that they have only been seen in  polar cases so far}. These vortices have a tendency to cause a segregation of the compositional field in their core, either expelling it or concentrating it (see Figure \ref{fig:LSV}) depending on the initial conditions selected. Denser cores drive stronger downward velocities, while lighter cores drive stronger upward velocities. This increase in the correlation between $w$ and $C$ causes a substantial increase in the vertical compositional flux compared with a non-rotating simulation at otherwise similar parameters. The effect is very substantial, and could potentially be invoked as a solution to the missing mixing problem in RGB stars (see Section  \ref{sec:fingapp}) \cite{SenguptaGaraud2018}. However, why these vortices form, and whether they would also form in stars (rather than in idealized simulations), remains to be determined. In particular, based on our results, and similar results obtained for rotating Rayleigh-B\'enard convection \cite{Guervilly2014,Favier2014}, it is not clear that they would form at lower latitudes \cite{MollGaraud2017}, or in computational domains that have different horizontal dimensions (i.e. $L_x \neq L_y$) \cite{Julien2018}. 

Finally, for rotationally dominated systems (with Ro $\ll 1$), large-scale vortices do not form. Instead, the dynamics of both fingering and ODDC are dominated by narrow tubular structures that are invariant along the axis of rotation. In that case, the periodicity of the domain becomes a limitation, and it is not clear that the numerical results obtained are reliable (see previous section). More remains to be done, using perhaps a different formalism (or much taller periodic domains) to characterize ODDC and fingering convection in that limit.

\subsubsection{Magnetic fields}

As discussed earlier, Charbonnel and Zahn \cite{CharbonnelZahn2007b} and Stevenson \cite{Stevenson1979} investigated the linear stability of magnetized fingering convection and magnetized ODDC, respectively, but very little is known so far about the effects of magnetic fields on double-diffusive convection in the nonlinear regime.

\begin{figure}[h]
\centering
\includegraphics[width=\textwidth]{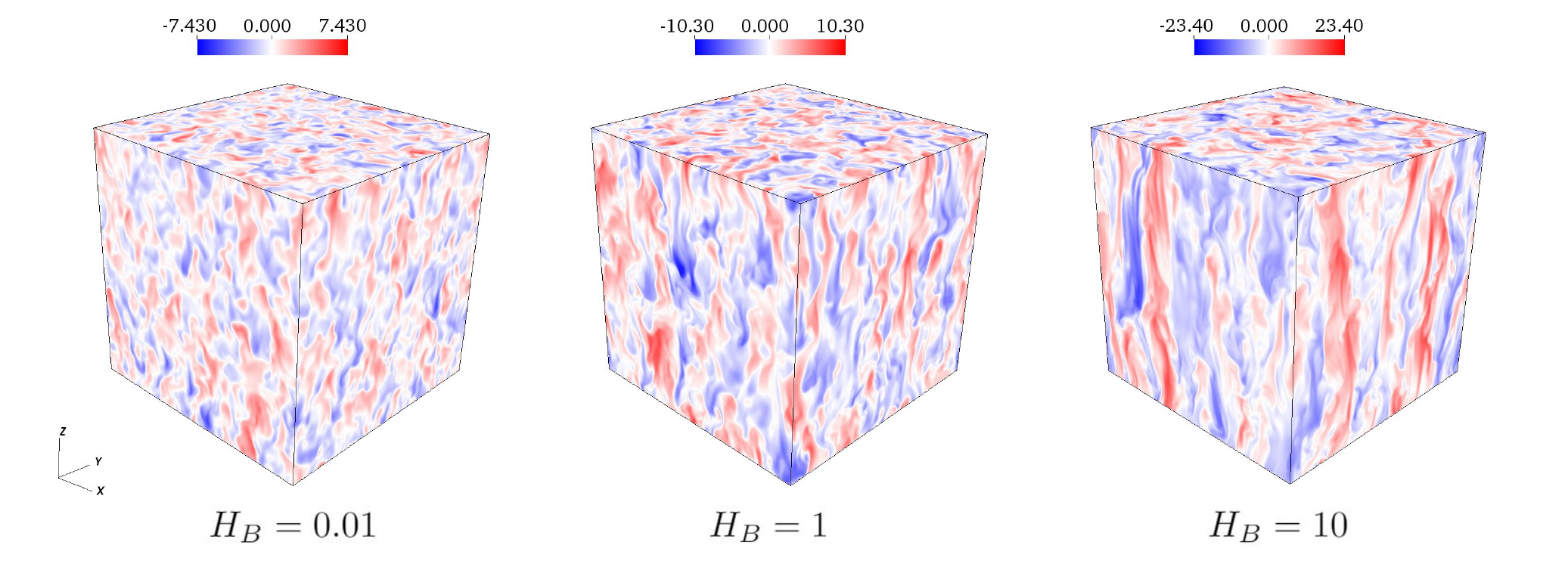}
\caption{Snapshots of the vertical velocity field in simulations of magnetized fingering convection with a vertical magnetic field, at ${\rm Pr} = \tau = 0.1$ and $R_0 = 1.45$. The vertical coherence of the fingers increases with the magnetic field strength (as measured by $H_B$), and so does the vertical velocity. Figure from \cite{HarringtonGaraud2019}.}\label{fig:magsnaps}       
\end{figure}

Beyond linear theory, the first numerical study of magnetized fingering convection has just been completed by Harrington and Garaud \cite{HarringtonGaraud2019}, in a model that assumes the presence of a {\it vertical}\footnote{The study of arbitrarily inclined fields is in progress.} large-scale background field of amplitude $B_0$. The impact of the field can be characterized by a new non-dimensional number, namely
\begin{equation}
H_B = \frac{B_0^2 d^2}{\rho_m \mu_0 \kappa_T^2}, 
\end{equation}
which is the square of the ratio of the Alfv\'en velocity associated with that field, to the selected unit velocity $\kappa_T / d$ (see Section \ref{sec:lineargeneral}).  Note however that since $\kappa_T / d$ is not always a good estimate for the actual fingering velocities (see Section \ref{sec:fingering}), especially at very low Prandtl numbers, $H_B \ll 1$ does not necessarily imply that the magnetic field is negligible (see below for more on this topic).
We have run a number of DNSs of fingering convection that include a vertical magnetic field, varying the quantity $H_B$, with otherwise fixed parameter values. We have found that vertical transport can be increased by orders of magnitude compared with non-magnetic simulations, because the field stabilizes the fingers partially against the parasitic shear instabilities that normally destroy them (see Section \ref{sec:fingering}).  As a result, larger vertical velocities can be achieved before the shear instability causes saturation (see Figure \ref{fig:magsnaps}), which in turn enhances the vertical fluxes. 

Our results can be quantitatively explained by a simple extension of the Brown et al. \cite{Brownal2013} model, that contains no additional free parameter. Recall that the mixing coefficient $D_{\rm fing}$ can be estimated using equation (\ref{eq:Dfingw}), where $K_B \simeq 1.24$ and where $\hat w_f$ is the anticipated vertical velocity within the finger at saturation obtained by matching the finger growth rate $\hat \lambda$ to the parasitic shear instability growth rate $\hat \sigma$. The latter decreases in the presence of a magnetic field \cite{Chandrasekhar1961}, because the field rigidifies the fingers. This effect can be captured using the approximate expression (see \cite{HarringtonGaraud2019} for detail): 
\begin{equation}
\hat \sigma \simeq 0.42 \hat w_f \hat k_h (0.5 - H_B \hat w_f^{-2})^{2/3} .
\end{equation}
This equation recovers the hydrodynamic limit studied by Brown et al. \cite{Brownal2013} when $H_B = 0$, as required. It also shows that when $H_B \neq 0$ the fingering elevator modes are unaffected by parasitic shear instabilities until their vertical velocities reach a threshold value of $\hat w_f = \sqrt{2H_B}$ (which corresponds to energy equipartition). Beyond that threshold, the finger velocities continue to grow until $\hat \sigma \simeq C_H \hat \lambda$, where $C_H \simeq 1.66$ is not a free parameter, but instead, is fixed by the requirement that our theory recovers the hydrodynamic limit when $H_B = 0$ (see \cite{HarringtonGaraud2019} for detail). 

All that remains is to solve the algebraic equation $C_H \hat \lambda = 0.42 \hat w_f \hat k_h (0.5 - H_B \hat w_f^{-2})^{2/3}$ for $\hat w_f$ numerically, then substitute the result in equation (\ref{eq:Dfingw}) to compute the mixing coefficient $D_{\rm fing}$. Note that for large $H_B$, we find that $\hat w_f \simeq \sqrt{2H_B}$ so using this expression in (\ref{eq:Dfingw}) yields
\begin{equation}
D_{\rm fing} \simeq \frac{2K_B H_B}{\hat \lambda + \tau \hat k_h^2} \kappa_T , 
\end{equation}
which recovers the aforementioned equation (\ref{eq:DfingB}) when $K_B \simeq 1.24$, $\tau \ll 1$, and if we use the approximation $\hat \lambda \simeq \sqrt{{\rm Pr}/(R_0-1)}$ (see equation \ref{eq:lambdaapprox}). 

Figure \ref{fig:Harrington} compares the new model for magnetized fingering convection with results of our DNSs including the vertical field, at moderate parameter values (${\rm Pr} =\tau = 0.1$, $R_0 = 1.45$), and confirms that it correctly predicts the variation of $D_{\rm fing}$ with magnetic field strength as $H_B$ increases. Applying the model to more stellar-like conditions (where the fingering growth rate is much smaller), we predict that magnetic fields become important for $H_B \sim 
10^{-4}$ in WDs, and $10^{-8}$ in RGB stars. With a very moderate magnetic field of the order of 300G, $D_{\rm fing}$ increases by two orders of magnitude compared with the non-magnetic case for both types of stars (see Figure \ref{fig:Harrington}). This strongly suggests that magnetic fields should be taken into account when modeling fingering convection in stars, and could be the answer the RGB star abundance conundrum described in Section \ref{sec:fingapp}.

 \begin{figure}[h]
\centering
\sidecaption
\includegraphics[width=0.5\textwidth]{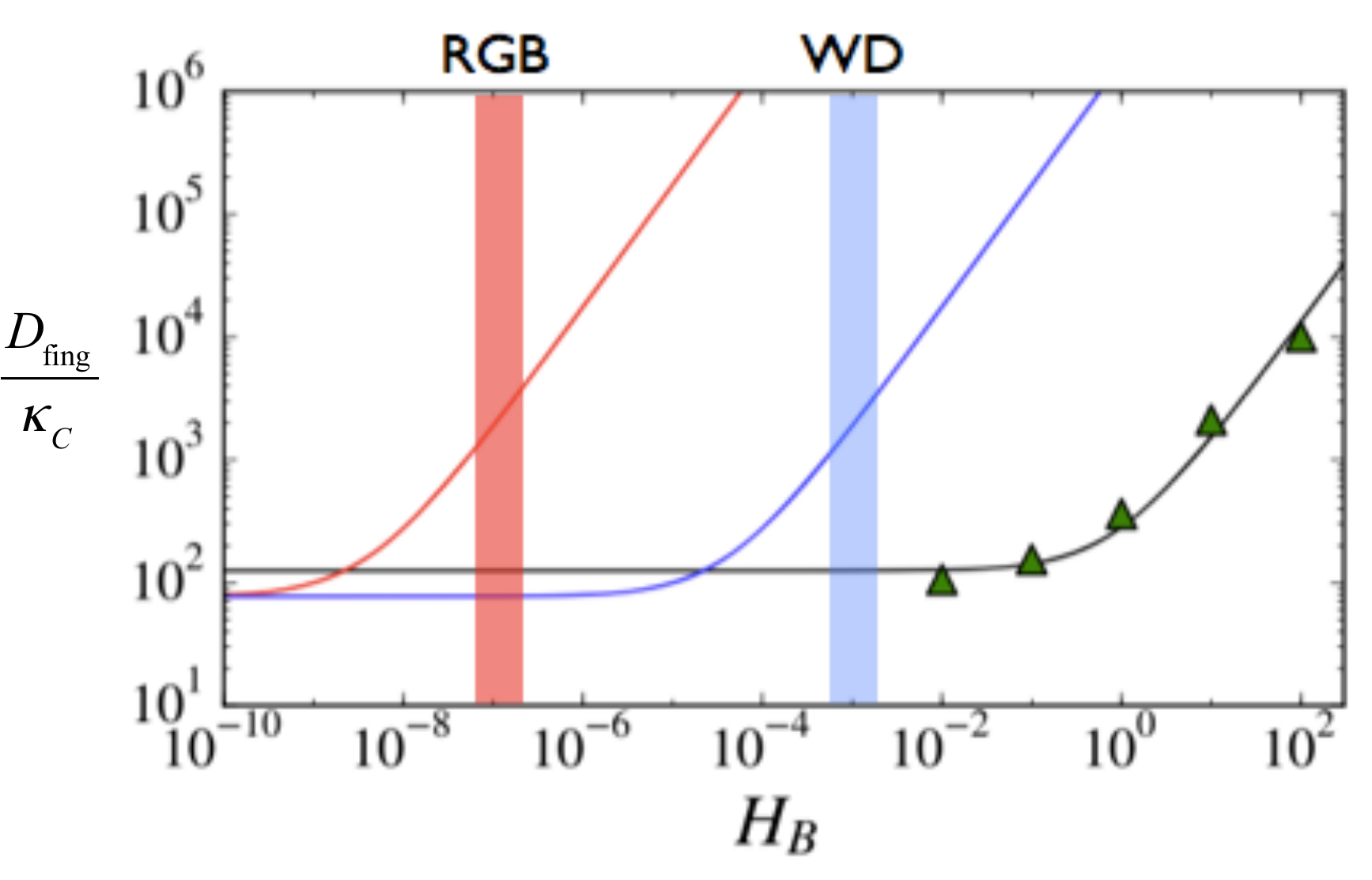}
\caption{$D_{\rm fing}/\kappa_C$ in simulations (symbols) and models (lines). The simulations were run at ${\rm Pr} =\tau = 0.1$ and $R_0 = 1.45$, and the model corresponding to these parameters is the black line. The red and blue lines are predictions for RGB and WD stars respectively, see \cite{HarringtonGaraud2019} for detail. The shaded areas are values of $H_B$ corresponding to $B_0 \sim 100$G. Figure adapted from \cite{HarringtonGaraud2019}. }
\label{fig:Harrington}       
\end{figure}

Of course, this conclusion remains to be tested in the more general case where the magnetic field is inclined with respect to gravity. Preliminary results suggest that transport is also increased in that case, but new effects also appear (such as a dynamo, see \cite{HarringtonGaraud2019}  and Harrington's MS thesis), whose impact needs to be better understood. 

%
%

\subsubsection{Shear}

The final physical process whose impact on {\it stellar} double-diffusive convection had not yet been investigated yet is shear, though numerical and experimental studies of sheared double-diffusive convection in geophysics exist, such as for instance \cite{Linden1974,KimuraSmyth2007,SmythKimura2010,Radkoal2015} for the fingering case, and \cite{Radko2016,BrownRadko2019} for ODDC. 

Although sheared ODDC in stars has not been studied yet, the effect of a moderate shear on fingering convection is now relatively well understood \cite{Garaudal2019}. While shear tends to suppress the development of fingering structures that vary in the streamwise direction, it has no effect on streamwise invariant perturbations. As a result, the initial instability takes the form of sheets aligned with the flow, rather than fingers \cite{Linden1974}. In the nonlinear regime, however, these sheets rapidly break up into 3D fingers, that are tilted in the direction of the shear. That tilt has two effects: to reduce the horizontal lengthscale of the fingers (by squeezing them together), and to add a horizontal component to the flow within the fingers. The Brown et al. \cite{Brownal2013} model for mixing by fingering convection can easily be extended to take both effects into account \cite{Garaudal2019}, and ultimately predicts that the mixing coefficient $D_{\rm fing}$ must be corrected to account for shear in the following way:
\begin{equation}
D_{\rm fing}(S) = D_{\rm fing}(S=0) \left( 1 + \chi^2 \left(\frac{S}{\lambda}\right)^2 \right)^{-2} , 
\label{eq:Dfingshear}
\end{equation}
where $D_{\rm fing}(S=0) $ is the value of $D_{\rm fing}$ in the absence of shear, $S$ is the local shearing rate, $\lambda$ is the growth rate of fastest-growing unsheared fingers, and $\chi$ is a number of order unity\footnote{In the simulations we have performed, which have a sinusoidal shear flow, $\chi = 1/3$, though this value could be different for linear shear flows.}. This prescription holds only for moderate shearing rates, i.e. $S  < 3\lambda$ or so. Beyond that, the shear itself begins to contribute to the vertical mixing as expected, but no quantitative model of the process exists (so far).

 \begin{figure}[h]
\centering
\sidecaption
\includegraphics[width=\textwidth]{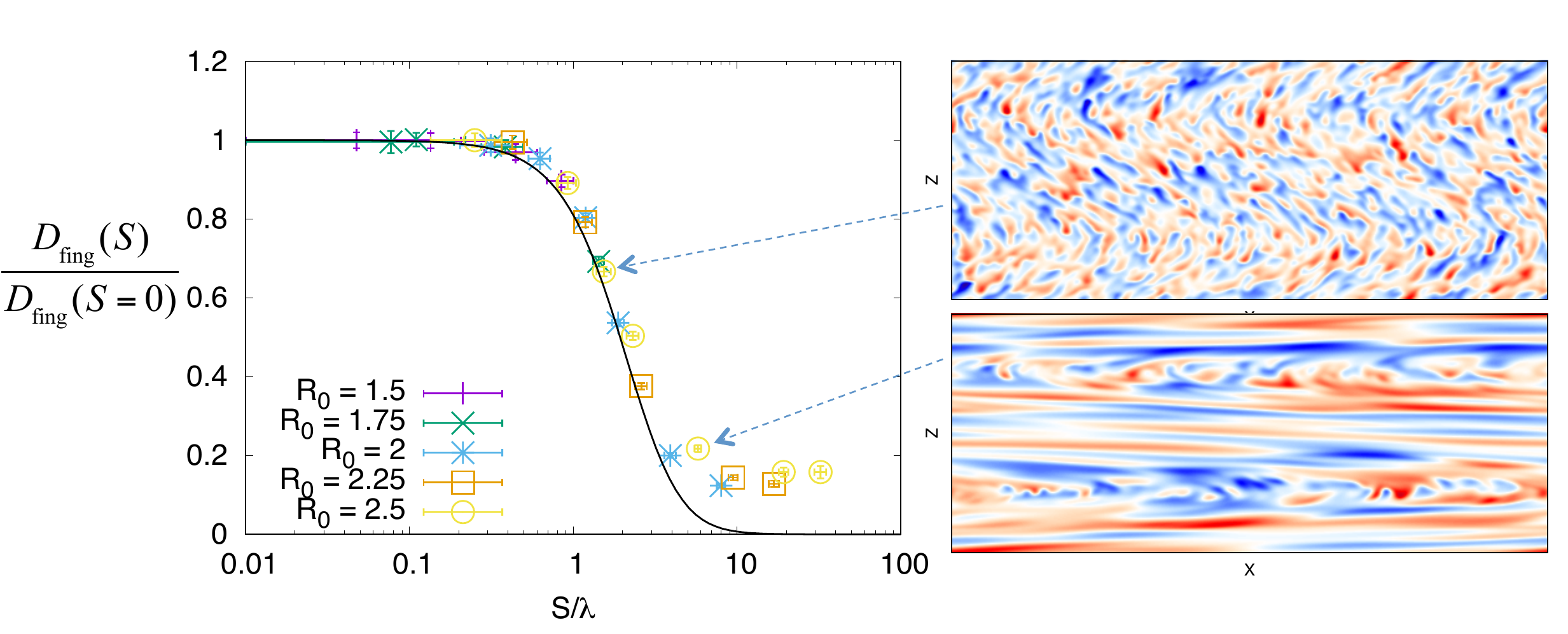}
\caption{Right: Illustrative snapshots of the compositional perturbations in sheared fingering convection, in two simulations where the background shear is sinusoidal. Parameters are ${\rm Pr} = \tau = 0.3$, and $R_0 = 2.5$ for both of them. The shearing rate is about 5 times larger in the bottom figure than in the top figure. Left: Comparison of the model prediction from equation (\ref{eq:Dfingshear}),  (black line) and the data, for $D_{\rm fing}(S) / D_{\rm fing}(S=0)$, as a function of $S / \lambda$ where $S$ is the maximum shearing rate in the domain. Figure adapted from \cite{Garaudal2019}. }
\label{fig:Harrington}       
\end{figure}

Finally, sheared fingering convection can also contribute to substantial momentum transport. We have found empirically that $\nu_{\rm turb} \simeq 0.25 D_{\rm fing}$ in the limit where $S < 3\lambda  $, growing to $\nu_{\rm turb} \simeq D_{\rm fing}$ for larger shearing rates. This can be used to inform models of angular momentum transport due to fingering convection in stars, which may (or may not) be useful in modeling the observed evolution of their interior rotation rates, as in \cite{Marques2013} for example.
\\
\\
{\it Acknowledgements:} The majority of this work was funded by the National Science Foundation through various grants over the past 10 year, including  NSF CBET-0933759, NSF AST-1211394,  NSF AST-1412951 and NSF CBET-1437275. These lectures were written up while on sabbatical at the Department of Applied Mathematics and Theoretical Physics at the University of Cambridge, supported by a visiting by-fellowship from Churchill College. I thank both institutions for their welcoming support during my visit. I would like to thank all my students, postdocs and collaborators on these projects, including Justin Brown, Nic Brummell, Jonathan Fortney, Peter Harrington, Anuj Kumar, Chris Mankovich, Michael Medrano, Giovanni Mirouh, Ryan Moll, Kevin Moore, Timour Radko, Erica Rosenblum, Sutirtha Sengupta, Josiah Schwab, Stephan Stellmach, Adrienne Traxler, Sylvie Vauclair, Toby Wood, and Varvara Zemskova. It has been fun and inspiring to work with them. I also thank Christopher Tout for his invaluable feedback on this manuscript and the organizers of the meeting, Michel Rieutord and Isabelle Baraffe for asking me to write these notes, which I hope will be useful to others. Finally, my special thanks go to Timour Radko, who inspired me to start working on this topic in the first place, Nic Brummell, with whom I always enjoy bouncing ideas, and to Stephan Stellmach, who developed the PADDI code that is at the heart of all this research.

\bibliographystyle{woc}
\bibliography{DDClowPr_references.bib}

\end{document}